\documentclass[sn-mathphys,Numbered]{sn-jnl}

\usepackage{graphicx}%
\usepackage{multirow}%
\usepackage{amsmath,amssymb,amsfonts}%
\usepackage{amsthm}%
\usepackage{mathrsfs}%
\usepackage[title]{appendix}%
\usepackage{xcolor}%
\usepackage{textcomp}%
\usepackage{manyfoot}%
\usepackage{booktabs}%
\usepackage{algorithm}%
\usepackage{algorithmicx}%
\usepackage{algpseudocode}%
\usepackage{listings}%
\usepackage{longtable}
\usepackage{natbib}
\usepackage{anyfontsize}
\usepackage[justification=Justified, font=scriptsize]{caption}

\theoremstyle{thmstyleone}%
\theoremstyle{thmstyletwo}%

\theoremstyle{thmstylethree}%

\begin{document}

\title[The X-IFU for NewAthena]{The X-ray Integral Field Unit at the end of the Athena reformulation phase}

\author[1]{\fnm{Philippe} \sur{Peille}}
\author[2]{\fnm{Didier} \sur{Barret}}
\author[1]{\fnm{Edoardo} \sur{Cucchetti}}
\author[1]{\fnm{Vincent} \sur{Albouys}}
\author[3]{\fnm{Luigi} \sur{Piro}}
\author[4]{\fnm{Aurora} \sur{Simionescu}}
\author[5]{\fnm{Massimo} \sur{Cappi}}
\author[1]{\fnm{Elise} \sur{Bellouard}}
\author[1]{\fnm{Céline} \sur{Cénac-Morthé}}
\author[1]{\fnm{Christophe} \sur{Daniel}}
% ...
\affil[1]{Centre National d'Etudes Spatiales, Centre spatial de Toulouse, 18 avenue Edouard Belin, 31401 Toulouse Cedex 9, France}
\affil[2]{Institut de Recherche en Astrophysique et Planétologie, Université de Toulouse, CNRS, UPS, CNES 9, Avenue du Colonel Roche, BP 44346, F-31028, Toulouse Cedex 4, France}
\affil[3]{INAF - Istituto di Astrofisica e Planetologia Spaziali, Via Fosso del Cavaliere 100, 00133, Roma, Italy}
\affil[4]{SRON, Netherlands Institute for Space Research, Niels Bohrweg 4, 2333 CA Leiden, The Netherlands}
\affil[5]{INAF, Osservatorio di Astrofisica e Scienza dello Spazio, via Gobetti 93/3, 40129, Bologna, Italy}
% ...




\abstract{The Athena mission entered a redefinition phase in July 2022, driven by the imperative to reduce the mission cost at completion for the European Space Agency below an acceptable target, while maintaining the flagship nature of its science return. This notably called for a complete redesign of the X-ray Integral Field Unit (X-IFU) cryogenic architecture towards a simpler active cooling chain. Passive cooling via successive radiative panels at spacecraft level is now used to provide a 50\,K thermal environment to an X-IFU owned cryostat. 4.5\,K cooling is achieved via a single remote active cryocooler unit, while a multi-stage Adiabatic Demagnetization Refrigerator ensures heat lift down to the 50\,mK required by the detectors. 
Amidst these changes, the core concept of the readout chain remains robust, employing Transition Edge Sensor microcalorimeters and a SQUID-based Time-Division Multiplexing scheme. Noteworthy is the introduction of a slower pixel. This enables an increase in the multiplexing factor (from 34 to 48) without compromising the instrument energy resolution, hence keeping significant system margins to the new 4\,eV resolution requirement. This allows reducing the number of channels by more than a factor two, and thus the resource demands on the system, while keeping a 4' field of view (compared to 5' before). 
In this article, we will give an overview of this new architecture, before detailing its anticipated performances. Finally, we will present the new X-IFU schedule, with its short term focus on demonstration activities towards a mission adoption in early 2027.}




\maketitle

\section{Introduction}\label{sec1}

The Athena X-ray Integral Field Unit (X-IFU) is the high resolution X-ray spectrometer \citep{Barret_2023ExA....55..373B}, studied since mid-2014 for operating in the mid-30s aboard the Athena space X-ray observatory of the European Space Agency (ESA). Athena is a versatile facility designed to address the Hot and Energetic Universe science theme \cite{2013arXiv1306.2307N,Barret_2013sf2a.conf..447B,barcons2015JPhCS.610a2008B,Barcons2017AN....338..153B,Barret2020AN....341..224B} that was selected in November 2013 by ESA's Survey Science Committee. For previous incarnations of X-IFU, when the cryostat and cooling chain were in the instrument Consortium perimeter (by opposition to being ESA/primes responsibilities), see \cite{Barret_2013arXiv1308.6784B,Ravera_2014SPIE.9144E..2LR,Barret_2016SPIE.9905E..2FB,Barret_2018SPIE10699E..1GB,Pajot_2018JLTP..193..901P}. X-IFU entered its System Requirement Review (SRR) in June 2022, at about the same time when ESA called for a mission redefinition (which eventually lead to a redesign of the X-IFU cryostat and the cooling chain), due to an unanticipated cost overrun of Athena at ESA. 

Along the reformulation of the Athena mission, the reduction of the X-IFU performances (e.g. spectral resolution, field of view, count rate capabilities) was scrutinized by the Athena Science Redefinition Team (SRDT), in view of the science objectives that had been defined previously for Athena, but also regarding some developments in the field over the last decade. It was found that most, of the original science goals would not be largely compromised, although in some cases, longer exposure times would be required to reach the same sensitivity limit. This obviously contributed to maintain the flagship status of NewAthena\footnote{Post reformulation name of the Athena mission.}. Building upon the scientific objectives of Athena \cite{2013arXiv1306.2307N} not repeated here, the SRDT promoted ten driving science goals for X-IFU, as listed below\footnote{Extracted from a presentation given by Mike Cruise and Matteo Guainazzi at the ESA Astronomy Working Group, as co-chairs of the SRDT in conclusion of the reformulation exercise. A publication of the SRDT is submitted to peer review, and is not made public at the time of this writing.} :
\begin{itemize}
    \item Determine how Active Galactic Nuclei (AGN) feedback affects its host galaxy,
    \item Determine the nature of the primary X-ray source in AGN and stellar compact objects,
    \item Probe the interstellar medium of high-redshift galaxies ($z\ge 7$), using gamma-ray bursts as background sources,
    \item Quantify how AGN feedback and gravitational energy are dissipated into bulk motions and turbulence of the hot gas in the galaxy clusters from the core to the halos,
    \item Map the cosmic baryons and probe their evolution and connection with the cosmic web,
    \item Quantify the chemical enrichment of the intracluster medium over cosmic time from the epoch of structure formation,
    \item Determine if and how the coronal activity of a star affects its planets,
    \item Constrain supernovae explosion mechanisms, progenitors and neutrino models,
    \item Probe particle acceleration in supernovae remnants.
\end{itemize}
These science goals are now translated into performance requirements for the X-IFU, through the elaboration of Science Requirement Document for the mission.

\begin{table}
\centering
\resizebox{\linewidth}{!}{
\begin{tabular}{| l | l |l |}
\hline
\textbf{Key performance parameter} & \textbf{Value and comments} & \textbf{Pre-rescope value} \\
\hline
Energy range & 0.2-12\,keV & Same\\
\hline
Spectral resolution & 4.0\,eV FWHM at 7\,keV & 2.5\,eV  \\
\hline
Energy scale calibration & 0.5\,eV (1\,$\sigma$) & 0.4\,eV (1\,$\sigma$) \\
\hline
Field of view & 4' (equivalent diameter) & 5'\\
\hline
Instrument efficiency at 0.35, 1, 7, 10\,keV & $>$ 12\%, 56\%, 63\%, 42\% & Similar\\
\hline
Non X-ray background & $< 5 \times 10$\textsuperscript{-3}\,cps/cm\textsuperscript{2}/keV (2-10\,keV) & Same \\
\hline
Relative time resolution & 10\,µs & Same \\
\hline
High res throughput (broadband, point source) & 80\% at 1\,mCrab (goal at 10\,mCrab) & Same \\
\hline
10\,eV throughput (5-8\,keV, point source) & 50\% at 1 Crab & Same\\
\hline
High res throughput (broadband, ext. source) & 80\% at $2\times 10^{-11}$\,ergs/s/cm$^{2}$/arcmin$^2$ & Same\\
\hline
Continuous cool time and duty cycle & Up to 28h with $>$ 75\% duty cycle & Same \\
\hline

\end{tabular}
}
\caption{The key X-IFU performance at the end of the reformulation phase. Formal requirements are pending the issue of the NewAthena Science Requirements Document based on the NewAthena mission driving science objectives and its flowdown towards the instrument.}
\label{tab:key_performance_requirements}
\end{table}

This paper provides a description of the X-ray Integral Field Unit in its configuration after the Athena mission reformulation activity that spanned from September 2022 to November 2023. The presented architecture, making use of passive cooling down to 50\,K, is the result of a full system redefinition between the X-IFU consortium, ESA, its international partners and the WFI\cite{Meidinger2020WFI} consortium (this last interface being directly handled by ESA). Its purpose was to bring the ESA Cost at Completion of the NewAthena mission below an acceptable target while maintaining the mission performance to flagship level. A necessary condition to reach this objective was a simplification of the interfaces between the instrument and the platform, with notably the instrument dewar being, as in its first incarnation, an X-IFU instrument consortium delivery to ESA. 

The first part of the paper gives an overview of the new instrument configuration, with a focus on the major changes related to the NewAthena mission rescope. The second provides a summary of the main instrument budgets. The third one describes the currently anticipated performance of the X-IFU instrument (see also Table \ref{tab:key_performance_requirements}). Finally, the last part gives an overview of the new project schedule and development logic.

\section{Main updates of the X-IFU design following the reformulation}

\begin{figure}
    \centering
    \includegraphics[width=1\linewidth]{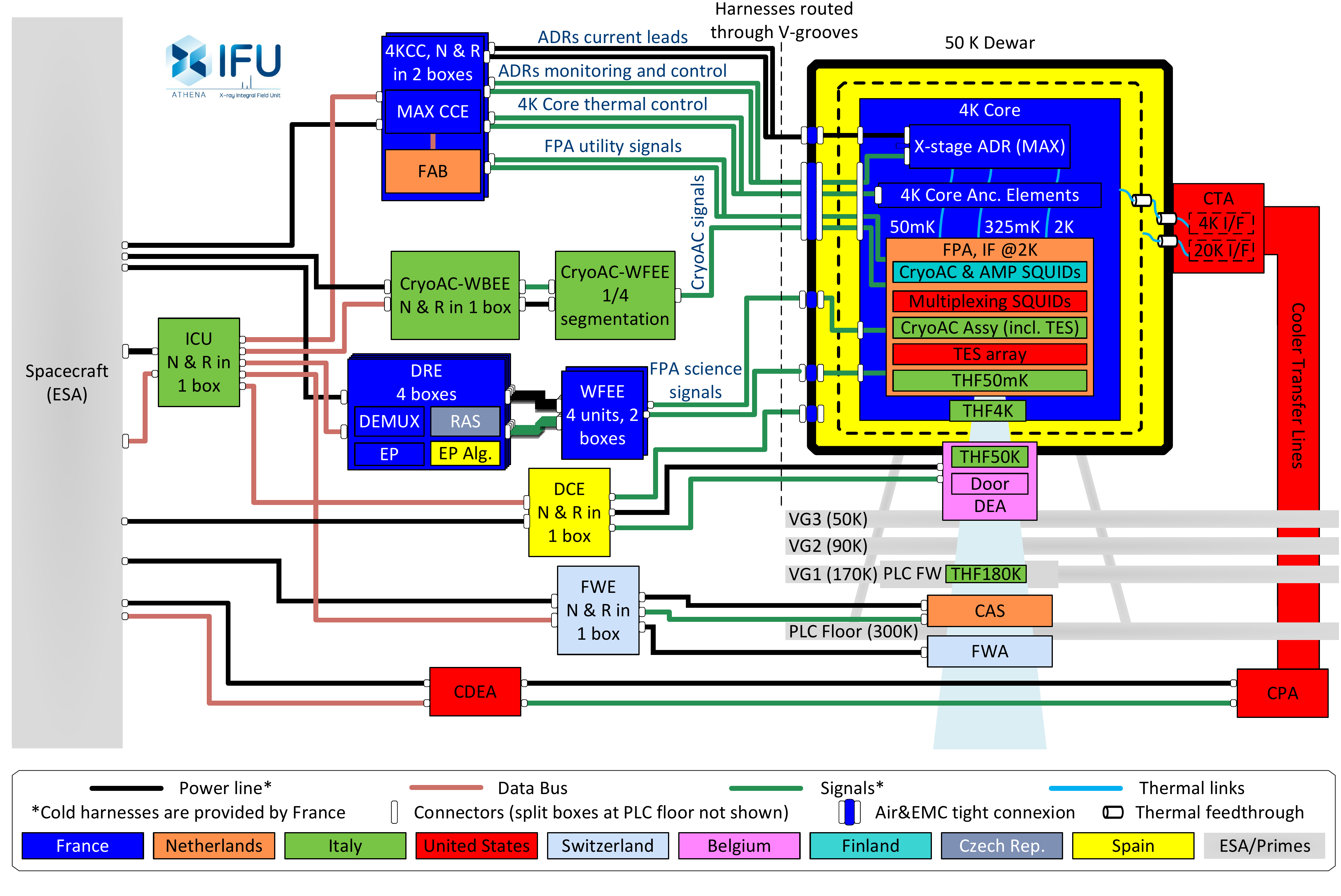}
    \caption{The Physical Breakdown of the X-IFU instrument following the reformulation. It highlights the main components of the instrument as well as the country responsible for its development.}
    \label{fig:new_phy_breakdown}
\end{figure}

Figure \ref{fig:new_phy_breakdown} provides an overview of the current instrument configuration, identifying the different subsystems and contributions. It is important to note that the current design still builds on significant heritage from the previous X-IFU study. This was an implied boundary condition for the success of the mission reformulation in order to limit the need for additional/complementary technological developments, as well as schedule slip. Concretely speaking, beyond the changes related to the cryogenic architecture described in the next two sections, virtually all other technological building blocks remain mostly identical to the previous concept. This is highlighted in Table \ref{tab:rescope-impact}.

\begin{table}
    \centering
    \resizebox{\linewidth}{!}{
    \begin{tabular}{p{3.1cm}p{2cm}p{6.1cm}}
    \hline
        Subsystem (\textit{Institute}) & Impact of reformulation & Change wrt. previous X-IFU design \\ \hline
        FPA (\textit{SRON}) & Low to medium & Reduction of FoV, change of segmentation \\ 
        TES array (\textit{NASA}) & Low & Reduction of FoV, change of segmentation \\ 
        MUX SQUIDs (\textit{NIST}) & Low & Only change of “aspect ratio” \\
        AMP SQUIDs (\textit{VTT}) & Low & No change \\ 
        Cold harness (\textit{CNES}) & High & Part of cold harness now outside of Dewar\\
        WFEE (\textit{APC}) & Low & Same electronics design, new units repartition \\ 
        DRE (\textit{IRAP}) & Low & Same electronics design, new units repartition \\
        DRE-DEMUX (\textit{IRAP}) & Low & Same electronics design \\
        DRE-RAS (\textit{IAP}) & Low & Same electronics design, one additional line \\
        DRE-EP (\textit{CEA-IRFU}) & Low & Slight reduction of processed pixels per module \\
        4\,K Core (\textit{CNES}) & Major & Complete redesign \\ 
        4KCC (\textit{CEA-IRFU}) & High & Significant redesign for 5 stage ADR \\ 
        FAB (\textit{SRON}) & Low & No change \\
        MAX (\textit{CEA-SBT}) & Major & Complete redesign (except 50\,mK stage) \\ 
        CryoAC chain (\textit{INAF}) & Low & Almost no impact (reduction of FoV) \\ 
        ICU (\textit{INAF}) & Low & Almost no impact (change of \# of SpW I/F) \\ 
        THF (\textit{Univ. Palermo}) & Medium & Change of diameters, new external filter \\ 
        50\,K Dewar (\textit{AVS}) & Major & New subsystem \\
        DCE (\textit{AVS}) & Major & New subsystem \\
        DEA (\textit{CSL}) & Major & New subsystem \\ 
        CCU (\textit{NASA}) & Major & New subsystem \\ 
        CAS (\textit{SRON}) & Low & Almost no impact (change of diameter) \\ 
        FW (\textit{Univ. Genova}) & Medium & Reduction to 5 positions \\ \hline
    \end{tabular}
    }
    \caption{Impact of the Athena mission and X-IFU instrument reformulation on the different X-IFU subsystems.}
    \label{tab:rescope-impact}
\end{table}

In particular, the overall detection chain concept (2-stage SQUID TDM readout of TES microcalorimeters operated by two stages of warm electronics - see more details in \cite{Barret_2023ExA....55..373B} and references therein) as well as its dimensioning rules were preserved. Similarly, the command-and-control and power distribution architecture of the instrument have been conserved across the rescope \cite{Barret_2023ExA....55..373B}. Finally, whereas the cryogenic anti-coincidence (CryoAC) detector design has recently evolved (see \cite{Macculi2023,Andrea2024}), it is only the result of a trade-off that was still open at the time of the 2022 pre-reformulation SRR and is not associated with the reformulation itself.

\subsection{New cryogenic architecture}

The most important modification of the instrument reformulation, the X-IFU cryogenic architecture, now relies on the following:
\begin{itemize}
    \item a passive cooling system implemented by the PayLoad Compartment (PLC, ESA/Primes responsibility), providing a 50\,K environment to the instrument,
    \item a Dewar vessel with an outer envelope at 50\,K (developed by AVS, Spain), which hosts the Instrument 4\,K Core containing the detector,
    \item a remote cryocooler (CryoCooler Unit, CCU, procured by NASA), providing active cooling to the instrument inner stages in the 20\,K and 4\,K ranges,
    \item a multi-stage Adiabatic Demagnetization Refrigerator (ADR, developed by CEA-SBT, France), able to cool the detector stage in the 50\,mK range starting from the 4\,K stage.
\end{itemize}

\begin{figure}
    \centering
    \includegraphics[width=\linewidth]{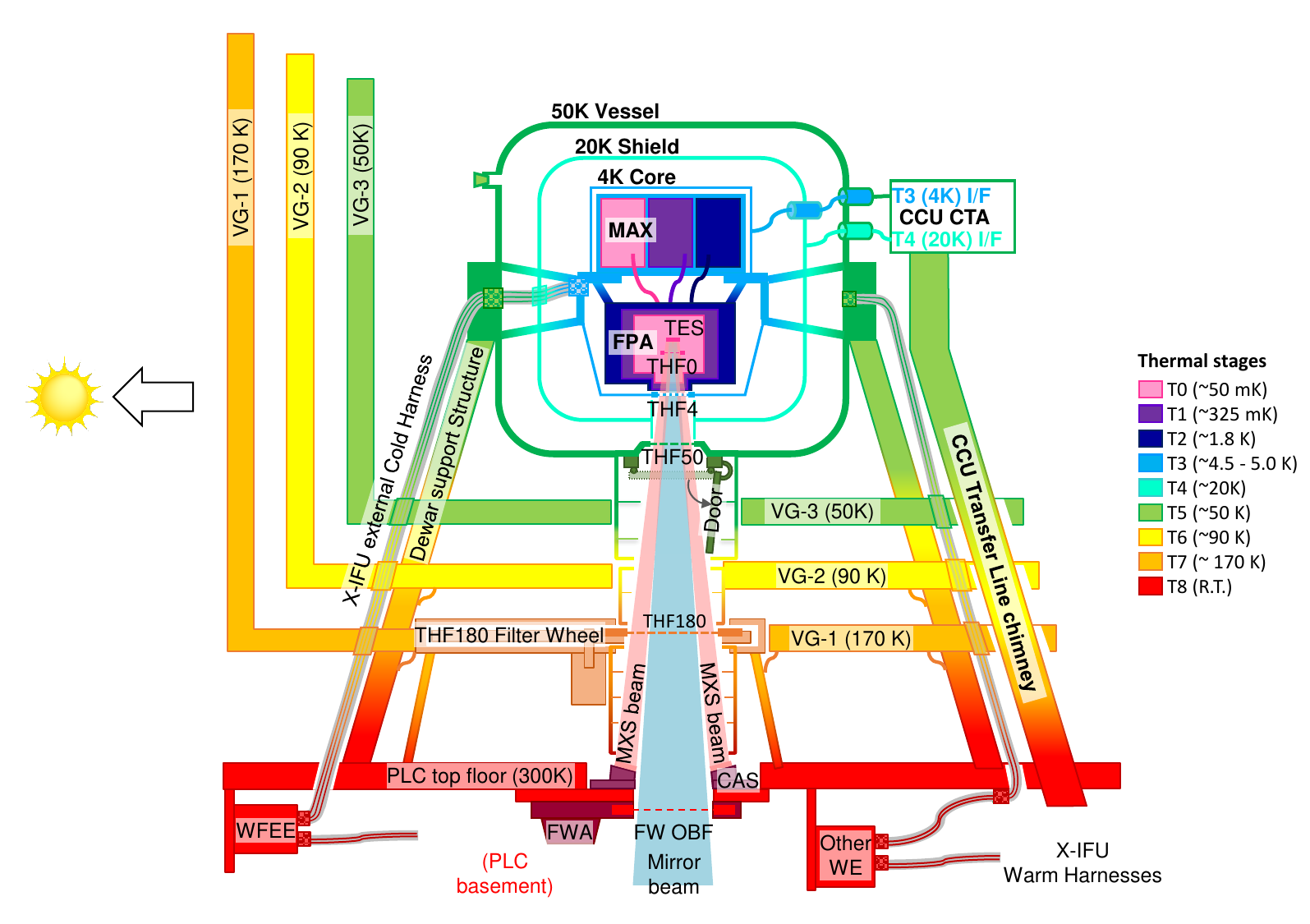}
    \caption{Schematic view of the new X-IFU cryogenic architecture in the “cold” PLC. The notional Sun direction is indicated on the left.}
    \label{fig:cryoarch}
\end{figure}

Figure \ref{fig:cryoarch} provides a schematic view of this new cryogenic architecture, with a representation in color of the different thermal stages. This configuration replaces a relatively complex, fully active (and redundant) cryogenic chain constituted of four 15\,K pulse tube refrigerators, two 4\,K Joule-Thomson coolers, two 2\,K Joule-Thomson coolers, and a sorption/ADR hybrid cooler (see more details on the previous configuration in \cite{Barret_2023ExA....55..373B}). With the mechanical cryogenic chain now remote and under a single responsibility, we expect the new design to ease the management of the interfaces (a key objective of the reformulation), notably in terms of electro-magnetic and micro-vibration compatibility.  


\subsubsection{Passive cooling from the Payload Compartment}
\label{subsec:PLC}

\begin{figure}
    \centering
    \includegraphics[width=0.33\linewidth]{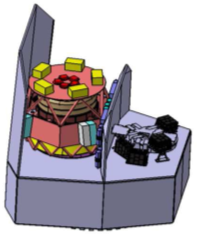} 
    \includegraphics[width=0.33\linewidth]{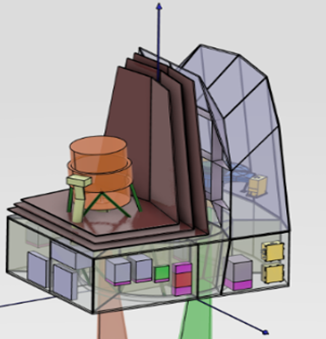}
    \caption{\emph{Left:} CAD model of the X-IFU in the Athena Science Instrument Module (SIM) as available at the end of phase A\protect\footnotemark. In this configuration, the SIM provided a closed 300 K environment (not shown here for visibility of the instruments) and the X-IFU featured a fully active cryogenic chain. \emph{Right:} CAD model of the NewAthena PLC as studied by ESA during the reformulation phase. This system is now being studied by two competitive European industrial primes. Figure credit: ESA.}
    \label{fig:CAD-PLC}
\end{figure}

Figure \ref{fig:CAD-PLC} (right) shows the CAD model of the PLC as studied by the ESA team during the reformulation phase (the left part of Figure \ref{fig:CAD-PLC} recalls the original phase A-B concept). Its main purpose is to accommodate both NewAthena instruments at the focal plane of the NewAthena mirror. Whereas the WFI remains located in a "300\,K" environment as before (with a detector stage passively cooled down to $\sim$ 180\,K), the X-IFU Dewar is now passively cooled down to 50\,K by a set of three L-shaped cryogenic radiators (called V-grooves), which can be seen on the left side of the image. The X-IFU electronics are located in the bottom part of the PLC (so-called "basement"), which remains at room temperature.
This is enabled by a modification of the allocated volume towards the WFI instrument, as well as a reduction of the mission Field of Regard (FoR) due to the increased Sun exclusion angle imposed by the V-grooves. 
Based on experience from previous missions/studies\footnote{JWST is another well-known example of passive cooling via subsequent radiative panels. With 5 deployable shields of much larger areas though, they result in a quite different temperature staging to the one expected to NewAthena.} such as Planck \cite{Planck} and Ariel\cite{Ariel}, the temperature staging is expected to lie in the range of 170\,K, 90\,K and 50\,K. With this concept, only limited heat lift capability is available at the intermediate radiators compared to concepts with active coolers, their main role being to shield the 50\,K stage, which itself provides $\sim$ 1 W of lift. The main benefit of this architecture of course lies in the relative simplicity of the required hardware, mainly at the expense of a complication of the testing environment on the ground. The full cryogenic performance indeed can now only be verified in thermal vacuum condition with a low and controlled thermal background on the V-grooves. 

We note that this concept is currently being re-assessed and further detailed by two European industrial primes, in a competitive process, in preparation of the mission Adoption expected in early 2027. 

\footnotetext{We show here this dated design due to the competitive nature of the SIM industrial studies performed in phase B with the X-IFU cryostat under the primes' perimeter.}

\subsubsection{The 50K Dewar}
The 50\,K Dewar is a mechanical and thermal assembly accommodated in the PLC 50\,K area, ensuring:
\begin{itemize}
    \item The mechanical support and thermal insulation of the X-IFU 4\,K Core from the PLC 50\,K environment,
    \item The mechanical support and thermal intercept in the 20\,K range of the X-IFU internal Cold Harnesses,
    \item The 4\,K and 20\,K thermal links (conductive links), including feedthrough functions, between the CCU cold tips and the instrument 4\,K and 20\,K stages,
    \item A vacuum-tight enclosure at room temperature during on-ground transport, storage, and integration activities, and during launch (together with other X-IFU subsystems),
    \item Residual venting capability in orbit, prior to Door opening and cool-down,
    \item A 50\,K Faraday cage (in combination with other X-IFU subsystems),
    \item Baffling functions along the sides of the photon path (protection from thermal radiations, straylight and contamination), between the Dewar entrance at 50\,K and the 4\,K Core entrance at 4\,K.
\end{itemize}

\subsubsection{The CryoCooler Unit}

Cooling down to 4.5\,K (with a required heat lift of 50 mW) is ensured by a remote cryocooler. It also provides 200 mW of heat lift around 20\,K to cool an internal radiative shield of the Dewar (see Figure \ref{fig:cryoarch}). For the sake of modularity, subsystems’ parallel verification, and integration ease, the interfaces between the cooler cold tips and the 50\,K Dewar are foreseen at the exterior of the Dewar main vessel, outside the vacuum enclosure and outside the Faraday cage. Dedicated feedthroughs are thus implemented inside the Dewar for the 4\,K and 20\,K thermal conductive links between the 4\,K Cooler cold tips, and the Instrument 4\,K and 20\,K stages. The 50\,K feedthroughs are both vacuum and EMC (electro-magnetic compatibility) tight such that the 50\,K Dewar constitutes a vacuum and EMC enclosure independent from the cooler (see Figure \ref{fig:cryoarch}).
 
At this stage, the CCU vendor remains to be selected and the exact cooler technology is thus not known. Its accommodation within the PLC however will follow these general principles (see also Figure \ref{fig:CAD-PLC}):
\begin{itemize}
    \item The cooler compressors and pre-coolers are remotely accommodated in the PLC basement, together with their control electronics.
    \item The cooler transfer line(s) and its(their) dedicated radiative shielding are routed to the Dewar across the PLC floor and the V-grooves and supported by the PLC.
    \item The 4\,K and 20\,K cold tips of the cooler interface with the 50\,K Dewar thermal feedthroughs.
\end{itemize}

\subsubsection{The 4K Core}

\begin{figure}
    \centering
    \includegraphics[width=1\linewidth]{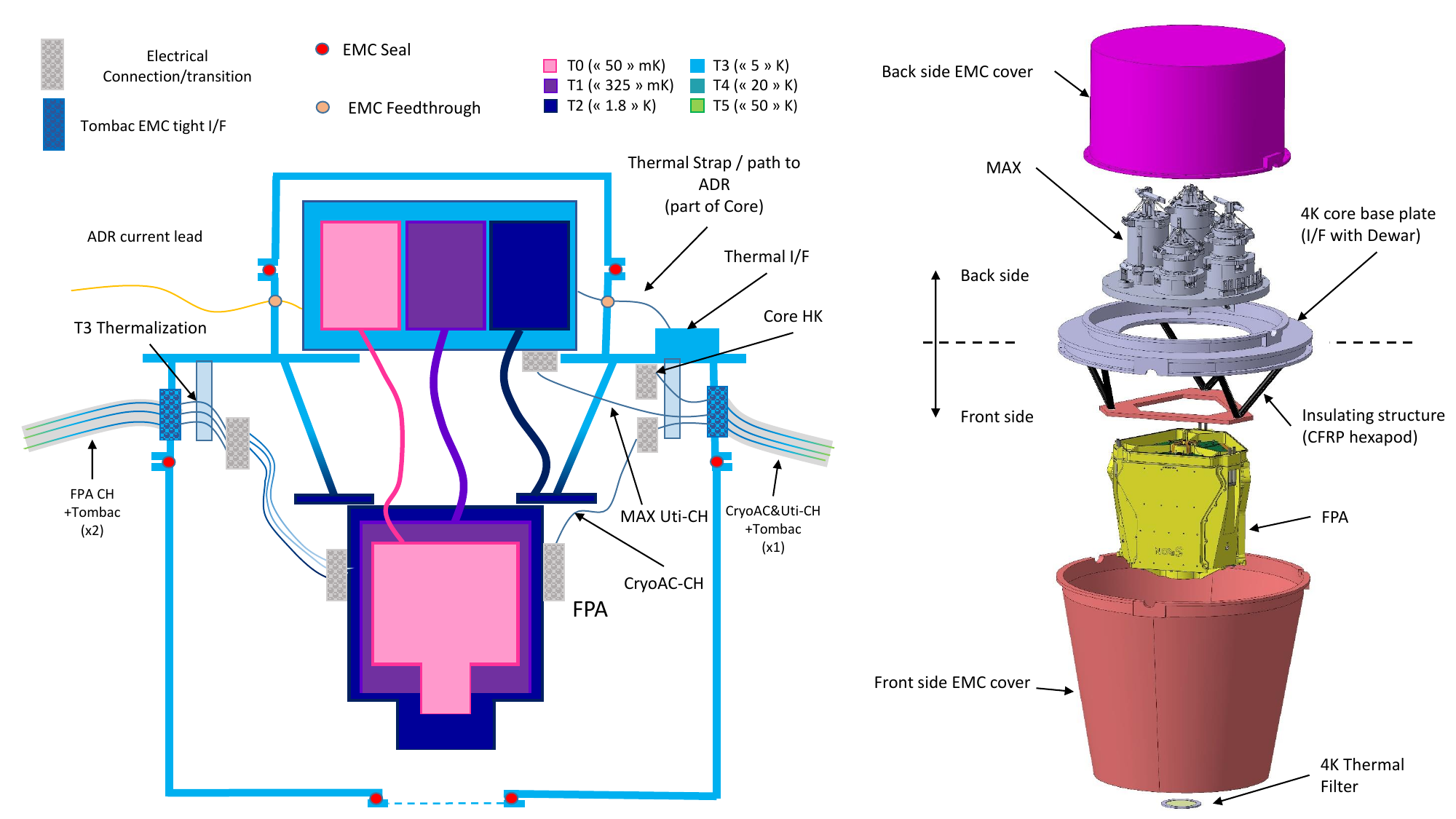}
    \caption{4\,K Core overview. \underline{Left:} schematic diagram, \underline{Right:} exploded CAD view}
    \label{fig:4Kcore}
\end{figure}

Accommodated at the heart of the 50\,K Dewar at the $\sim$ 4.5\,K stage, the Instrument 4\,K Core primarily hosts the Focal Plane Assembly (FPA, \cite{Jackson_2016SPIE.9905E..2IJ}, \cite{Weers2024}) and the Multi-stage Adiabatic demagnetization refrigerator for X-IFU (MAX, \cite{Duval2024}). Figure \ref{fig:4Kcore} gives an overview of the current 4\,K Core concept. The FPA is suspended at $\sim$ 2\,K thanks to low-conductivity carbon-fiber-reinforced polymer (CFRP) struts, whereas MAX is directly mounted onto the $\sim$ 4.5\,K 4\,K Core baseplate, which provides the main mechanical interface to the Dewar internal supports. EMC covers are implemented on both sides (a radio-frequency tight thermal filter is implemented at the optical entrance) in order to close the 4\,K Faraday cage.

This architecture was preferred to alternatives featuring an FPA interface temperature at 4.5 - 5\,K (to account for a pessimistic case gradient between the CCU cold tip and the 4\,K Core interface) as it provides a smoother temperature staging and is thus considered more robust from a cryogenics perspective. It further keeps the FPA temperature to its pre-rescope value and thus maximizes the heritage from the previous study. 

\begin{figure}
    \centering
    \includegraphics[width=\linewidth]{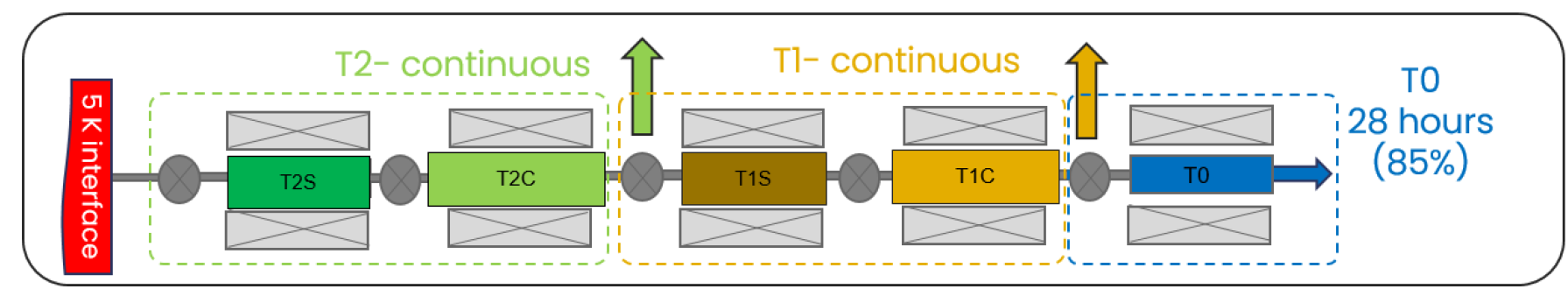}
    \caption{MAX’s 5-stages configuration. The three arrows indicate thermal interfaces at T0 ($\sim$ 50\,mK), T1 ($\sim$ 325\,mK) and T2 ($\sim$ 1.8\,K) at which MAX provides heat lift. Figure taken from \cite{Duval2024}.}
    \label{fig:MAX-principle}
\end{figure}

\begin{figure}
    \centering
    \includegraphics[width=1\linewidth]{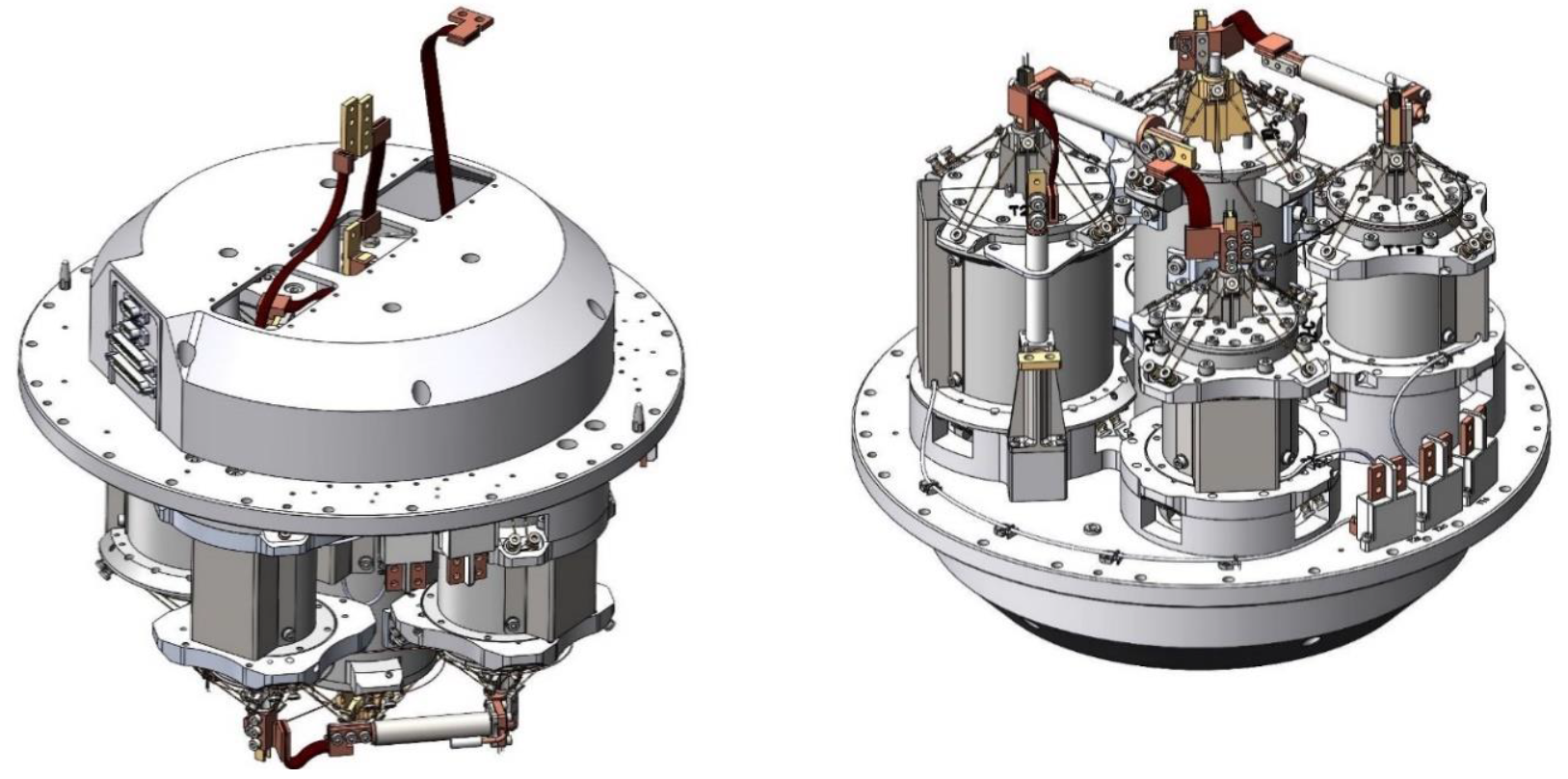}
    \caption{CAD design of the to-be-manufactured demonstration model of MAX. This detailed design is still evolving. Figure taken from \cite{Duval2024}.}
    \label{fig:ADR-DM}
\end{figure}

The last stage cooler (MAX) is a 5-stage ADR that provides three thermal interfaces at T0 ($\sim$ 50\,mK), T1 ($\sim$ 325\,mK) and T2 ($\sim$ 1.8\,K) and rejects heat to its warm interface at T3 ($\lesssim$ 5.0\,K). The 5 ADR stages are in series as seen in Figure \ref{fig:MAX-principle}. It provides 28-hours of cooling at 50\,mK and continuous cooling at T1 and T2. A pair of ADR stages is necessary for each continuous cooling interface.
Each ADR stage comprises a superconducting coil and its ferromagnetic shield. At its center, a core of magnetocaloric materials is suspended using Kevlar suspensions. Heat switches are connecting the stages to either thermal disconnect or transfer heat from one stage to another.
The preliminary demonstration model (DM) CAD concept (see Figure \ref{fig:ADR-DM}) is built on a compact assembly of the 5 stages on a common plane with heat switches located on both sides of the supporting plate. The design optimization focuses on reducing volume, mass and parasitic induced magnetic field to the FPA. Overall, in order to provide heat lifts of 0.9 µW, 10 µW and 1 mW at T0, T1 and T2 respectively, MAX is expected to reject around 15 mW peak at T3 (with margins).

\subsection{The new X-IFU optical configuration}

With part of the instrument aperture now outside of the Dewar and going through the V-grooves, the X-IFU optical configuration had to be reviewed for what concerns the cold part of the system. In contrast, the elements located at room temperature, namely the Filter Wheel (FW) and the Calibration Assembly (CAS, with its modulated X-ray sources, MXS, for in-flight gain tracking), remain mostly unchanged. Beyond increases of diameters due to the larger distance between the focal plane and the 300\,K environment, the only significant modification with respect to the previous X-IFU incarnation is the reduction of the number of FW positions from seven to five in order to keep similar wheel inertia and motorization technology as the previous design. The new design now features a single optical blocking filter and a single high count-rate filter, instead of two of each type previously (see \cite{Bozzo_2016arXiv160903776B}). 

\subsubsection{Thermal filters}
\label{subsub:thf}

\begin{figure}
    \centering
    \includegraphics[width=1\linewidth]{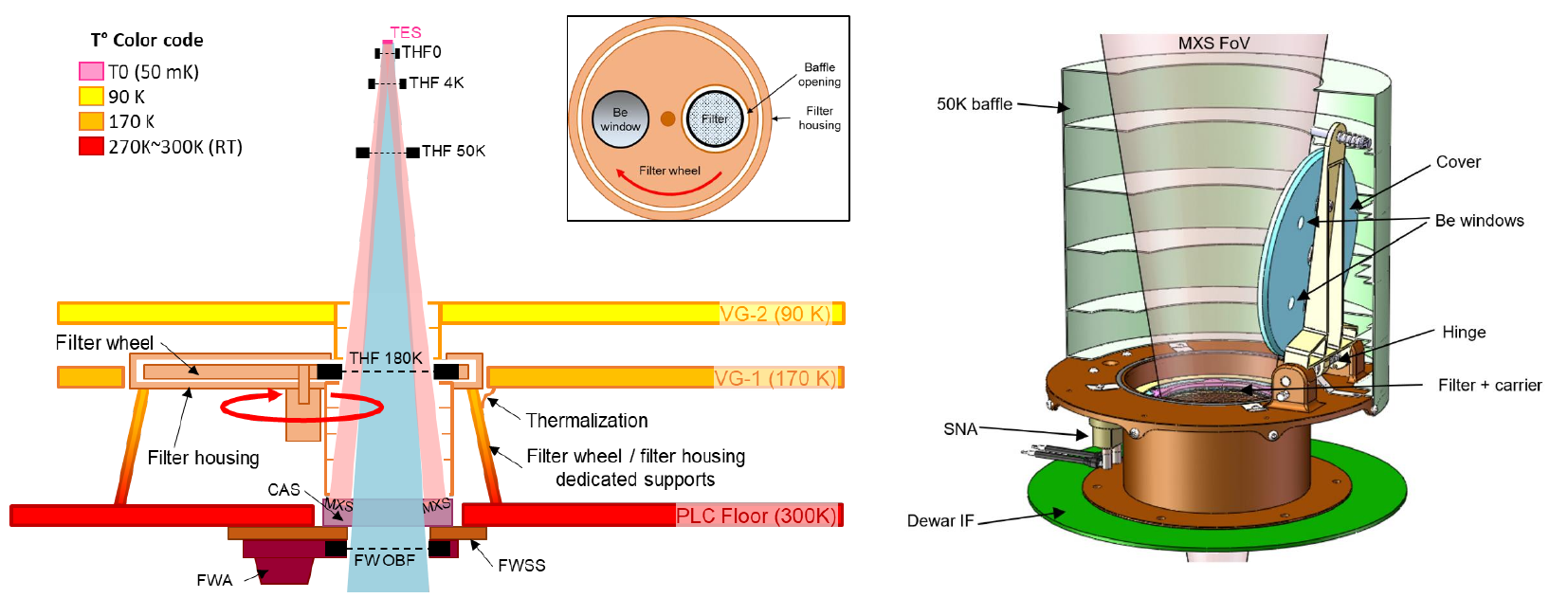}
    \caption{\underline{Left:} Configuration of the X-IFU aperture, showing the different THFs, the FW and the CAS. The top-right corner contains a schematic diagram (top view) of the filter housing and filter wheel hosting the external thermal filter. \underline{Right:} Dewar Entrance Assembly preliminary concept.}
    \label{fig:Optic-conf}
\end{figure}

The new baseline configuration accommodates 4 thermal filters (THF) to reduce the radiation heat-load from warmer surfaces onto the instrument cold stages, as well as the IR/optical load onto the detectors, that creates shot noise:
\begin{itemize}
    \item Three internal filters located at the 50\,mK (inside the FPA), 4\,K (on a 4\,K Core EMC cover), and 50\,K (inside the DEA, see below) stages. Whereas the final filter design is still being re-iterated accounting for the new diameters and new mechanical environments, the filters are expected to keep a similar design to the SRR one (thicknesses subject to potential small evolution) : a thin film of $\sim$ 45\,nm of polyimide and $\sim$ 30\,nm of aluminum ($\sim$ 7\,nm of which get oxidized to Al2O3) supported by a low blocking factor (few percents) metallic mesh. The 4\,K and 50\,K filters also ensure EMC tightness of their respective Faraday cage.
    \item One thicker (150\,nm of polyimide instead of 45\,nm) external filter (E-THF) located in the V-Groove 1 area (170\,K). This filter will be maintained at 180\,K in operation to be kept above the water saturation point temperature at low residual pressure (prevention of icing). As shown in Figure \ref{fig:Optic-conf}, this filter is assumed to be hosted by a filter wheel (or other mechanism) allowing placing the Thermal Filter inside of a non-evacuated housing during launch, to provide protection from acoustic loads. During launch and the mission early phase, a thick beryllium filter (or similar) would be put in its place in the optical path in order to catch contamination from the main early outgasing phase of the spacecraft. 
\end{itemize}
Overall, the number of thermal filters has thus been reduced with respect to the previous design, with only three internal filters and one external one, compared to five internal filters before. This results in an overall lower total aluminum thickness ($\sim 120$\,nm including oxidized thickness) but larger total polyimide thickness ($\sim 285$\,nm, dominated by the external filter). This configuration was verified to be consistent with the required shot noise attenuation through preliminary simulations, but as mentioned above, optimization remains ongoing.

\subsubsection{The Dewar Entrance Assembly}
The Dewar Entrance Assembly (DEA) is one single integrated unit composed of the following three main sub-assemblies (see Figure \ref{fig:Optic-conf}, right):
\begin{itemize}
    \item The 50\,K thermal filter and its carrier.
    \item The Dewar Door.
    \item The external 50\,K baffle.
\end{itemize}

The DEA is mounted onto the Dewar via a conductive interface. The Door is vacuum tight at ambient temperature through the use of a gasket onto which the cover is pressed. Vacuum tightness is ensured during launch and warm ground testing conditions. The one-shot Door is spring actuated and opens inside the 50\,K baffle, using a redundant non-explosive separation nut actuator (SNA) once the system is cold. Heaters will be implemented to maintain the door around 80K during actuation in order to simplify the mechanism verification/qualification. Together with the heaters, the DEA will carry the necessary housekeeping thermal sensors. EMC gaskets are foreseen at the Dewar interface and at the filter and carrier interfaces to ensure the 50\,K Faraday cage. Labyrinths are implemented to vent the filter cavity when pumping/re-pressurizing the Dewar while limiting contamination. 

The cover of the Door includes 6 (on the external part of the door, aligned with the MXS sources) + 1 (at the center, aligned with the optical axis) X-ray transparent Be windows to allow X-ray illumination while the door is closed during ground testing (notably at PLC level) and early in-flight verification.

The 50\,K baffle will be adequately coated and will include vanes as needed to limit the straylight towards the 50\,K filter and the radiative heat load from the PLC and upper V-grooves. By reducing the gap with the 50\,K and 90\,K V-grooves, the 50\,K baffle will also limit the contamination of the 50\,K filter from the PLC.

\subsection{Update of the detection chain configuration}

\begin{figure}
    \centering
    \includegraphics[width=1\linewidth]{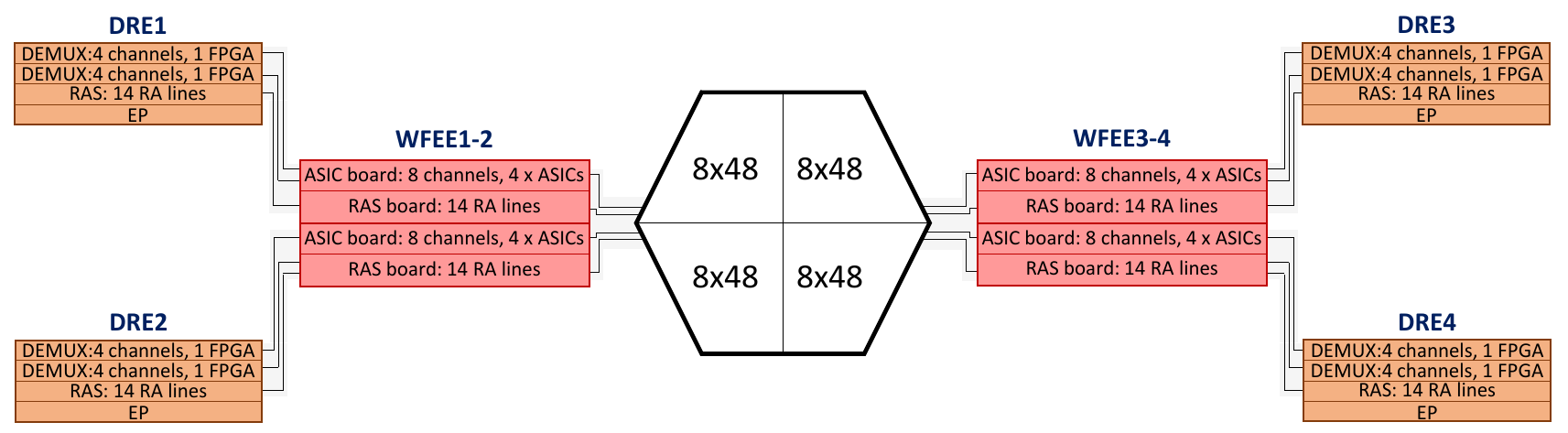}
    \caption{The new X-IFU detection chain architecture. Whereas the focal plane and detectors remain hexagonal, the readout is now performed with four independent sets of electronics. Note that whereas a multiplexing factor of 48 is employed, only 32x47=1504 TES detector populate the field of view. One readout row per column is indeed occupied by a pure resistor (not sensitive to X-rays, temperature or magnetic field) used to monitor drifts of the readout electronics.}
    \label{fig:det2}
\end{figure}

As mentioned earlier, the X-IFU detection chain architecture has been mostly preserved during the mission reformulation. It still utilizes two stages of Superconducting Quantum Interference Devices (Multiplexing SQUIDs\cite{Durkin_2021ITAS...3165279D,Durkin2023} at 50\,mK and Amplifier SQUIDs\cite{Kiviranta_2021ITAS...3160356K} at 2\,K) to readout a Transition Edge Sensors array \cite{Smith_2021ITAS...3161918S,Wakeham2023} operated at 50\,mK using a Time Domain Multiplexing scheme. This readout chain remains commanded by two sets of room temperature electronics (see more details in \cite{Barret_2023ExA....55..373B, Geoffray2024}): the Warm Front-End Electronics (WFEE, \cite{Prele:2024msq}) and Digital Readout Electronics (DRE, \cite{Murat2024}), located in the PLC basement (see Section \ref{subsec:PLC}). Cold harnesses, routed along the 50\,K Dewar mechanical supports, connect these electronics (as well as the the 4\,K Core Controller, 4KCC, and Dewar Control Electronics, DCE) to the 50\,K Dewar (see Figure \ref{fig:cryoarch}).

The instrument rescope activities, however, lead to a significant reduction of the number of channels from 72 at the pre-reformulation SRR to 32 in the current concept. This change was accompanied by a modification of the instrument segmentation logic, which now presents a 4-fold symmetry as illustrated in Figure \ref{fig:det2} (by opposition with the previous 6-fold symmetric design). Both revisions lead to a reduction of the required instrument hardware as well as of the overall X-IFU needs from the spacecraft, in terms of dissipation, volume and thermal losses from harnesses (see instrument budgets in the next section). In contrast, the readout multiplexing factor was increased from 34 to 48, which allowed retaining a field of view of 4' equivalent diameter populated by 1504 TES detectors. This was made possible by a new, slower and thus easier to readout, detector design (see more details \cite{Wakeham2023,Smith2024}). This is an optimization that was already on-going when the reformulation started. With a $\sim$ 40 \% lower slew rate, despite the higher MUX factor, the same readout noise target (and noise requirements from the electronics) could be preserved and a similar resolution performance at detection chain level is expected. The main adverse consequence of this optimization is a degradation of the instrument count rate capability, but as emphasized in \cite{Peille_2018JLTP..193..940P,Peille_2018SPIE10699E..4KP} and Section \ref{subsec:ctr}, the X-IFU has always retained significant count rate margins since the introduction of mirror defocusing. The associated science impact is thus expected to be minimal.

\section{X-IFU mass and power budgets in the new architecture}

The budgets presented below correspond to the estimates provided to ESA in the new X-IFU Payload Definition Document (PDD) in November 2023 as an input to the NewAthena spacecraft industrial studies. These are preliminary and subject to change across the project lifetime, in particular those regarding the cryogenic elements, which were heavily impacted by the mission reformulation.

\subsection{Mass budget}

The X-IFU mass budget is given in Table \ref{tab:mass-budget}. The presented numbers include Design Maturity Margins (DMM), as needed at this phase of the project, as well as some specific contingencies to cover already identified technical risks (e.g. mechanical loads environment, volume of cryogenic elements, ...). This budget however does not include the CCU, whose mass is formally a direct interface between NASA and ESA/primes, nor the instrument warm harnesses that are expected to be provided by the primes. 

With the cryogenic chain concept completely revisited and the responsibilities changing, a direct comparison with the previous concept cannot be done at instrument level. The V-groove system indeed effectively replaces the warmer stages of the X-IFU cryostat. One can nonetheless note a significant decrease of $\sim$ 40\,kg of warm electronics (see budget in \cite{Barret_2023ExA....55..373B}). This is explained by the simplification of the readout chain, a gain that is partially offsetted by the increase of the 4KCC mass (which now controls five ADR stages) with respect to the pre-reformulation equivalent electronics box.

\begin{table}
    \centering
    \begin{tabular}{ll}
        Unit & Mass (kg) \\ \hline
        FPA & 11.0 \\ 
        MAX & 19.4 \\ 
        4\,K Core Ancillary Elements & 15.5 \\ 
        \textbf{Sub-total: 4\,K Core} & \textbf{45.9} \\ \hline
        Internal Cold Harnesses & 10.5 \\ 
        50\,K Dewar & 100.0 \\ 
        Dewar Entrance Assembly & 5.1 \\ 
        \textbf{Sub-total: Dewar Assembly} & \textbf{161.5} \\ \hline
        External Cold Harnesses & 19.3 \\ 
        External Thermal Filter THF180 & 0.4 \\ 
        \textbf{Sub-total: Subsystems within PLC V-grooves} & 19.7 \\ \hline
        CAS & 7.0 \\ 
        FWA & 13.0 \\ 
        WFEEs (2 boxes) & 19.0 \\ 
        DREs (4 boxes) & 57.6 \\ 
        CryoAC WFEE & 2.0 \\ 
        CryoAC WBEE & 4.0 \\ 
        4KCCs (2 boxes) & 32.6 \\ 
        FWE & 3.2 \\ 
        ICU & 15.0 \\ 
        DCE & 4.0 \\ 
        \textbf{Sub-total: Subsystems in PLC basement} & \textbf{157.4} \\ \hline
        \textbf{TOTAL INSTRUMENT MASS}  & \textbf{338.6} \\ \hline
    \end{tabular}
    \caption{The X-IFU mass bugdet as provided in the November 2023 Payload Definition Document.}
    \label{tab:mass-budget}
\end{table}

\subsection{Power budget}

Table \ref{tab:power-budget} presents the X-IFU power budget. As for the mass budget, this corresponds to the November 2023 reference and includes DMMs as needed. We note that only primary supplied units are accounted here. As illustrated in Figure \ref{fig:new_phy_breakdown}, the WFEEs are powered by the DREs, the CryoAC WFEE by its Warm Back-End Electronics (WBEE), and the Filter Wheel Assembly (FWA) and CAS by the Filter Wheel Electronics (FWE).

For comparison, the X-IFU power budget at the pre-reformulation SRR ammounted to $\sim$ 820 W \cite{Barret_2023ExA....55..373B}, depending on the instrument mode. The reduction by $\sim$ 300 W illustrates plainly the readout chain simplification operated during the mission rescope to ease the system spacecraft design.   

\begin{table}
    \centering
    \begin{tabular}{ll}
        Primary power supplied unit & Average power consumption (W) \\ \hline
        Digital Readout Electronics (4 boxes) & 372 \\ 
        CryoAC Warm Backend Electronics & 30 \\ 
        Dewar Control Electronics & 33 \\ 
        4\,K Core Controller (2 boxes) & 59 \\ 
        Filter Wheel Electronics & 18 \\ 
        Instrument Control Unit & 20 \\ \hline
        \textbf{TOTAL POWER} & \textbf{532} \\ \hline
    \end{tabular}
    \caption{The X-IFU power budget. It is presented with the filter wheel not being rotated. For reference, an increase of $\sim$ 19 W is expected in this mode.}
    \label{tab:power-budget}
\end{table}

\section{Anticipated X-IFU performance}

\subsection{Main changes to instrument performance following reformulation}

As mentioned above, the overall readout chain architecture has mostly been preserved across the mission reformulation phase. This ensures that the overall performance of the instrument is similar to its previous configuration and is still driven by the fundamental goals listed in Section \ref{sec1}. For a detailed description of the X-IFU performance targets, we refer the reader to \cite{Barret_2023ExA....55..373B} and references therein. We list here the main differences with respect to the former design:
\begin{itemize}
    \item A more conservative target for energy resolution has been selected by ESA, in particular to ensure larger margins for system level interference and therefore de-risk the overall project. A short discussion is provided in Section~\ref{subsec:eres}.
    \item The most notable change in the instrument performance is related to the decrease of the mirror collecting area and updates to the overall instrument efficiency. This aspect is detailed in Section~\ref{subsec:effar}.
    \item The previous point, together with the update of the pixel design, cause a direct effect on the count rate capabilities of the instrument. The anticipated performances are discussed in Section~\ref{subsec:ctr}.
    \item Finally, the change in the number of pixels has a direct effect on the field-of-view of the instrument. The smaller field of view means that an update of the mock observing plan will be required. This topic is not addressed in this paper.
\end{itemize}

\subsection{Energy resolution target}
\label{subsec:eres}

\begin{figure}
    \centering
    \includegraphics[width=0.75\linewidth]{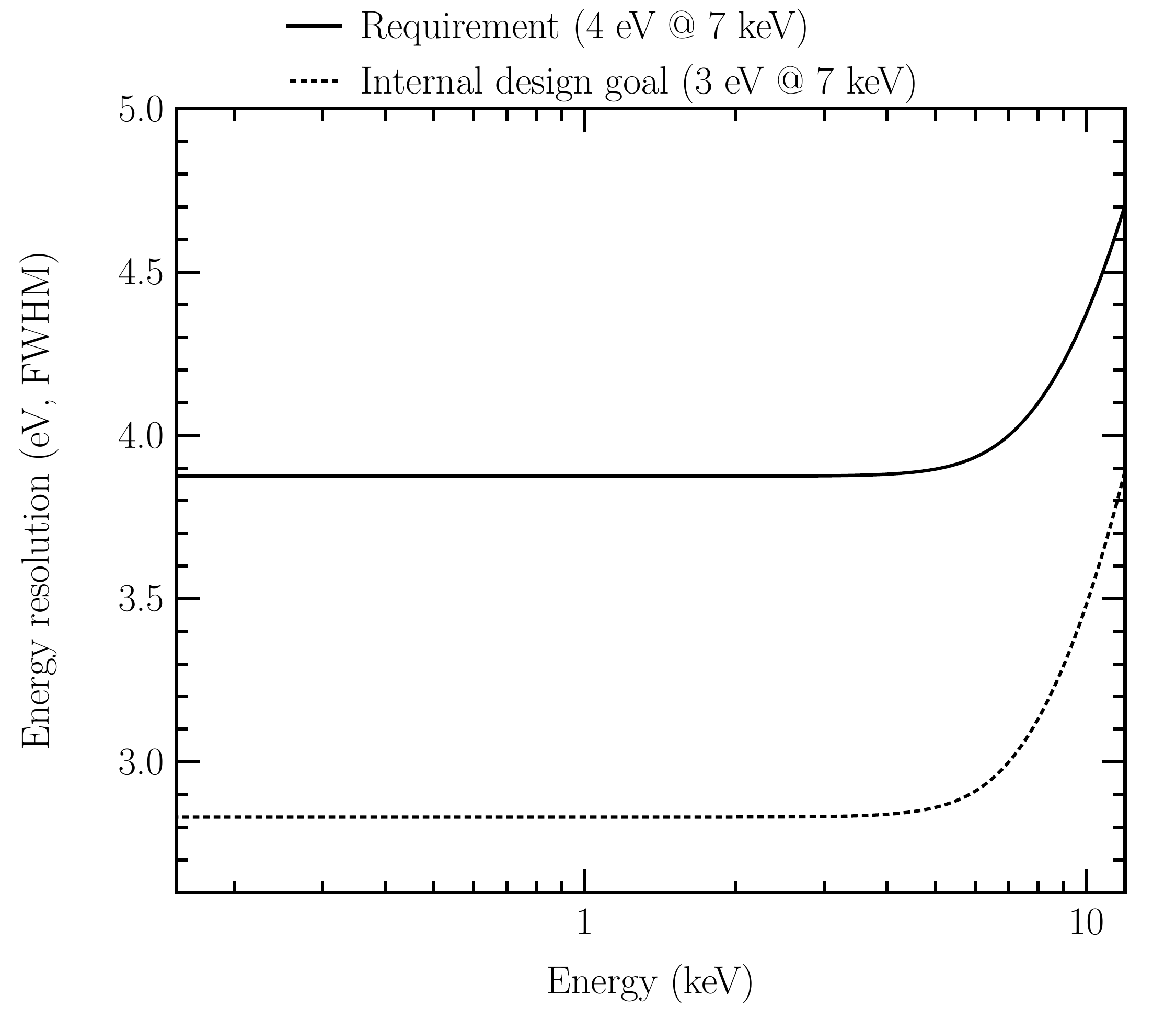}
    \caption{Energy resolution variation as a function of energy for the required (solid line) energy resolution of 4\,eV at 7\,keV, and for the internal design goal (dotted line) of 3\,eV at 7\,keV. A constant (with energy) quadrature degradation has been added to the measured pixel level resolution to match each respective target at 7\,keV.}
    \label{fig:xifu_energy_resolution}
\end{figure}

In the new instrument design, most parts of the readout chain -- and therefore its subsequent performance -- have been maintained. Additionally, both the smaller sensitivity of TES pixels to magnetic field \cite{Wakeham2023,Smith2024} and an increased distance between the cryostat and the cryogenic machines forebode beneficial effects on electromagnetic and micro-vibration perturbations, which are two of the main contributors to the final X-IFU performance. 

However, to maintain a higher system-level margin and in particular simplify ground testing requirements, the previous energy resolution target of 2.5\,eV FWHM at 7\,keV has been raised, in agreement with ESA. Two configurations are considered: the baseline case -- aiming for a 4\,eV FWHM at 7\,keV -- and a so-called internal design goal -- aiming for a 3\,eV FWHM at 7\,keV. The latter corresponds to the budget target used at instrument level to derive subsystem level requirements. It does include a moderate level of margin (0.8\,eV in quadrature at the time of writing), such that it should cover for expected uncertainties related to instrument development and constitutes the current estimate of the instrument resolution. Margin with respect to the formal 4\,eV requirement is thus reserved for "unknown unknowns" at system level.

In both cases, the energy resolution behavior as a function of energy is derived from measurements performed at NASA GSFC on previous pixel configurations, as shown in Figure~\ref{fig:xifu_energy_resolution}. We notice in particular that for lower energies a resolution of $\sim$\,2.8\,eV FWHM is expected.

\subsection{X-IFU instrument efficiency and effective area}
\label{subsec:effar}

\subsubsection{X-IFU instrument efficiency}

\begin{figure}
    \centering
    \includegraphics[width=0.8\linewidth]{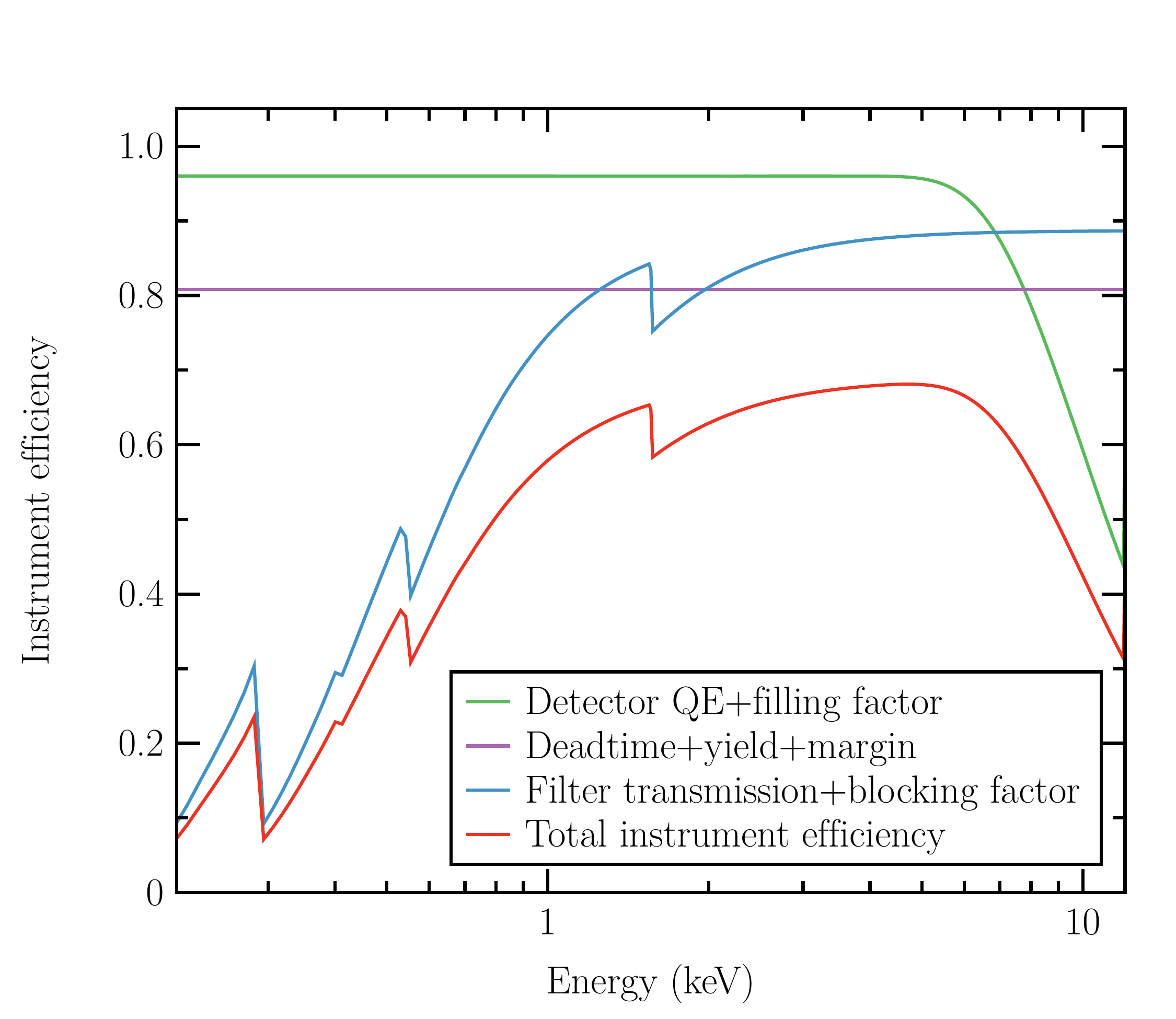}
    \caption{Total instrument efficiency of the X-IFU (red) as a function of energy. The main contributors to the performance are also shown: the thermal filters (blue), the detector (green) and the contributions of dead time, margins and overall yield (purple).}
    \label{fig:ie_full}
\end{figure}

Despite the change in the thermal filters design, most of the instrument efficiency previously available \cite{Barret_2023ExA....55..373B} has been maintained. The increase of the total polyimide thickness (now $\sim 285$\,nm, dominated by the external filter) is compensated by a decrease of the total Al thickness (now $\sim 120$\,nm including oxidized thickness - see Section~\ref{subsub:thf} and limitations therein). Detector quantum efficiency and filling factor remain identical to the previous configuration. The total instrument efficiency is provided in Figure~\ref{fig:ie_full}. A 2\,\% dead-time effect (1\,\% from the CryoAC and 1\,\% for the exclusion of cosmic ray generated thermal transients\footnote{Time dedicated to MXS calibration is accounted in the mission calibration time.}), an expected yield of 91.6\,\% across the field of view, as well as an overall system-level margin of 10\% at all energies are added. We notice that most of the low-energy instrument efficiency is driven by the thermal filters, while the high-energy part of the spectrum is driven by the detector quantum efficiency. 

No contamination is included in the curves in Figure \ref{fig:ie_full}. Contamination  will cause an energy-dependent effect, whose spectral signature depends on the contaminants’ composition. The current X-IFU performance budgets allow up to 20\,\% total transmission loss at end-of-life at 0.35\,keV. This corresponds to a maximum equivalent carbon surface density of 6.2\,µg/cm$^2$.

\subsubsection{X-IFU effective area}

\begin{figure}
    \centering
    \includegraphics[width=0.48\linewidth]{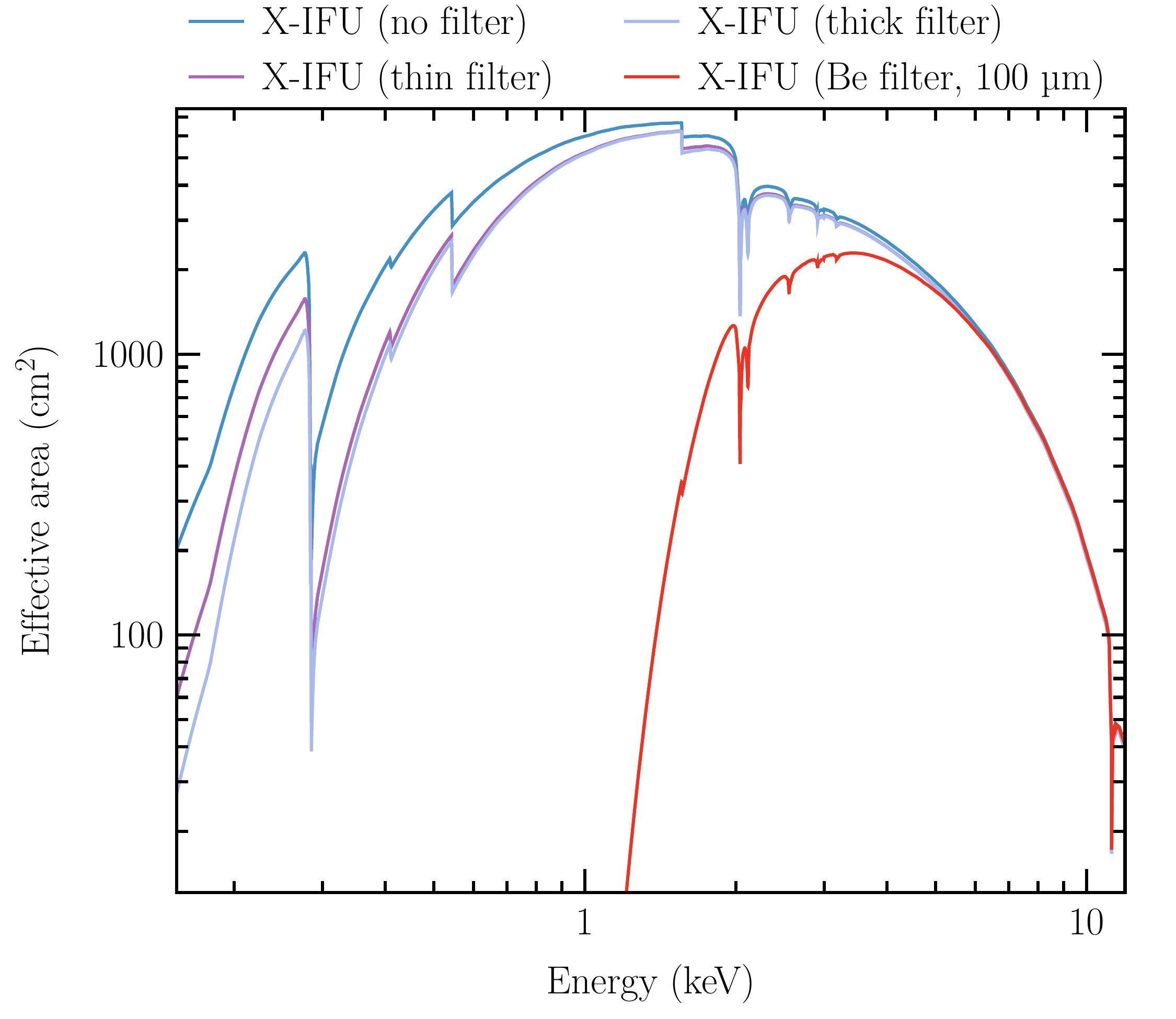}
    \includegraphics[width=0.48\linewidth]{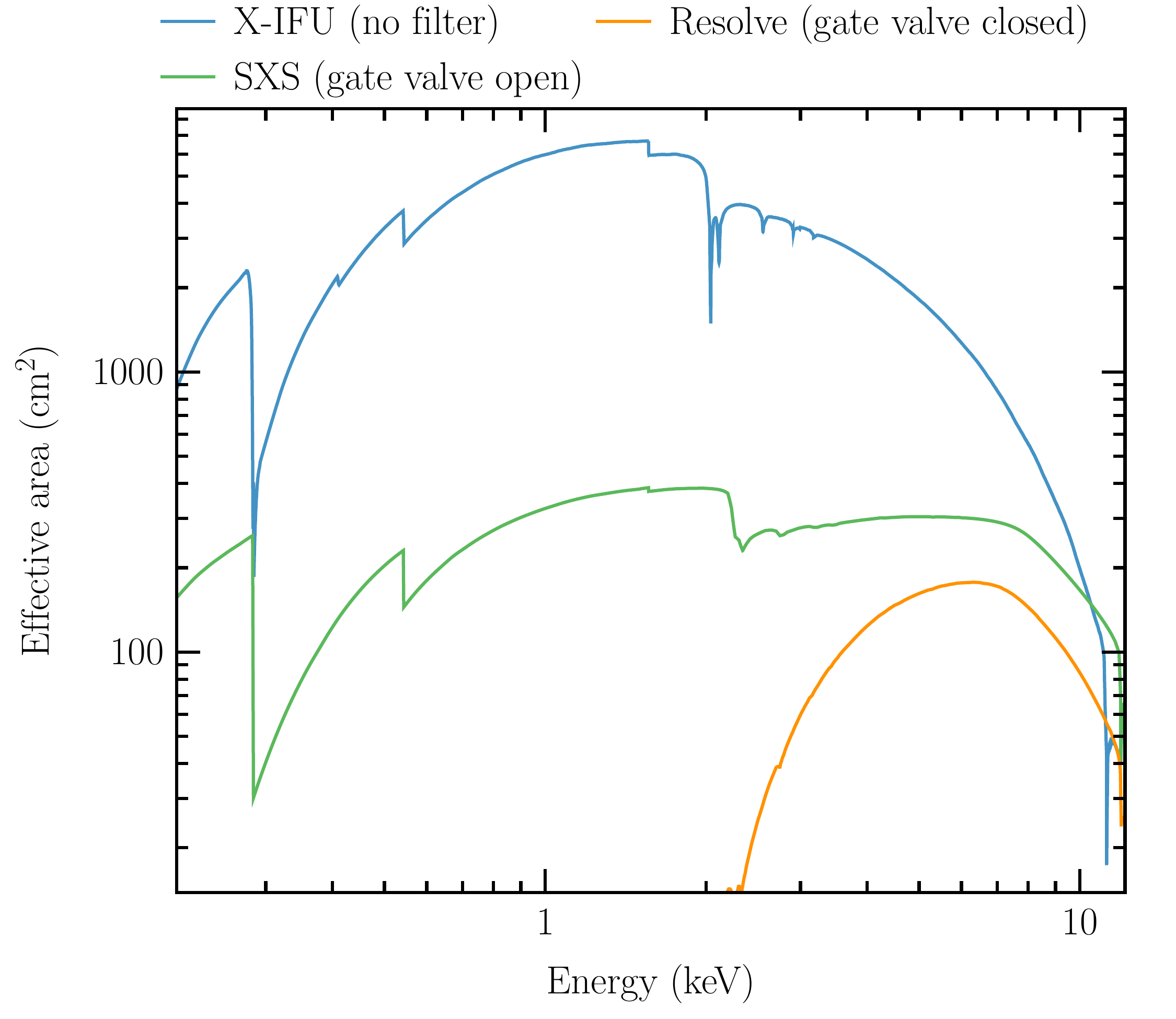}
    \caption{\emph{Left: }X-IFU effective area (cm$^2$) for the 13 rows mirror configuration with the thin (200 nm polyimide, 30 nm aluminum - 7 of which oxidized - and a 4 \% blocking factor mesh) and thick (same as thin but with 60 nm of aluminum) optical filters, as well as with a 100\,µm Beryllium filter. \emph{Right: }Comparison between the X-IFU response in the no filter configuration, with the XRISM/Resolve response with the gate valve closed, but also with the Hitomi/SXS response, corresponding to the gate valve open.}
    \label{fig:xifu_arfs}
\end{figure}

The main change on the overall instrument effective area comes from the decrease of the mirror collective area -- in particular at lower energies -- related to the reduction from 15 to 13 mirror rows and the change in mirror coating\footnote{See \url{https://www.cosmos.esa.int/web/athena/resources-by-esa}}. The effective area curves for the new XIFU baseline are shown in Figure \ref{fig:xifu_arfs} (left). This includes the no-filter configuration and three possible filter wheel configurations (thin, thick and Be filter)\footnote{The current filter wheel design only accounts for a single optical blocking filter. The trade between a thin or thick filter, or an intermediate solution, remains to be done. The full list of assumptions, the effective area curves for this configuration and details on their generation are available on the X-IFU resource portal \url{https://x-ifu.irap.omp.eu/resources}}. The peak effective area is reached between 1 and 2\,keV and saturates at a value at about 6700\,cm$^2$. Compared to the previous X-IFU configuration \cite{Barret_2023ExA....55..373B}, a loss of $\sim$ 30\,\% in total count rate is expected, in particular at lower energies. This loss mostly affects the observation of extended sources (e.g., hot clusters or supernov\ae~remnants) which have line-rich spectra below 2\,keV, and isolated stars. 

In Figure~\ref{fig:xifu_arfs} (right), we compare the X-IFU response with the XRISM/Resolve\cite{Tashiro2020,Ishisaki2022} response with the gate valve closed, but also with the Hitomi/SXS \cite{Kelley2016} response which would correspond to the case of the gate valve being eventually opened (it has been blocked in closed position for now for Resolve). In the later configuration, X-IFU would have a peak effective area $\sim 15 $ times larger. At 6\,keV, this ratio drops to $\sim$ 7, with Resolve with the gate valve closed. This comparison shows that even with a decreased effective area, the X-IFU will remain a transformational instrument.

\subsection{Count rate performance}
\label{subsec:ctr}

Using the previous assumptions on the X-IFU effective area, we can provide an updated estimate of the count rate capabilities of the new instrument configuration. We define the throughput as the ratio between the number of valid events (e.g., high-resolution) observed over the total number of events received. Table~\ref{tab:ctrate} provides an estimate of the expected count rate values seen by the focal plane array for the most extreme point and extended sources. 

\begin{table}[h]
\resizebox{\textwidth}{!}{
\begin{tabular}{c||c|c|c|c}
\hline

Source type                                                                                     & Flux                                                                & \begin{tabular}[c]{@{}c@{}}Count rate \\ (X-IFU entrance)\end{tabular}      & \begin{tabular}[c]{@{}c@{}}Count rate \\ (X-IFU detector)\end{tabular}      & Throughput  \\
\hline \hline
\begin{tabular}[c]{@{}c@{}}Defocused point source\\ \textless 3\,eV resolution\end{tabular}  & \begin{tabular}[c]{@{}c@{}}1 mCrab\\ {[}10\,mCrab{]}\end{tabular}    & \begin{tabular}[c]{@{}c@{}}70\,cts/s\\ {[}700\,cts/s{]}\end{tabular}           & \begin{tabular}[c]{@{}c@{}}39\,cts/s\\ {[}390\,cts/s{]}\end{tabular}           & \textgreater 80\,\% \\
\hline
\begin{tabular}[c]{@{}c@{}}Defocused point source\\ \textless 10\,eV\,resolution\end{tabular} & 1\,Crab                                                              & 70\,kcts/s                                                                    & \begin{tabular}[c]{@{}c@{}}6.3 kcts/s \\ (with Be filter)\end{tabular}       & \textgreater 50\,\% \\
\hline
\begin{tabular}[c]{@{}c@{}}In focus point source\\ \textless 3\,eV resolution\end{tabular}   & 0.25\,mCrab                                                          & 17\,cts/s                                                                     & 10\,cts/s                                                                     & \textgreater 80\,\% \\
\hline
\begin{tabular}[c]{@{}c@{}}Extended source\\ \textless 3\,eV resolution\end{tabular}         & \begin{tabular}[c]{@{}c@{}}Perseus\\ {[}Cassiopea A{]}\end{tabular} & \begin{tabular}[c]{@{}c@{}}0.40\,cts/s/pix\\ {[}1.2\,cts/s/pix{]}\end{tabular} & \begin{tabular}[c]{@{}c@{}}0.33\,cts/s/pix\\ {[}1.0\,cts/s/pix{]}\end{tabular} & \textgreater 80\,\% \\
\hline
\end{tabular}}
\caption{X-IFU count rate capability targets and expected energy resolution (i.e. based on 3 eV design goal and not for now on 4 eV requirement - higher throughput may be obtained at the expense of degrading the spectral resolution, but this trade remains to be performed depending on the level of margin in both resolution and throughput required at system level). For different sources, the equivalent count rate at the X-IFU entrance (mirror-only) and at the X-IFU detector are provided. The values in brackets represent goals, as they were defined at the time of the 2022 SRR.}
\label{tab:ctrate}
\end{table}
For each of the cases provided, the count rate capability is estimated using \texttt{XSPEC} \cite{Arnaud1996} models of the corresponding astrophysical sources as inputs of the \texttt{SIXTE} instrument simulator \cite{2023WilmsSIXTE}.  Two main effects cause a loss of throughput \cite{Peille2016,Peille_2018SPIE10699E..4KP}:
\begin{itemize}
    \item The first effect is grading: the energy resolution and the energy estimate of a given pulse depend on the separation of this pulse with respect to its preceding and succeeding event. A so-called `high-resolution' pulse (corresponding to the highest performance of the instrument) requires a separation of 24\,ms with its preceding event, and 63\,ms with its succeeding pulse. For high enough count rates, this means that a non-negligible fraction of events is either rejected, or expected to have a lower accuracy in reconstructed energy.
    \item The second effect is crosstalk: when the instrument observes high count rate sources, photons can reach the detector closely in time either spatially or within the same readout column. This gives rise to crosstalk, affecting the events' reconstructed energy. This effect is simulated in \texttt{SIXTE} using predefined crosstalk look-up tables computed via the detailed instrument simulator \texttt{XIFUSIM} \cite{Kirsch2022}.
\end{itemize}

\begin{figure}[t]
    \centering
    \includegraphics[width=0.5\linewidth]{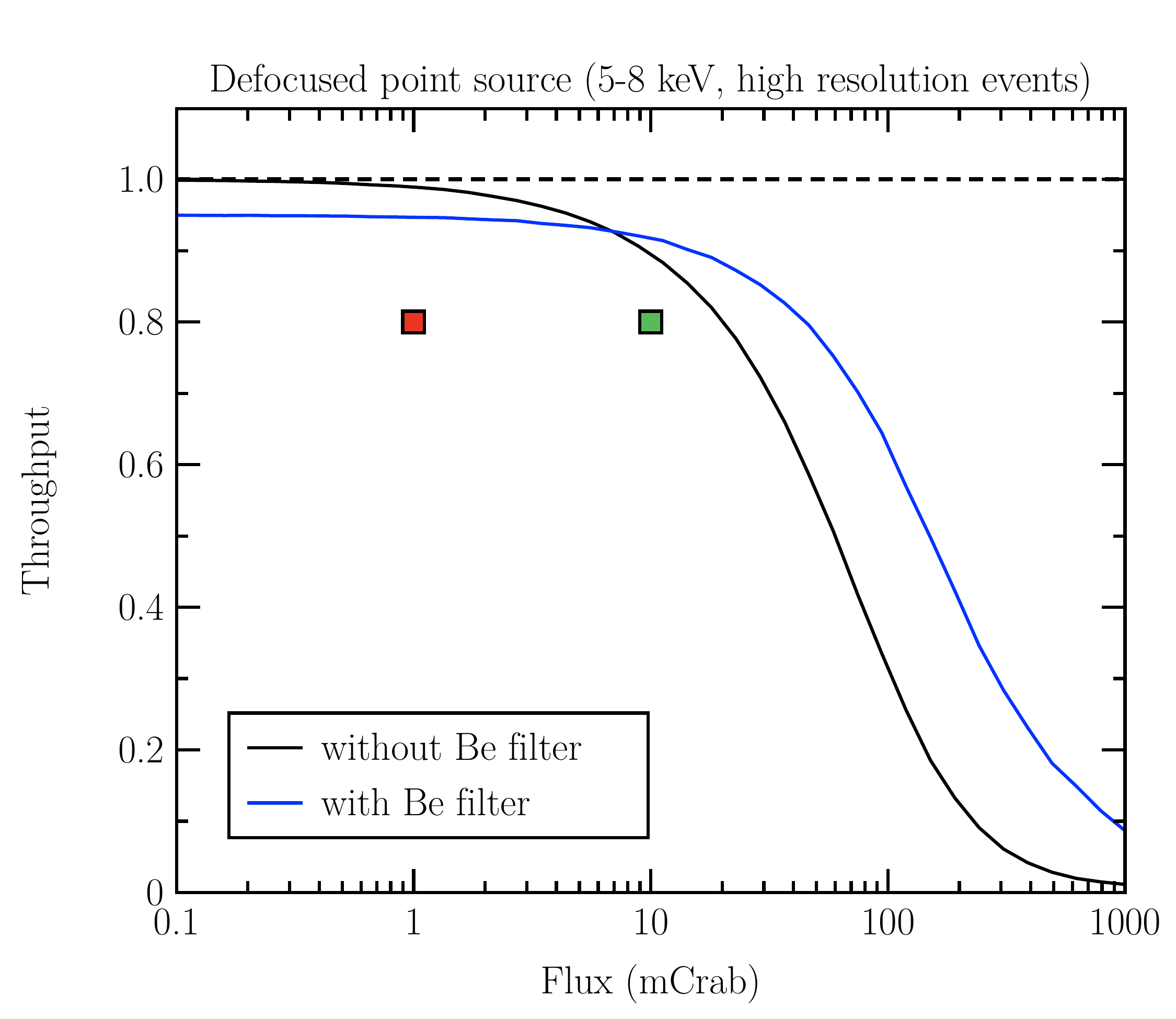}\includegraphics[width=0.5\linewidth]{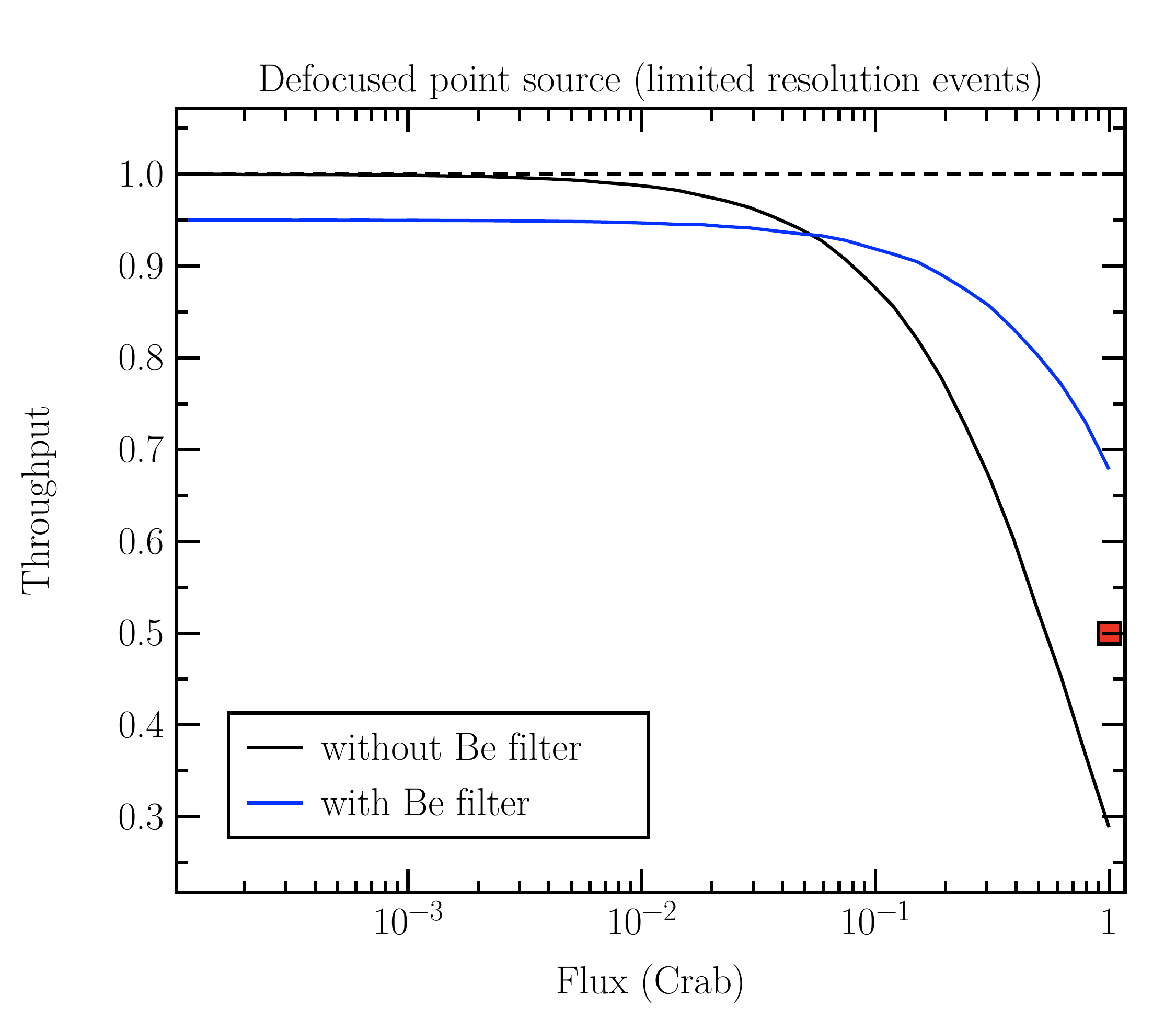}
    \includegraphics[width=0.5\linewidth]{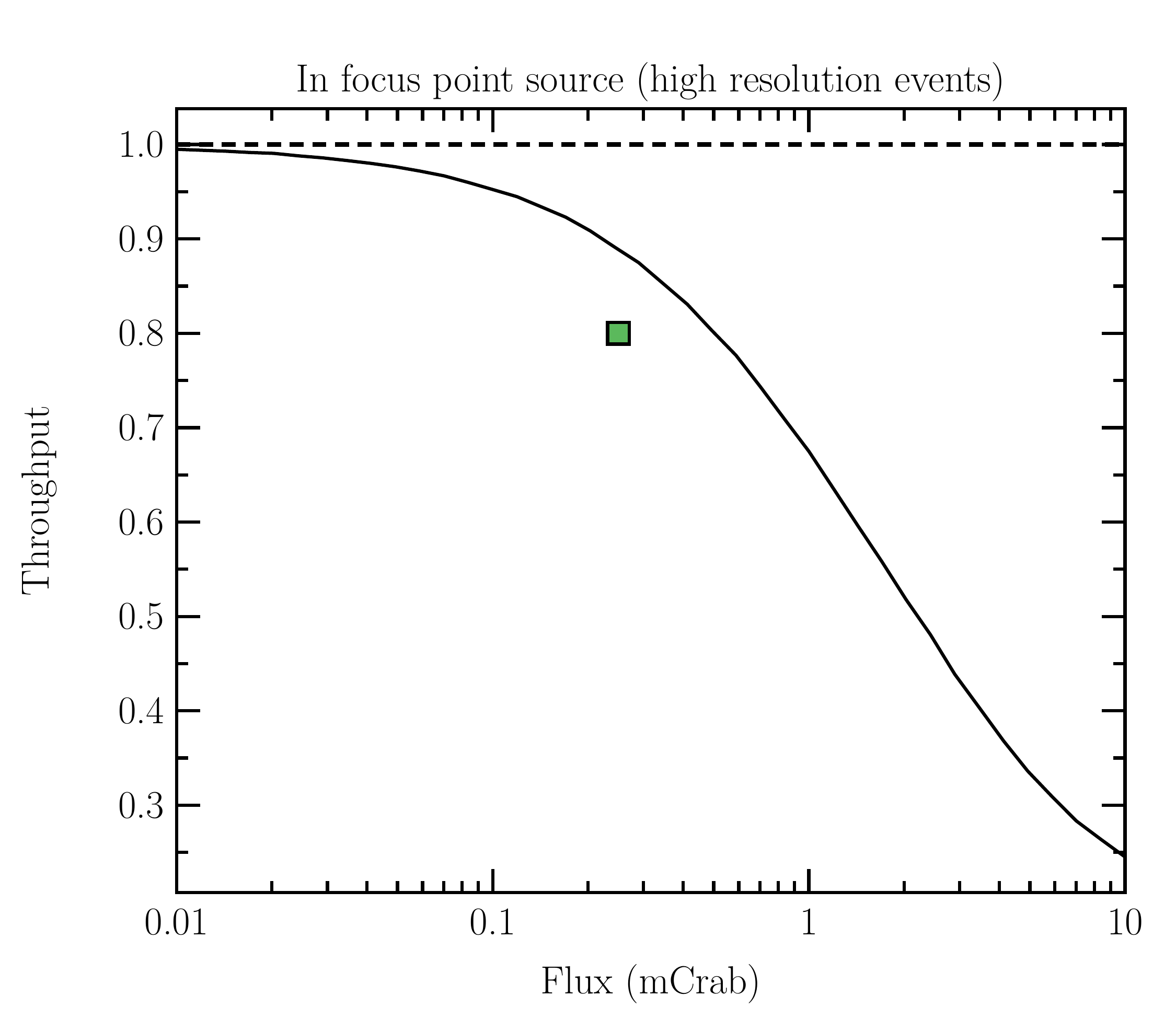}\includegraphics[width=0.5\linewidth]{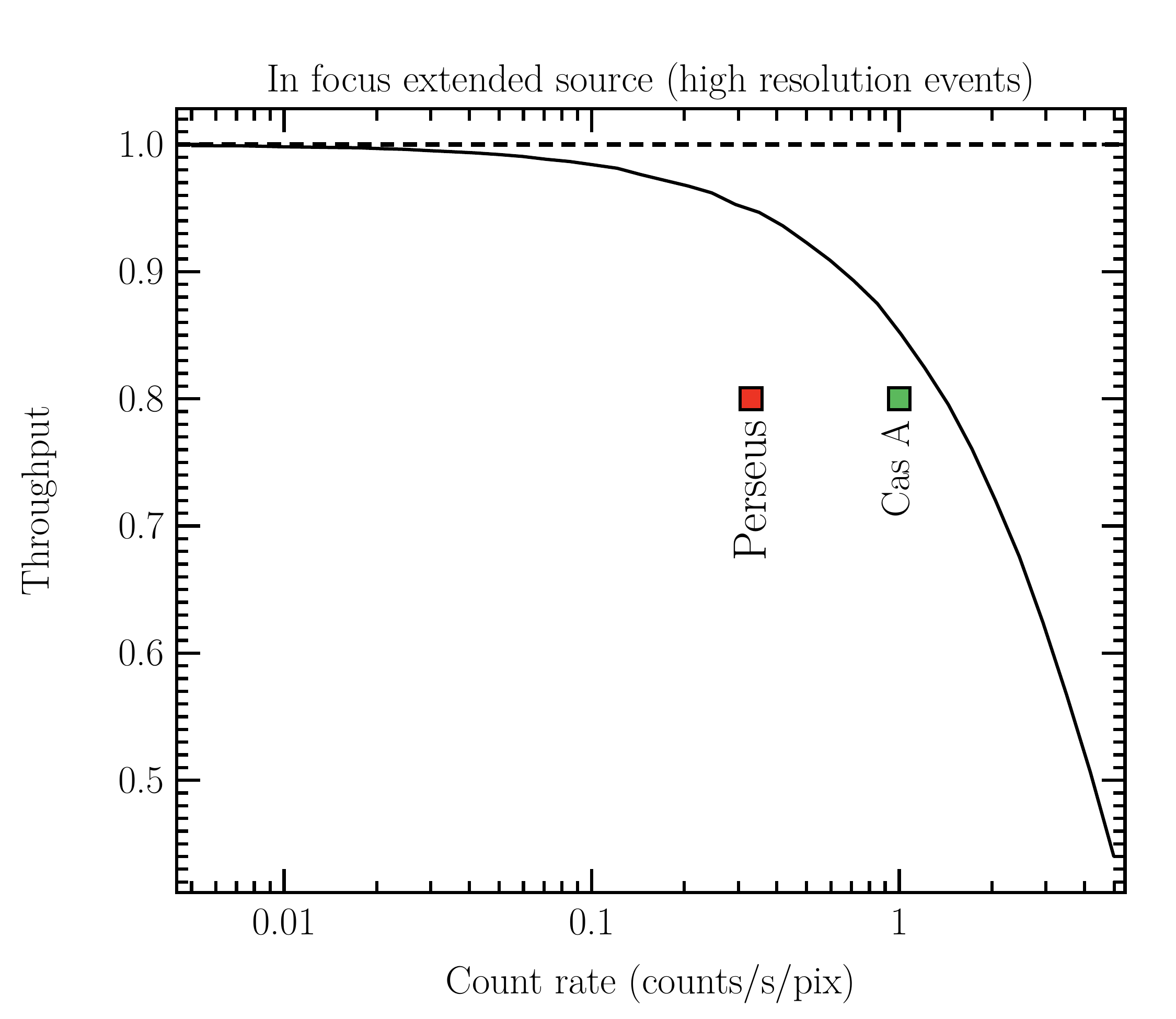}
    \caption{Count rate capabilities of the X-IFU accounting for grading and crosstalk. \emph{Top:} Throughput as a function of the flux for defocused point sources with (blue) or without (black) Be filter. The plot is provided for the high-resolution grade (3\,eV, \emph{left}) and the limited-resolution grade (10\,eV, \emph{right}) over the 5-8\,keV band. \emph{Bottom:} High-resolution throughput for in focus point sources (\emph{left}) and for extended sources (\emph{right}) over the full spectrum as a function of the flux and the count rate at detector level (cts/s/pix) respectively. The red and green squares report the requirement and the goal respectively (see Table~\ref{tab:ctrate}).}
    \label{fig:xifu_ctr_summary}
\end{figure}

The throughput results are summarized in Table~\ref{tab:ctrate} for the various cases illustrated in Figure~\ref{fig:xifu_ctr_summary}. Requirements and goals are also reminded. For extended sources, the new instrument configuration is able to reach its throughput requirement (Perseus cluster) and even the goal (Cassiopea A supernova remnant) with 10\,\% margin. For point sources, fluxes up to $\sim$~0.5\,mCrab can in principle be observed in focus. For brighter sources, a defocussing of the mirror (35\,mm) is planned to spread the photon counts over more pixels, allowing the observation of up to a few tens of mCrab. If even brighter sources are observed, the additional Be filter\footnote{We note that an alternative design based on thick polyimide and aluminum with similar transmission in the X-ray band is being considered.} can be used to strongly decrease the count rate below $\sim$\,4\,keV. Over the 5 to 8\,keV band, this results in a $\sim$\,5\,\% loss in count rate (see blue curves Figure~\ref{fig:xifu_ctr_summary}) but allows the observation of sources up to the Crab with an energy resolution better than 10\,eV FWHM at 7\,keV using shorter optimal filters (lower grading options).

We see that the X-IFU throughput performance remains similar to the previous configuration (see \cite{Barret_2023ExA....55..373B}). This is due to two compensating effects. On one hand, the decrease of mirror effective area leads to a lower count rate at the focal plane level (and thus higher throughput) at a given source flux. On the other hand, the new slower pixel gives a lower throughput at a given count rate. We however emphasize that the total valid output count rate of the X-IFU for bright sources will remain lower than before as conserving a similar throughput performance will not compensate the loss of effective area.

\section{The new X-IFU schedule and development logic}

\begin{figure}
    \centering
    \includegraphics[width=1\linewidth]{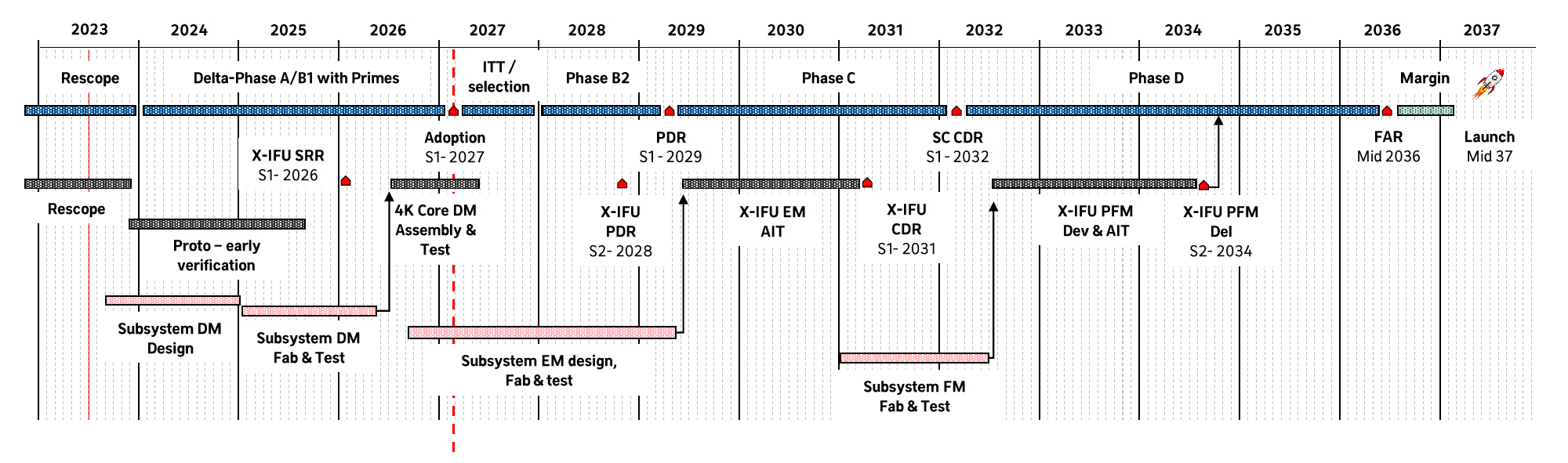}
    \caption{Latest development schedule of the X-IFU instrument. The main phases of the mission/spacecraft ESA studies are represented in blue and at the top of the figure. The corresponding X-IFU instrument level and subsystem activities are shown in grey and pink in the middle and bottom parts of the schedule respectively.}
    \label{fig:schedule}
\end{figure}

Figure \ref{fig:schedule} gives the latest X-IFU high level schedule as iterated with ESA, in coherence with the mission level development logic. The NewAthena mission adoption is now positioned in early 2027, for a launch date in mid-2037. Accounting for $\sim$ one year of contractualization/industrial prime selection phase at ESA, this leads to an overall implementation phase at system level of 9 to 10 years. This duration is consistent with the objective that ESA gave to limit the schedule in order to control the mission cost at completion. 

At X-IFU level, the development plan before mission adoption remains focused towards technology development. Table \ref{tab:demo} lists all of the activities currently scheduled at the level of the different subsystems. Whereas most of them are the direct (or slightly adapted) continuation of the efforts from the previous instrument study, a number of key additions can be noted to address the new  architecture:
\begin{table}
    \centering
    \includegraphics[width=1\linewidth]{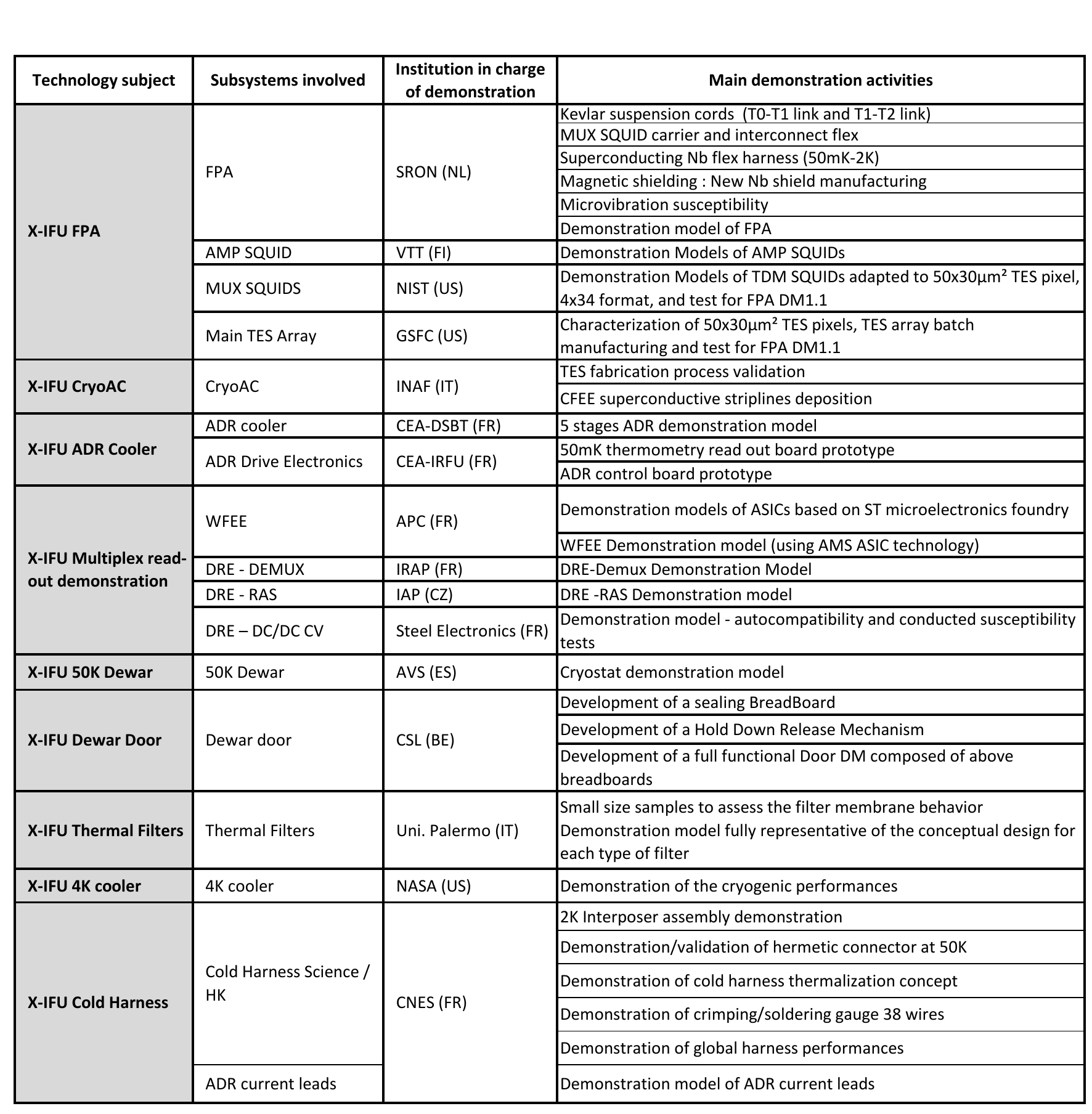}
    \caption{Summary of demonstration activities planned at X-IFU level in preparation of the mission adoption}
    \label{tab:demo}
\end{table}
\begin{itemize}
    \item A new FPA demonstration model (FPA DM1.1) is going to be built and tested by SRON (see \cite{Weers2024}). Whereas this model will re-use most of the thermo-mechanical hardware of the FPA DM1.0, it will now implement a full-sized TES array based on the latest pixel design as well as representative TDM cold electronics (MUX SQUIDs at 50\,mK, AMP SQUIDs at 2\,K, as well as interface networks at 2\,K representative of the foreseen flight design).
    \item CEA/dSBT will develop and test a full 5-stage ADR DM model to demonstrate the cryogenic performance required by the new architecture.
    \item A cryostat DM is foreseen by AVS to demonstrate the key elements of the 50\,K Dewar architecture.
    \item Breadboards of the key technologies of the Dewar Door (for the sealing principle, and for the opening mechanism) are planned by CSL. These will then be combined in a full functional DM model.
    \item NASA is conducting a technology demonstration program with potential US cryocooler vendors. Its prime objective is to demonstrate the heat lifts at 4.5 and 20 K required by X-IFU within the NewAthena PLC constraints (notably input power).
\end{itemize}

\begin{figure}
    \centering
    \includegraphics[width=1\linewidth]{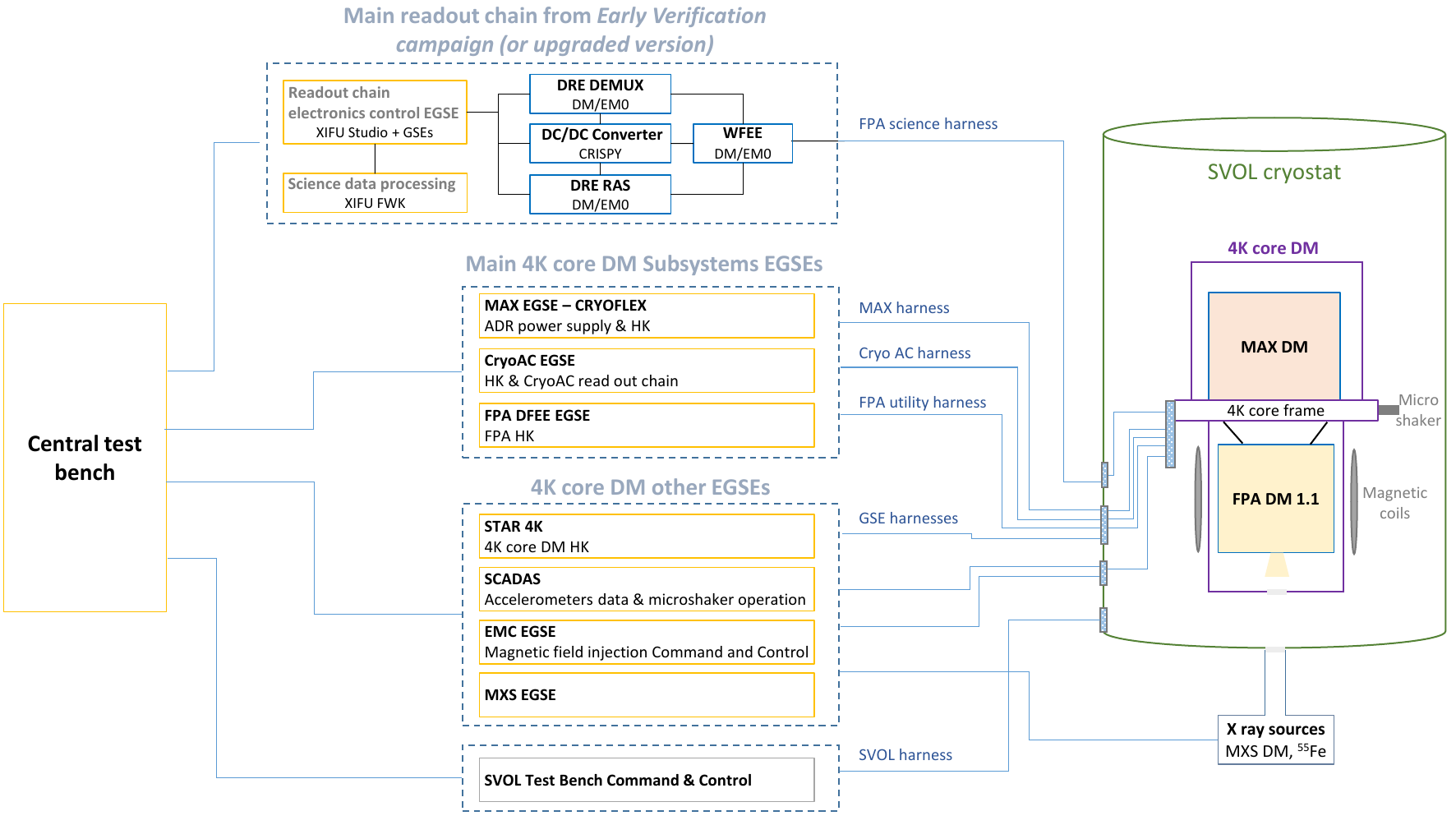}
    \caption{Preliminary test configuration for the 4\,K core DM.}
    \label{fig:4KcoreDM}
\end{figure}

Beyond these subsystem level demonstrations, two system level activities are planned to de-risk the X-IFU instrument development:
\begin{itemize}
    \item The detection chain early verification campaign (activity already scheduled before the mission reformulation): the WFEE DM, DRE-DEMUX DM, and DRE-RAS DM will be coupled together and ultimately to XIFU-like balanced differential cold electronics and detector in a test cryostat at IRAP (see \cite{Geoffray2024,Castellani2022,Beaumont2022}). The aim of this campaign is to consolidate early, well in advance of the EM campaign, the main TDM detection chain design choices and overall validate its main functions, interfaces and preliminary performances.
    \item The 4\,K core DM (see Figure \ref{fig:4KcoreDM}): The FPA DM1.1 and ADR DM will both be integrated in a demonstration model of the 4\,K core and tested inside a laboratory cryostat at CNES (SVOL). Laboratory electronics will be used to operate most of the setup (multi-stage ADR control, FPA and 4\,K core thermometry, cryoAC readout), whereas the main TDM readout will be performed with demonstration models of the WFEE and DRE (either the current DMs and/or upgraded versions) in order to operate a minimum 4 readout channels, with a goal of implementing 8. The main objectives of this activity are 1) to validate design choices and manufacturing of specific items for the 4\,K core in advance of the EM, 2) to consolidate the integration sequence of the FPA and ADR in the 4\,K core structure, 3) to test and optimize the regulation schemes of all the thermal stages of the 4\,K core (50\,mK, 325\,mK, 1.8\,K, and 4.5\,K), 4) to gain confidence in the thermal (and to a lesser extent mechanical - representativeness on this point is not a primary objective) mathematical modeling of the 4\,K core to inform the design of later models, 5) to perform a functional verification of the 4\,K core (including magnetic compatibility between the ADR and the FPA) and assess the sensitivity of its performance to external perturbations (µvibrations and magnetic fields), 6) to verify the compatibility between the 4\,K core and the X-IFU readout chain. As shown in Figure \ref{fig:schedule}, the 4\,K core DM integration and test is scheduled for mid-2026 to mid-2027, such that full testing is not expected to be completed before mission adoption, but will be done in advance of the EM development. 
\end{itemize}

After adoption, the instrument development follows a relatively classical Engineering Model (EM) and Proto-Flight Model (PFM) approach. Worthy of note, with the X-IFU Dewar now in the instrument perimeter, instrument verification, qualification and calibration will be performed at CNES (see \cite{Molin2024} for a description of the X-IFU calibration plan). This will require the development of a dedicated ground cryostat (Thermal Ground Support Equipment) in order to test the 50\,K Dewar Assembly and the X-IFU readout and control electronics. In this configuration, the 50\,K Dewar and 4\,K core will be cooled down conductively by ground cryo-coolers. Whereas special care will be taken to replicate and test the main interfaces of the instrument with the PLC and the CCU, this will naturally limit in particular the representativeness of the 50\,K Dewar, DEA and cold harness radiative environment. This element of the NewAthena design can only be verified at PLC level, a level at which coupling between the CCU and X-IFU will also be assessed. For that purpose, the X-IFU EM model will be delivered to ESA in 2031 in order to support a PLC Cryogenic Qualification Model test in a vacuum chamber. The X-IFU PFM model is then expected to be delivered in 2034, for integration and test at higher level, before a launch in 2037. 

\section{Environmental considerations} A Life Cycle Assessment (LCA) of the X-IFU, prior to entering the reformulation phase, was performed \cite{Barret2024ExA....57...19B}. It was found that the most significant environmental impacts arise from:
\begin{itemize}
    \item testing activities, related to energy consumption in clean rooms,
    \item office work, related to energy consumption in office buildings,
    \item and instrument manufacturing, related to the use of mineral and metal resources.
\end{itemize}  
In addition, travels remain an area of concern, despite the policy to reduce flying adopted by the Consortium. As described above, the instrument design has evolved, so as the perimeter of responsibility of the Consortium. A new LCA is being computed, in house, through the use of the SIMAPRO modeling software, fed by a more thorough collection of data, involving all subsystems. This will confirm the hot spots previously identified, further consolidating the areas over which to take actions, as to reduce the environmental footprint of X-IFU, while complying with the stringent requirements in terms of performance and risk management imposed to space projects. 

\section{Conclusions}
The X-IFU passed the reformulation of the Athena mission with minimal degradation of its anticipated performance, and contributed to maintain the flagship status of the NewAthena mission. Yet its simplified architecture and the transfer of some responsibilities from ESA to NASA (e.g. the 4\,K cooler) and to the X-IFU consortium (e.g. the 50\,K Dewar and the downstream AIT/AIV activities) enabled a reduction in the overall cost of the mission, to an envelope affordable with the ESA science program. As a consequence, the industrial activities on NewAthena have now started. The success of these studies condition the start of the mission adoption review process, expected to kick-off in June 2026. In parallel, ramping up the critical technologies is also required to reach the highest Technology Readiness Level at adoption, and demonstrate that all critical items of X-IFU are under control, to secure its implementation schedule. It is the prime commitment of the X-IFU Consortium to make everything possible to secure the adoption of NewAthena in the ESA science program early 2027, so that the implementation of the mission can start towards the goal of launching NewAthena in 2037. 
\section{Acknowledgments}

It is a pleasure to thank all the members of the Athena community and beyond, for their support and guidance along the reformulation of the Athena mission. A special thank to the Science Redefinition Team and its Chairs, Mike Cruise and Matteo Guainazzi, for their constructive and positive interactions with the X-IFU consortium leads. Thank you also the mission reformulation team, involving ESA, NASA, JAXA and the WFI and X-IFU consortia, for leading the successful reformulation of Athena. Finally, we would like to re-express our gratitude to the lead funding agencies supporting X-IFU for their unfailing and vigorous support, at a time when the future of the mission was not secured, and from now on, towards the goal of adopting NewAthena in 2027. 

The Italian contribution to XIFU is supported by ASI (Italian Space Agency) through Contract No. 2019-27-HH.0. The X-IFU consortium members working at Instituto de Física de Cantabria acknowledge Grant PID2021\_122955OB-C41 funded by MCIN/AEI/10.13039/501100011033 and by “ERDF A way of making Europe".

\bibliography{biblio.bib}


\begin{thebibliography}{42}
\ifx \bisbn   \undefined \def \bisbn  #1{ISBN #1}\fi
\ifx \binits  \undefined \def \binits#1{#1}\fi
\ifx \bauthor  \undefined \def \bauthor#1{#1}\fi
\ifx \batitle  \undefined \def \batitle#1{#1}\fi
\ifx \bjtitle  \undefined \def \bjtitle#1{#1}\fi
\ifx \bvolume  \undefined \def \bvolume#1{\textbf{#1}}\fi
\ifx \byear  \undefined \def \byear#1{#1}\fi
\ifx \bissue  \undefined \def \bissue#1{#1}\fi
\ifx \bfpage  \undefined \def \bfpage#1{#1}\fi
\ifx \blpage  \undefined \def \blpage #1{#1}\fi
\ifx \burl  \undefined \def \burl#1{\textsf{#1}}\fi
\ifx \doiurl  \undefined \def \doiurl#1{\url{https://doi.org/#1}}\fi
\ifx \betal  \undefined \def \betal{\textit{et al.}}\fi
\ifx \binstitute  \undefined \def \binstitute#1{#1}\fi
\ifx \binstitutionaled  \undefined \def \binstitutionaled#1{#1}\fi
\ifx \bctitle  \undefined \def \bctitle#1{#1}\fi
\ifx \beditor  \undefined \def \beditor#1{#1}\fi
\ifx \bpublisher  \undefined \def \bpublisher#1{#1}\fi
\ifx \bbtitle  \undefined \def \bbtitle#1{#1}\fi
\ifx \bedition  \undefined \def \bedition#1{#1}\fi
\ifx \bseriesno  \undefined \def \bseriesno#1{#1}\fi
\ifx \blocation  \undefined \def \blocation#1{#1}\fi
\ifx \bsertitle  \undefined \def \bsertitle#1{#1}\fi
\ifx \bsnm \undefined \def \bsnm#1{#1}\fi
\ifx \bsuffix \undefined \def \bsuffix#1{#1}\fi
\ifx \bparticle \undefined \def \bparticle#1{#1}\fi
\ifx \barticle \undefined \def \barticle#1{#1}\fi
\bibcommenthead
\ifx \bconfdate \undefined \def \bconfdate #1{#1}\fi
\ifx \botherref \undefined \def \botherref #1{#1}\fi
\ifx \url \undefined \def \url#1{\textsf{#1}}\fi
\ifx \bchapter \undefined \def \bchapter#1{#1}\fi
\ifx \bbook \undefined \def \bbook#1{#1}\fi
\ifx \bcomment \undefined \def \bcomment#1{#1}\fi
\ifx \oauthor \undefined \def \oauthor#1{#1}\fi
\ifx \citeauthoryear \undefined \def \citeauthoryear#1{#1}\fi
\ifx \endbibitem  \undefined \def \endbibitem {}\fi
\ifx \bconflocation  \undefined \def \bconflocation#1{#1}\fi
\ifx \arxivurl  \undefined \def \arxivurl#1{\textsf{#1}}\fi
\csname PreBibitemsHook\endcsname

\bibitem[\protect\citeauthoryear{{Barret} et~al.}{2023}]{Barret_2023ExA....55..373B}
\begin{barticle}
\bauthor{\bsnm{{Barret}}, \binits{D.}},
\bauthor{\bsnm{{Albouys}}, \binits{V.}},
\bauthor{\bsnm{{Herder}}, \binits{J.-W.d.}},
\bauthor{\bsnm{{Piro}}, \binits{L.}},
\bauthor{\bsnm{{Cappi}}, \binits{M.}},
\bauthor{\bsnm{{Huovelin}}, \binits{J.}},
\bauthor{\bsnm{{Kelley}}, \binits{R.}},
\bauthor{\bsnm{{Mas-Hesse}}, \binits{J.M.}},
\bauthor{\bsnm{{Paltani}}, \binits{S.}},
\bauthor{\bsnm{{Rauw}}, \binits{G.}},
\bauthor{\bsnm{{Rozanska}}, \binits{A.}},
\bauthor{\bsnm{{Svoboda}}, \binits{J.}},
\bauthor{\bsnm{{Wilms}}, \binits{J.}},
\bauthor{\bsnm{{Yamasaki}}, \binits{N.}},
\bauthor{\bsnm{{Audard}}, \binits{M.}},
\bauthor{\bsnm{{Bandler}}, \binits{S.}},
\bauthor{\bsnm{{Barbera}}, \binits{M.}},
\bauthor{\bsnm{{Barcons}}, \binits{X.}},
\bauthor{\bsnm{{Bozzo}}, \binits{E.}},
\bauthor{\bsnm{{Ceballos}}, \binits{M.T.}},
\bauthor{\bsnm{{Charles}}, \binits{I.}},
\bauthor{\bsnm{{Costantini}}, \binits{E.}},
\bauthor{\bsnm{{Dauser}}, \binits{T.}},
\bauthor{\bsnm{{Decourchelle}}, \binits{A.}},
\bauthor{\bsnm{{Duband}}, \binits{L.}},
\bauthor{\bsnm{{Duval}}, \binits{J.-M.}},
\bauthor{\bsnm{{Fiore}}, \binits{F.}},
\bauthor{\bsnm{{Gatti}}, \binits{F.}},
\bauthor{\bsnm{{Goldwurm}}, \binits{A.}},
\bauthor{\bsnm{{Hartog}}, \binits{R.d.}},
\bauthor{\bsnm{{Jackson}}, \binits{B.}},
\bauthor{\bsnm{{Jonker}}, \binits{P.}},
\bauthor{\bsnm{{Kilbourne}}, \binits{C.}},
\bauthor{\bsnm{{Korpela}}, \binits{S.}},
\bauthor{\bsnm{{Macculi}}, \binits{C.}},
\bauthor{\bsnm{{Mendez}}, \binits{M.}},
\bauthor{\bsnm{{Mitsuda}}, \binits{K.}},
\bauthor{\bsnm{{Molendi}}, \binits{S.}},
\bauthor{\bsnm{{Pajot}}, \binits{F.}},
\bauthor{\bsnm{{Pointecouteau}}, \binits{E.}},
\bauthor{\bsnm{{Porter}}, \binits{F.}},
\bauthor{\bsnm{{Pratt}}, \binits{G.W.}},
\bauthor{\bsnm{{Pr{\^e}le}}, \binits{D.}},
\bauthor{\bsnm{{Ravera}}, \binits{L.}},
\bauthor{\bsnm{{Sato}}, \binits{K.}},
\bauthor{\bsnm{{Schaye}}, \binits{J.}},
\bauthor{\bsnm{{Shinozaki}}, \binits{K.}},
\bauthor{\bsnm{{Skup}}, \binits{K.}},
\bauthor{\bsnm{{Soucek}}, \binits{J.}},
\bauthor{\bsnm{{Thibert}}, \binits{T.}},
\bauthor{\bsnm{{Vink}}, \binits{J.}},
\bauthor{\bsnm{{Webb}}, \binits{N.}},
\bauthor{\bsnm{{Chaoul}}, \binits{L.}},
\bauthor{\bsnm{{Raulin}}, \binits{D.}},
\bauthor{\bsnm{{Simionescu}}, \binits{A.}},
\bauthor{\bsnm{{Torrejon}}, \binits{J.M.}},
\bauthor{\bsnm{{Acero}}, \binits{F.}},
\bauthor{\bsnm{{Branduardi-Raymont}}, \binits{G.}},
\bauthor{\bsnm{{Ettori}}, \binits{S.}},
\bauthor{\bsnm{{Finoguenov}}, \binits{A.}},
\bauthor{\bsnm{{Grosso}}, \binits{N.}},
\bauthor{\bsnm{{Kaastra}}, \binits{J.}},
\bauthor{\bsnm{{Mazzotta}}, \binits{P.}},
\bauthor{\bsnm{{Miller}}, \binits{J.}},
\bauthor{\bsnm{{Miniutti}}, \binits{G.}},
\bauthor{\bsnm{{Nicastro}}, \binits{F.}},
\bauthor{\bsnm{{Sciortino}}, \binits{S.}},
\bauthor{\bsnm{{Yamaguchi}}, \binits{H.}},
\bauthor{\bsnm{{Beaumont}}, \binits{S.}},
\bauthor{\bsnm{{Cucchetti}}, \binits{E.}},
\bauthor{\bsnm{{D'Andrea}}, \binits{M.}},
\bauthor{\bsnm{{Eckart}}, \binits{M.}},
\bauthor{\bsnm{{Ferrando}}, \binits{P.}},
\bauthor{\bsnm{{Kammoun}}, \binits{E.}},
\bauthor{\bsnm{{Lotti}}, \binits{S.}},
\bauthor{\bsnm{{Mesnager}}, \binits{J.-M.}},
\bauthor{\bsnm{{Natalucci}}, \binits{L.}},
\bauthor{\bsnm{{Peille}}, \binits{P.}},
\bauthor{\bsnm{{de Plaa}}, \binits{J.}},
\bauthor{\bsnm{{Ardellier}}, \binits{F.}},
\bauthor{\bsnm{{Argan}}, \binits{A.}},
\bauthor{\bsnm{{Bellouard}}, \binits{E.}},
\bauthor{\bsnm{{Carron}}, \binits{J.}},
\bauthor{\bsnm{{Cavazzuti}}, \binits{E.}},
\bauthor{\bsnm{{Fiorini}}, \binits{M.}},
\bauthor{\bsnm{{Khosropanah}}, \binits{P.}},
\bauthor{\bsnm{{Martin}}, \binits{S.}},
\bauthor{\bsnm{{Perry}}, \binits{J.}},
\bauthor{\bsnm{{Pinsard}}, \binits{F.}},
\bauthor{\bsnm{{Pradines}}, \binits{A.}},
\bauthor{\bsnm{{Rigano}}, \binits{M.}},
\bauthor{\bsnm{{Roelfsema}}, \binits{P.}},
\bauthor{\bsnm{{Schwander}}, \binits{D.}},
\bauthor{\bsnm{{Torrioli}}, \binits{G.}},
\bauthor{\bsnm{{Ullom}}, \binits{J.}},
\bauthor{\bsnm{{Vera}}, \binits{I.}},
\bauthor{\bsnm{{Villegas}}, \binits{E.M.}},
\bauthor{\bsnm{{Zuchniak}}, \binits{M.}},
\bauthor{\bsnm{{Brachet}}, \binits{F.}},
\bauthor{\bsnm{{Cicero}}, \binits{U.L.}},
\bauthor{\bsnm{{Doriese}}, \binits{W.}},
\bauthor{\bsnm{{Durkin}}, \binits{M.}},
\bauthor{\bsnm{{Fioretti}}, \binits{V.}},
\bauthor{\bsnm{{Geoffray}}, \binits{H.}},
\bauthor{\bsnm{{Jacques}}, \binits{L.}},
\bauthor{\bsnm{{Kirsch}}, \binits{C.}},
\bauthor{\bsnm{{Smith}}, \binits{S.}},
\bauthor{\bsnm{{Adams}}, \binits{J.}},
\bauthor{\bsnm{{Gloaguen}}, \binits{E.}},
\bauthor{\bsnm{{Hoogeveen}}, \binits{R.}},
\bauthor{\bsnm{{van der Hulst}}, \binits{P.}},
\bauthor{\bsnm{{Kiviranta}}, \binits{M.}},
\bauthor{\bsnm{{van der Kuur}}, \binits{J.}},
\bauthor{\bsnm{{Ledot}}, \binits{A.}},
\bauthor{\bsnm{{van Leeuwen}}, \binits{B.-J.}},
\bauthor{\bsnm{{van Loon}}, \binits{D.}},
\bauthor{\bsnm{{Lyautey}}, \binits{B.}},
\bauthor{\bsnm{{Parot}}, \binits{Y.}},
\bauthor{\bsnm{{Sakai}}, \binits{K.}},
\bauthor{\bsnm{{van Weers}}, \binits{H.}},
\bauthor{\bsnm{{Abdoelkariem}}, \binits{S.}},
\bauthor{\bsnm{{Adam}}, \binits{T.}},
\bauthor{\bsnm{{Adami}}, \binits{C.}},
\bauthor{\bsnm{{Aicardi}}, \binits{C.}},
\bauthor{\bsnm{{Akamatsu}}, \binits{H.}},
\bauthor{\bsnm{{Alonso}}, \binits{P.E.M.}},
\bauthor{\bsnm{{Amato}}, \binits{R.}},
\bauthor{\bsnm{{Andr{\'e}}}, \binits{J.}},
\bauthor{\bsnm{{Angelinelli}}, \binits{M.}},
\bauthor{\bsnm{{Anon-Cancela}}, \binits{M.}},
\bauthor{\bsnm{{Anvar}}, \binits{S.}},
\bauthor{\bsnm{{Atienza}}, \binits{R.}},
\bauthor{\bsnm{{Attard}}, \binits{A.}},
\bauthor{\bsnm{{Auricchio}}, \binits{N.}},
\bauthor{\bsnm{{Balado}}, \binits{A.}},
\bauthor{\bsnm{{Bancel}}, \binits{F.}},
\bauthor{\bsnm{{Barusso}}, \binits{L.F.}},
\bauthor{\bsnm{{Bascu{\~n}an}}, \binits{A.}},
\bauthor{\bsnm{{Bernard}}, \binits{V.}},
\bauthor{\bsnm{{Berrocal}}, \binits{A.}},
\bauthor{\bsnm{{Blin}}, \binits{S.}},
\bauthor{\bsnm{{Bonino}}, \binits{D.}},
\bauthor{\bsnm{{Bonnet}}, \binits{F.}},
\bauthor{\bsnm{{Bonny}}, \binits{P.}},
\bauthor{\bsnm{{Boorman}}, \binits{P.}},
\bauthor{\bsnm{{Boreux}}, \binits{C.}},
\bauthor{\bsnm{{Bounab}}, \binits{A.}},
\bauthor{\bsnm{{Boutelier}}, \binits{M.}},
\bauthor{\bsnm{{Boyce}}, \binits{K.}},
\bauthor{\bsnm{{Brienza}}, \binits{D.}},
\bauthor{\bsnm{{Bruijn}}, \binits{M.}},
\bauthor{\bsnm{{Bulgarelli}}, \binits{A.}},
\bauthor{\bsnm{{Calarco}}, \binits{S.}},
\bauthor{\bsnm{{Callanan}}, \binits{P.}},
\bauthor{\bsnm{{Campello}}, \binits{A.P.}},
\bauthor{\bsnm{{Camus}}, \binits{T.}},
\bauthor{\bsnm{{Canourgues}}, \binits{F.}},
\bauthor{\bsnm{{Capobianco}}, \binits{V.}},
\bauthor{\bsnm{{Cardiel}}, \binits{N.}},
\bauthor{\bsnm{{Castellani}}, \binits{F.}},
\bauthor{\bsnm{{Cheatom}}, \binits{O.}},
\bauthor{\bsnm{{Chervenak}}, \binits{J.}},
\bauthor{\bsnm{{Chiarello}}, \binits{F.}},
\bauthor{\bsnm{{Clerc}}, \binits{L.}},
\bauthor{\bsnm{{Clerc}}, \binits{N.}},
\bauthor{\bsnm{{Cobo}}, \binits{B.}},
\bauthor{\bsnm{{Coeur-Joly}}, \binits{O.}},
\bauthor{\bsnm{{Coleiro}}, \binits{A.}},
\bauthor{\bsnm{{Colonges}}, \binits{S.}},
\bauthor{\bsnm{{Corcione}}, \binits{L.}},
\bauthor{\bsnm{{Coriat}}, \binits{M.}},
\bauthor{\bsnm{{Coynel}}, \binits{A.}},
\bauthor{\bsnm{{Cuttaia}}, \binits{F.}},
\bauthor{\bsnm{{D'Ai}}, \binits{A.}},
\bauthor{\bsnm{{D'anca}}, \binits{F.}},
\bauthor{\bsnm{{Dadina}}, \binits{M.}},
\bauthor{\bsnm{{Daniel}}, \binits{C.}},
\bauthor{\bsnm{{Dauner}}, \binits{L.}},
\bauthor{\bsnm{{DeNigris}}, \binits{N.}},
\bauthor{\bsnm{{Dercksen}}, \binits{J.}},
\bauthor{\bsnm{{DiPirro}}, \binits{M.}},
\bauthor{\bsnm{{Doumayrou}}, \binits{E.}},
\bauthor{\bsnm{{Dubbeldam}}, \binits{L.}},
\bauthor{\bsnm{{Dupieux}}, \binits{M.}},
\bauthor{\bsnm{{Dupourqu{\'e}}}, \binits{S.}},
\bauthor{\bsnm{{Durand}}, \binits{J.L.}},
\bauthor{\bsnm{{Eckert}}, \binits{D.}},
\bauthor{\bsnm{{Eiriz}}, \binits{V.}},
\bauthor{\bsnm{{Ercolani}}, \binits{E.}},
\bauthor{\bsnm{{Etcheverry}}, \binits{C.}},
\bauthor{\bsnm{{Finkbeiner}}, \binits{F.}},
\bauthor{\bsnm{{Fiocchi}}, \binits{M.}},
\bauthor{\bsnm{{Fossecave}}, \binits{H.}},
\bauthor{\bsnm{{Franssen}}, \binits{P.}},
\bauthor{\bsnm{{Frericks}}, \binits{M.}},
\bauthor{\bsnm{{Gabici}}, \binits{S.}},
\bauthor{\bsnm{{Gant}}, \binits{F.}},
\bauthor{\bsnm{{Gao}}, \binits{J.-R.}},
\bauthor{\bsnm{{Gastaldello}}, \binits{F.}},
\bauthor{\bsnm{{Genolet}}, \binits{L.}},
\bauthor{\bsnm{{Ghizzardi}}, \binits{S.}},
\bauthor{\bsnm{{Gil}}, \binits{M.A.A.}},
\bauthor{\bsnm{{Giovannini}}, \binits{E.}},
\bauthor{\bsnm{{Godet}}, \binits{O.}},
\bauthor{\bsnm{{Gomez-Elvira}}, \binits{J.}},
\bauthor{\bsnm{{Gonzalez}}, \binits{R.}},
\bauthor{\bsnm{{Gonzalez}}, \binits{M.}},
\bauthor{\bsnm{{Gottardi}}, \binits{L.}},
\bauthor{\bsnm{{Granat}}, \binits{D.}},
\bauthor{\bsnm{{Gros}}, \binits{M.}},
\bauthor{\bsnm{{Guignard}}, \binits{N.}},
\bauthor{\bsnm{{Hieltjes}}, \binits{P.}},
\bauthor{\bsnm{{Hurtado}}, \binits{A.J.}},
\bauthor{\bsnm{{Irwin}}, \binits{K.}},
\bauthor{\bsnm{{Jacquey}}, \binits{C.}},
\bauthor{\bsnm{{Janiuk}}, \binits{A.}},
\bauthor{\bsnm{{Jaubert}}, \binits{J.}},
\bauthor{\bsnm{{Jim{\'e}nez}}, \binits{M.}},
\bauthor{\bsnm{{Jolly}}, \binits{A.}},
\bauthor{\bsnm{{Jourdan}}, \binits{T.}},
\bauthor{\bsnm{{Julien}}, \binits{S.}},
\bauthor{\bsnm{{Kedziora}}, \binits{B.}},
\bauthor{\bsnm{{Korb}}, \binits{A.}},
\bauthor{\bsnm{{Kreykenbohm}}, \binits{I.}},
\bauthor{\bsnm{{K{\"o}nig}}, \binits{O.}},
\bauthor{\bsnm{{Langer}}, \binits{M.}},
\bauthor{\bsnm{{Laudet}}, \binits{P.}},
\bauthor{\bsnm{{Laurent}}, \binits{P.}},
\bauthor{\bsnm{{Laurenza}}, \binits{M.}},
\bauthor{\bsnm{{Lesrel}}, \binits{J.}},
\bauthor{\bsnm{{Ligori}}, \binits{S.}},
\bauthor{\bsnm{{Lorenz}}, \binits{M.}},
\bauthor{\bsnm{{Luminari}}, \binits{A.}},
\bauthor{\bsnm{{Maffei}}, \binits{B.}},
\bauthor{\bsnm{{Maisonnave}}, \binits{O.}},
\bauthor{\bsnm{{Marelli}}, \binits{L.}},
\bauthor{\bsnm{{Massonet}}, \binits{D.}},
\bauthor{\bsnm{{Maussang}}, \binits{I.}},
\bauthor{\bsnm{{Melchor}}, \binits{A.G.}},
\bauthor{\bsnm{{Le Mer}}, \binits{I.}},
\bauthor{\bsnm{{Millan}}, \binits{F.J.S.}},
\bauthor{\bsnm{{Millerioux}}, \binits{J.-P.}},
\bauthor{\bsnm{{Mineo}}, \binits{T.}},
\bauthor{\bsnm{{Minervini}}, \binits{G.}},
\bauthor{\bsnm{{Molin}}, \binits{A.}},
\bauthor{\bsnm{{Monestes}}, \binits{D.}},
\bauthor{\bsnm{{Montinaro}}, \binits{N.}},
\bauthor{\bsnm{{Mot}}, \binits{B.}},
\bauthor{\bsnm{{Murat}}, \binits{D.}},
\bauthor{\bsnm{{Nagayoshi}}, \binits{K.}},
\bauthor{\bsnm{{Naz{\'e}}}, \binits{Y.}},
\bauthor{\bsnm{{Nogu{\`e}s}}, \binits{L.}},
\bauthor{\bsnm{{Pailot}}, \binits{D.}},
\bauthor{\bsnm{{Panessa}}, \binits{F.}},
\bauthor{\bsnm{{Parodi}}, \binits{L.}},
\bauthor{\bsnm{{Petit}}, \binits{P.}},
\bauthor{\bsnm{{Piconcelli}}, \binits{E.}},
\bauthor{\bsnm{{Pinto}}, \binits{C.}},
\bauthor{\bsnm{{Plaza}}, \binits{J.M.E.}},
\bauthor{\bsnm{{Plaza}}, \binits{B.}},
\bauthor{\bsnm{{Poyatos}}, \binits{D.}},
\bauthor{\bsnm{{Prouv{\'e}}}, \binits{T.}},
\bauthor{\bsnm{{Ptak}}, \binits{A.}},
\bauthor{\bsnm{{Puccetti}}, \binits{S.}},
\bauthor{\bsnm{{Puccio}}, \binits{E.}},
\bauthor{\bsnm{{Ramon}}, \binits{P.}},
\bauthor{\bsnm{{Reina}}, \binits{M.}},
\bauthor{\bsnm{{Rioland}}, \binits{G.}},
\bauthor{\bsnm{{Rodriguez}}, \binits{L.}},
\bauthor{\bsnm{{Roig}}, \binits{A.}},
\bauthor{\bsnm{{Rollet}}, \binits{B.}},
\bauthor{\bsnm{{Roncarelli}}, \binits{M.}},
\bauthor{\bsnm{{Roudil}}, \binits{G.}},
\bauthor{\bsnm{{Rudnicki}}, \binits{T.}},
\bauthor{\bsnm{{Sanisidro}}, \binits{J.}},
\bauthor{\bsnm{{Sciortino}}, \binits{L.}},
\bauthor{\bsnm{{Silva}}, \binits{V.}},
\bauthor{\bsnm{{Sordet}}, \binits{M.}},
\bauthor{\bsnm{{Soto-Aguilar}}, \binits{J.}},
\bauthor{\bsnm{{Spizzi}}, \binits{P.}},
\bauthor{\bsnm{{Surace}}, \binits{C.}},
\bauthor{\bsnm{{Fern{\'a}ndez S{\'a}nchez}}, \binits{M.}},
\bauthor{\bsnm{{Taralli}}, \binits{E.}},
\bauthor{\bsnm{{Terrasa}}, \binits{G.}},
\bauthor{\bsnm{{Terrier}}, \binits{R.}},
\bauthor{\bsnm{{Todaro}}, \binits{M.}},
\bauthor{\bsnm{{Ubertini}}, \binits{P.}},
\bauthor{\bsnm{{Uslenghi}}, \binits{M.}},
\bauthor{\bsnm{{de Vaate}}, \binits{J.G.B.}},
\bauthor{\bsnm{{Vaccaro}}, \binits{D.}},
\bauthor{\bsnm{{Varisco}}, \binits{S.}},
\bauthor{\bsnm{{Varni{\`e}re}}, \binits{P.}},
\bauthor{\bsnm{{Vibert}}, \binits{L.}},
\bauthor{\bsnm{{Vidriales}}, \binits{M.}},
\bauthor{\bsnm{{Villa}}, \binits{F.}},
\bauthor{\bsnm{{Vodopivec}}, \binits{B.M.}},
\bauthor{\bsnm{{Volpe}}, \binits{A.}},
\bauthor{\bsnm{{de Vries}}, \binits{C.}},
\bauthor{\bsnm{{Wakeham}}, \binits{N.}},
\bauthor{\bsnm{{Walmsley}}, \binits{G.}},
\bauthor{\bsnm{{Wise}}, \binits{M.}},
\bauthor{\bsnm{{de Wit}}, \binits{M.}},
\bauthor{\bsnm{{Wo{\'z}niak}}, \binits{G.}}:
\batitle{{The Athena X-ray Integral Field Unit: a consolidated design for the system requirement review of the preliminary definition phase}}.
\bjtitle{Experimental Astronomy}
\bvolume{55}(\bissue{2}),
\bfpage{373}--\blpage{426}
(\byear{2023})
\doiurl{10.1007/s10686-022-09880-7}
\end{barticle}
\endbibitem

\bibitem[\protect\citeauthoryear{{Nandra} et~al.}{2013}]{2013arXiv1306.2307N}
\begin{botherref}
\oauthor{\bsnm{{Nandra}}, \binits{K.}},
\oauthor{\bsnm{{Barret}}, \binits{D.}},
\oauthor{\bsnm{{Barcons}}, \binits{X.}},
\oauthor{\bsnm{{Fabian}}, \binits{A.}},
\oauthor{\bsnm{{den Herder}}, \binits{J.-W.}},
\oauthor{\bsnm{{Piro}}, \binits{L.}},
\oauthor{\bsnm{{Watson}}, \binits{M.}},
\oauthor{\bsnm{{Adami}}, \binits{C.}},
\oauthor{\bsnm{{Aird}}, \binits{J.}},
\oauthor{\bsnm{{Afonso}}, \binits{J.M.}},
\oauthor{\bsnm{{Alexander}}, \binits{D.}},
\oauthor{\bsnm{{Argiroffi}}, \binits{C.}},
\oauthor{\bsnm{{Amati}}, \binits{L.}},
\oauthor{\bsnm{{Arnaud}}, \binits{M.}},
\oauthor{\bsnm{{Atteia}}, \binits{J.-L.}},
\oauthor{\bsnm{{Audard}}, \binits{M.}},
\oauthor{\bsnm{{Badenes}}, \binits{C.}},
\oauthor{\bsnm{{Ballet}}, \binits{J.}},
\oauthor{\bsnm{{Ballo}}, \binits{L.}},
\oauthor{\bsnm{{Bamba}}, \binits{A.}},
\oauthor{\bsnm{{Bhardwaj}}, \binits{A.}},
\oauthor{\bsnm{{Stefano Battistelli}}, \binits{E.}},
\oauthor{\bsnm{{Becker}}, \binits{W.}},
\oauthor{\bsnm{{De Becker}}, \binits{M.}},
\oauthor{\bsnm{{Behar}}, \binits{E.}},
\oauthor{\bsnm{{Bianchi}}, \binits{S.}},
\oauthor{\bsnm{{Biffi}}, \binits{V.}},
\oauthor{\bsnm{{B{\^\i}rzan}}, \binits{L.}},
\oauthor{\bsnm{{Bocchino}}, \binits{F.}},
\oauthor{\bsnm{{Bogdanov}}, \binits{S.}},
\oauthor{\bsnm{{Boirin}}, \binits{L.}},
\oauthor{\bsnm{{Boller}}, \binits{T.}},
\oauthor{\bsnm{{Borgani}}, \binits{S.}},
\oauthor{\bsnm{{Borm}}, \binits{K.}},
\oauthor{\bsnm{{Bouch{\'e}}}, \binits{N.}},
\oauthor{\bsnm{{Bourdin}}, \binits{H.}},
\oauthor{\bsnm{{Bower}}, \binits{R.}},
\oauthor{\bsnm{{Braito}}, \binits{V.}},
\oauthor{\bsnm{{Branchini}}, \binits{E.}},
\oauthor{\bsnm{{Branduardi-Raymont}}, \binits{G.}},
\oauthor{\bsnm{{Bregman}}, \binits{J.}},
\oauthor{\bsnm{{Brenneman}}, \binits{L.}},
\oauthor{\bsnm{{Brightman}}, \binits{M.}},
\oauthor{\bsnm{{Br{\"u}ggen}}, \binits{M.}},
\oauthor{\bsnm{{Buchner}}, \binits{J.}},
\oauthor{\bsnm{{Bulbul}}, \binits{E.}},
\oauthor{\bsnm{{Brusa}}, \binits{M.}},
\oauthor{\bsnm{{Bursa}}, \binits{M.}},
\oauthor{\bsnm{{Caccianiga}}, \binits{A.}},
\oauthor{\bsnm{{Cackett}}, \binits{E.}},
\oauthor{\bsnm{{Campana}}, \binits{S.}},
\oauthor{\bsnm{{Cappelluti}}, \binits{N.}},
\oauthor{\bsnm{{Cappi}}, \binits{M.}},
\oauthor{\bsnm{{Carrera}}, \binits{F.}},
\oauthor{\bsnm{{Ceballos}}, \binits{M.}},
\oauthor{\bsnm{{Christensen}}, \binits{F.}},
\oauthor{\bsnm{{Chu}}, \binits{Y.-H.}},
\oauthor{\bsnm{{Churazov}}, \binits{E.}},
\oauthor{\bsnm{{Clerc}}, \binits{N.}},
\oauthor{\bsnm{{Corbel}}, \binits{S.}},
\oauthor{\bsnm{{Corral}}, \binits{A.}},
\oauthor{\bsnm{{Comastri}}, \binits{A.}},
\oauthor{\bsnm{{Costantini}}, \binits{E.}},
\oauthor{\bsnm{{Croston}}, \binits{J.}},
\oauthor{\bsnm{{Dadina}}, \binits{M.}},
\oauthor{\bsnm{{D'Ai}}, \binits{A.}},
\oauthor{\bsnm{{Decourchelle}}, \binits{A.}},
\oauthor{\bsnm{{Della Ceca}}, \binits{R.}},
\oauthor{\bsnm{{Dennerl}}, \binits{K.}},
\oauthor{\bsnm{{Dolag}}, \binits{K.}},
\oauthor{\bsnm{{Done}}, \binits{C.}},
\oauthor{\bsnm{{Dovciak}}, \binits{M.}},
\oauthor{\bsnm{{Drake}}, \binits{J.}},
\oauthor{\bsnm{{Eckert}}, \binits{D.}},
\oauthor{\bsnm{{Edge}}, \binits{A.}},
\oauthor{\bsnm{{Ettori}}, \binits{S.}},
\oauthor{\bsnm{{Ezoe}}, \binits{Y.}},
\oauthor{\bsnm{{Feigelson}}, \binits{E.}},
\oauthor{\bsnm{{Fender}}, \binits{R.}},
\oauthor{\bsnm{{Feruglio}}, \binits{C.}},
\oauthor{\bsnm{{Finoguenov}}, \binits{A.}},
\oauthor{\bsnm{{Fiore}}, \binits{F.}},
\oauthor{\bsnm{{Galeazzi}}, \binits{M.}},
\oauthor{\bsnm{{Gallagher}}, \binits{S.}},
\oauthor{\bsnm{{Gandhi}}, \binits{P.}},
\oauthor{\bsnm{{Gaspari}}, \binits{M.}},
\oauthor{\bsnm{{Gastaldello}}, \binits{F.}},
\oauthor{\bsnm{{Georgakakis}}, \binits{A.}},
\oauthor{\bsnm{{Georgantopoulos}}, \binits{I.}},
\oauthor{\bsnm{{Gilfanov}}, \binits{M.}},
\oauthor{\bsnm{{Gitti}}, \binits{M.}},
\oauthor{\bsnm{{Gladstone}}, \binits{R.}},
\oauthor{\bsnm{{Goosmann}}, \binits{R.}},
\oauthor{\bsnm{{Gosset}}, \binits{E.}},
\oauthor{\bsnm{{Grosso}}, \binits{N.}},
\oauthor{\bsnm{{Guedel}}, \binits{M.}},
\oauthor{\bsnm{{Guerrero}}, \binits{M.}},
\oauthor{\bsnm{{Haberl}}, \binits{F.}},
\oauthor{\bsnm{{Hardcastle}}, \binits{M.}},
\oauthor{\bsnm{{Heinz}}, \binits{S.}},
\oauthor{\bsnm{{Alonso Herrero}}, \binits{A.}},
\oauthor{\bsnm{{Herv{\'e}}}, \binits{A.}},
\oauthor{\bsnm{{Holmstrom}}, \binits{M.}},
\oauthor{\bsnm{{Iwasawa}}, \binits{K.}},
\oauthor{\bsnm{{Jonker}}, \binits{P.}},
\oauthor{\bsnm{{Kaastra}}, \binits{J.}},
\oauthor{\bsnm{{Kara}}, \binits{E.}},
\oauthor{\bsnm{{Karas}}, \binits{V.}},
\oauthor{\bsnm{{Kastner}}, \binits{J.}},
\oauthor{\bsnm{{King}}, \binits{A.}},
\oauthor{\bsnm{{Kosenko}}, \binits{D.}},
\oauthor{\bsnm{{Koutroumpa}}, \binits{D.}},
\oauthor{\bsnm{{Kraft}}, \binits{R.}},
\oauthor{\bsnm{{Kreykenbohm}}, \binits{I.}},
\oauthor{\bsnm{{Lallement}}, \binits{R.}},
\oauthor{\bsnm{{Lanzuisi}}, \binits{G.}},
\oauthor{\bsnm{{Lee}}, \binits{J.}},
\oauthor{\bsnm{{Lemoine-Goumard}}, \binits{M.}},
\oauthor{\bsnm{{Lobban}}, \binits{A.}},
\oauthor{\bsnm{{Lodato}}, \binits{G.}},
\oauthor{\bsnm{{Lovisari}}, \binits{L.}},
\oauthor{\bsnm{{Lotti}}, \binits{S.}},
\oauthor{\bsnm{{McCharthy}}, \binits{I.}},
\oauthor{\bsnm{{McNamara}}, \binits{B.}},
\oauthor{\bsnm{{Maggio}}, \binits{A.}},
\oauthor{\bsnm{{Maiolino}}, \binits{R.}},
\oauthor{\bsnm{{De Marco}}, \binits{B.}},
\oauthor{\bsnm{{de Martino}}, \binits{D.}},
\oauthor{\bsnm{{Mateos}}, \binits{S.}},
\oauthor{\bsnm{{Matt}}, \binits{G.}},
\oauthor{\bsnm{{Maughan}}, \binits{B.}},
\oauthor{\bsnm{{Mazzotta}}, \binits{P.}},
\oauthor{\bsnm{{Mendez}}, \binits{M.}},
\oauthor{\bsnm{{Merloni}}, \binits{A.}},
\oauthor{\bsnm{{Micela}}, \binits{G.}},
\oauthor{\bsnm{{Miceli}}, \binits{M.}},
\oauthor{\bsnm{{Mignani}}, \binits{R.}},
\oauthor{\bsnm{{Miller}}, \binits{J.}},
\oauthor{\bsnm{{Miniutti}}, \binits{G.}},
\oauthor{\bsnm{{Molendi}}, \binits{S.}},
\oauthor{\bsnm{{Montez}}, \binits{R.}},
\oauthor{\bsnm{{Moretti}}, \binits{A.}},
\oauthor{\bsnm{{Motch}}, \binits{C.}},
\oauthor{\bsnm{{Naz{\'e}}}, \binits{Y.}},
\oauthor{\bsnm{{Nevalainen}}, \binits{J.}},
\oauthor{\bsnm{{Nicastro}}, \binits{F.}},
\oauthor{\bsnm{{Nulsen}}, \binits{P.}},
\oauthor{\bsnm{{Ohashi}}, \binits{T.}},
\oauthor{\bsnm{{O'Brien}}, \binits{P.}},
\oauthor{\bsnm{{Osborne}}, \binits{J.}},
\oauthor{\bsnm{{Oskinova}}, \binits{L.}},
\oauthor{\bsnm{{Pacaud}}, \binits{F.}},
\oauthor{\bsnm{{Paerels}}, \binits{F.}},
\oauthor{\bsnm{{Page}}, \binits{M.}},
\oauthor{\bsnm{{Papadakis}}, \binits{I.}},
\oauthor{\bsnm{{Pareschi}}, \binits{G.}},
\oauthor{\bsnm{{Petre}}, \binits{R.}},
\oauthor{\bsnm{{Petrucci}}, \binits{P.-O.}},
\oauthor{\bsnm{{Piconcelli}}, \binits{E.}},
\oauthor{\bsnm{{Pillitteri}}, \binits{I.}},
\oauthor{\bsnm{{Pinto}}, \binits{C.}},
\oauthor{\bsnm{{de Plaa}}, \binits{J.}},
\oauthor{\bsnm{{Pointecouteau}}, \binits{E.}},
\oauthor{\bsnm{{Ponman}}, \binits{T.}},
\oauthor{\bsnm{{Ponti}}, \binits{G.}},
\oauthor{\bsnm{{Porquet}}, \binits{D.}},
\oauthor{\bsnm{{Pounds}}, \binits{K.}},
\oauthor{\bsnm{{Pratt}}, \binits{G.}},
\oauthor{\bsnm{{Predehl}}, \binits{P.}},
\oauthor{\bsnm{{Proga}}, \binits{D.}},
\oauthor{\bsnm{{Psaltis}}, \binits{D.}},
\oauthor{\bsnm{{Rafferty}}, \binits{D.}},
\oauthor{\bsnm{{Ramos-Ceja}}, \binits{M.}},
\oauthor{\bsnm{{Ranalli}}, \binits{P.}},
\oauthor{\bsnm{{Rasia}}, \binits{E.}},
\oauthor{\bsnm{{Rau}}, \binits{A.}},
\oauthor{\bsnm{{Rauw}}, \binits{G.}},
\oauthor{\bsnm{{Rea}}, \binits{N.}},
\oauthor{\bsnm{{Read}}, \binits{A.}},
\oauthor{\bsnm{{Reeves}}, \binits{J.}},
\oauthor{\bsnm{{Reiprich}}, \binits{T.}},
\oauthor{\bsnm{{Renaud}}, \binits{M.}},
\oauthor{\bsnm{{Reynolds}}, \binits{C.}},
\oauthor{\bsnm{{Risaliti}}, \binits{G.}},
\oauthor{\bsnm{{Rodriguez}}, \binits{J.}},
\oauthor{\bsnm{{Rodriguez Hidalgo}}, \binits{P.}},
\oauthor{\bsnm{{Roncarelli}}, \binits{M.}},
\oauthor{\bsnm{{Rosario}}, \binits{D.}},
\oauthor{\bsnm{{Rossetti}}, \binits{M.}},
\oauthor{\bsnm{{Rozanska}}, \binits{A.}},
\oauthor{\bsnm{{Rovilos}}, \binits{E.}},
\oauthor{\bsnm{{Salvaterra}}, \binits{R.}},
\oauthor{\bsnm{{Salvato}}, \binits{M.}},
\oauthor{\bsnm{{Di Salvo}}, \binits{T.}},
\oauthor{\bsnm{{Sanders}}, \binits{J.}},
\oauthor{\bsnm{{Sanz-Forcada}}, \binits{J.}},
\oauthor{\bsnm{{Schawinski}}, \binits{K.}},
\oauthor{\bsnm{{Schaye}}, \binits{J.}},
\oauthor{\bsnm{{Schwope}}, \binits{A.}},
\oauthor{\bsnm{{Sciortino}}, \binits{S.}},
\oauthor{\bsnm{{Severgnini}}, \binits{P.}},
\oauthor{\bsnm{{Shankar}}, \binits{F.}},
\oauthor{\bsnm{{Sijacki}}, \binits{D.}},
\oauthor{\bsnm{{Sim}}, \binits{S.}},
\oauthor{\bsnm{{Schmid}}, \binits{C.}},
\oauthor{\bsnm{{Smith}}, \binits{R.}},
\oauthor{\bsnm{{Steiner}}, \binits{A.}},
\oauthor{\bsnm{{Stelzer}}, \binits{B.}},
\oauthor{\bsnm{{Stewart}}, \binits{G.}},
\oauthor{\bsnm{{Strohmayer}}, \binits{T.}},
\oauthor{\bsnm{{Str{\"u}der}}, \binits{L.}},
\oauthor{\bsnm{{Sun}}, \binits{M.}},
\oauthor{\bsnm{{Takei}}, \binits{Y.}},
\oauthor{\bsnm{{Tatischeff}}, \binits{V.}},
\oauthor{\bsnm{{Tiengo}}, \binits{A.}},
\oauthor{\bsnm{{Tombesi}}, \binits{F.}},
\oauthor{\bsnm{{Trinchieri}}, \binits{G.}},
\oauthor{\bsnm{{Tsuru}}, \binits{T.G.}},
\oauthor{\bsnm{{Ud-Doula}}, \binits{A.}},
\oauthor{\bsnm{{Ursino}}, \binits{E.}},
\oauthor{\bsnm{{Valencic}}, \binits{L.}},
\oauthor{\bsnm{{Vanzella}}, \binits{E.}},
\oauthor{\bsnm{{Vaughan}}, \binits{S.}},
\oauthor{\bsnm{{Vignali}}, \binits{C.}},
\oauthor{\bsnm{{Vink}}, \binits{J.}},
\oauthor{\bsnm{{Vito}}, \binits{F.}},
\oauthor{\bsnm{{Volonteri}}, \binits{M.}},
\oauthor{\bsnm{{Wang}}, \binits{D.}},
\oauthor{\bsnm{{Webb}}, \binits{N.}},
\oauthor{\bsnm{{Willingale}}, \binits{R.}},
\oauthor{\bsnm{{Wilms}}, \binits{J.}},
\oauthor{\bsnm{{Wise}}, \binits{M.}},
\oauthor{\bsnm{{Worrall}}, \binits{D.}},
\oauthor{\bsnm{{Young}}, \binits{A.}},
\oauthor{\bsnm{{Zampieri}}, \binits{L.}},
\oauthor{\bsnm{{In't Zand}}, \binits{J.}},
\oauthor{\bsnm{{Zane}}, \binits{S.}},
\oauthor{\bsnm{{Zezas}}, \binits{A.}},
\oauthor{\bsnm{{Zhang}}, \binits{Y.}},
\oauthor{\bsnm{{Zhuravleva}}, \binits{I.}}:
{The Hot and Energetic Universe: A White Paper presenting the science theme motivating the Athena+ mission}.
arXiv e-prints,
1306--2307
(2013)
{\href{https://arxiv.org/abs/1306.2307}{{arXiv:1306.2307}}}
{[astro-ph.HE]}
\end{botherref}
\endbibitem

\bibitem[\protect\citeauthoryear{{Barret} et~al.}{2013}]{Barret_2013sf2a.conf..447B}
\begin{bchapter}
\bauthor{\bsnm{{Barret}}, \binits{D.}},
\bauthor{\bsnm{{Nandra}}, \binits{K.}},
\bauthor{\bsnm{{Barcons}}, \binits{X.}},
\bauthor{\bsnm{{Fabian}}, \binits{A.}},
\bauthor{\bsnm{{den Herder}}, \binits{J.W.}},
\bauthor{\bsnm{{Piro}}, \binits{L.}},
\bauthor{\bsnm{{Watson}}, \binits{M.}},
\bauthor{\bsnm{{Aird}}, \binits{J.}},
\bauthor{\bsnm{{Branduardi-Raymont}}, \binits{G.}},
\bauthor{\bsnm{{Cappi}}, \binits{M.}},
\bauthor{\bsnm{{Carrera}}, \binits{F.}},
\bauthor{\bsnm{{Comastri}}, \binits{A.}},
\bauthor{\bsnm{{Costantini}}, \binits{E.}},
\bauthor{\bsnm{{Croston}}, \binits{J.}},
\bauthor{\bsnm{{Decourchelle}}, \binits{A.}},
\bauthor{\bsnm{{Done}}, \binits{C.}},
\bauthor{\bsnm{{Dovciak}}, \binits{M.}},
\bauthor{\bsnm{{Ettori}}, \binits{S.}},
\bauthor{\bsnm{{Finoguenov}}, \binits{A.}},
\bauthor{\bsnm{{Georgakakis}}, \binits{A.}},
\bauthor{\bsnm{{Jonker}}, \binits{P.}},
\bauthor{\bsnm{{Kaastra}}, \binits{J.}},
\bauthor{\bsnm{{Matt}}, \binits{G.}},
\bauthor{\bsnm{{Motch}}, \binits{C.}},
\bauthor{\bsnm{{O'Brien}}, \binits{P.}},
\bauthor{\bsnm{{Pareschi}}, \binits{G.}},
\bauthor{\bsnm{{Pointecouteau}}, \binits{E.}},
\bauthor{\bsnm{{Pratt}}, \binits{G.}},
\bauthor{\bsnm{{Rauw}}, \binits{G.}},
\bauthor{\bsnm{{Reiprich}}, \binits{T.}},
\bauthor{\bsnm{{Sanders}}, \binits{J.}},
\bauthor{\bsnm{{Sciortino}}, \binits{S.}},
\bauthor{\bsnm{{Willingale}}, \binits{R.}},
\bauthor{\bsnm{{Wilms}}, \binits{J.}}:
\bctitle{{Athena+: The first Deep Universe X-ray Observatory}}.
In: \beditor{\bsnm{{Cambresy}}, \binits{L.}},
\beditor{\bsnm{{Martins}}, \binits{F.}},
\beditor{\bsnm{{Nuss}}, \binits{E.}},
\beditor{\bsnm{{Palacios}}, \binits{A.}} (eds.)
\bbtitle{SF2A-2013: Proceedings of the Annual Meeting of the French Society of Astronomy and Astrophysics},
pp. \bfpage{447}--\blpage{453}
(\byear{2013})
\end{bchapter}
\endbibitem

\bibitem[\protect\citeauthoryear{{Barcons} et~al.}{2015}]{barcons2015JPhCS.610a2008B}
\begin{bchapter}
\bauthor{\bsnm{{Barcons}}, \binits{X.}},
\bauthor{\bsnm{{Nandra}}, \binits{K.}},
\bauthor{\bsnm{{Barret}}, \binits{D.}},
\bauthor{\bsnm{{den Herder}}, \binits{J.-W.}},
\bauthor{\bsnm{{Fabian}}, \binits{A.C.}},
\bauthor{\bsnm{{Piro}}, \binits{L.}},
\bauthor{\bsnm{{Watson}}, \binits{M.G.}},
\bauthor{\bsnm{{the Athena Team}}}:
\bctitle{{Athena: the X-ray observatory to study the hot and energetic Universe}}.
In: \bbtitle{Journal of Physics Conference Series}.
\bsertitle{Journal of Physics Conference Series},
vol. \bseriesno{610},
p. \bfpage{012008}
(\byear{2015}).
\doiurl{10.1088/1742-6596/610/1/012008}
\end{bchapter}
\endbibitem

\bibitem[\protect\citeauthoryear{{Barcons} et~al.}{2017}]{Barcons2017AN....338..153B}
\begin{barticle}
\bauthor{\bsnm{{Barcons}}, \binits{X.}},
\bauthor{\bsnm{{Barret}}, \binits{D.}},
\bauthor{\bsnm{{Decourchelle}}, \binits{A.}},
\bauthor{\bsnm{{den Herder}}, \binits{J.W.}},
\bauthor{\bsnm{{Fabian}}, \binits{A.C.}},
\bauthor{\bsnm{{Matsumoto}}, \binits{H.}},
\bauthor{\bsnm{{Lumb}}, \binits{D.}},
\bauthor{\bsnm{{Nandra}}, \binits{K.}},
\bauthor{\bsnm{{Piro}}, \binits{L.}},
\bauthor{\bsnm{{Smith}}, \binits{R.K.}},
\bauthor{\bsnm{{Willingale}}, \binits{R.}}:
\batitle{{Athena: ESA's X-ray observatory for the late 2020s}}.
\bjtitle{Astronomische Nachrichten}
\bvolume{338}(\bissue{153}),
\bfpage{153}--\blpage{158}
(\byear{2017})
\doiurl{10.1002/asna.201713323}
\end{barticle}
\endbibitem

\bibitem[\protect\citeauthoryear{{Barret} et~al.}{2020}]{Barret2020AN....341..224B}
\begin{barticle}
\bauthor{\bsnm{{Barret}}, \binits{D.}},
\bauthor{\bsnm{{Decourchelle}}, \binits{A.}},
\bauthor{\bsnm{{Fabian}}, \binits{A.}},
\bauthor{\bsnm{{Guainazzi}}, \binits{M.}},
\bauthor{\bsnm{{Nandra}}, \binits{K.}},
\bauthor{\bsnm{{Smith}}, \binits{R.}},
\bauthor{\bsnm{{den Herder}}, \binits{J.-W.}}:
\batitle{{The Athena space X-ray observatory and the astrophysics of hot plasma}}.
\bjtitle{Astronomische Nachrichten}
\bvolume{341}(\bissue{2}),
\bfpage{224}--\blpage{235}
(\byear{2020})
\doiurl{10.1002/asna.202023782}
{\href{https://arxiv.org/abs/1912.04615}{{arXiv:1912.04615}}}
{[astro-ph.IM]}
\end{barticle}
\endbibitem

\bibitem[\protect\citeauthoryear{{Barret} et~al.}{2013}]{Barret_2013arXiv1308.6784B}
\begin{botherref}
\oauthor{\bsnm{{Barret}}, \binits{D.}},
\oauthor{\bsnm{{den Herder}}, \binits{J.W.}},
\oauthor{\bsnm{{Piro}}, \binits{L.}},
\oauthor{\bsnm{{Ravera}}, \binits{L.}},
\oauthor{\bsnm{{Den Hartog}}, \binits{R.}},
\oauthor{\bsnm{{Macculi}}, \binits{C.}},
\oauthor{\bsnm{{Barcons}}, \binits{X.}},
\oauthor{\bsnm{{Page}}, \binits{M.}},
\oauthor{\bsnm{{Paltani}}, \binits{S.}},
\oauthor{\bsnm{{Rauw}}, \binits{G.}},
\oauthor{\bsnm{{Wilms}}, \binits{J.}},
\oauthor{\bsnm{{Ceballos}}, \binits{M.}},
\oauthor{\bsnm{{Duband}}, \binits{L.}},
\oauthor{\bsnm{{Gottardi}}, \binits{L.}},
\oauthor{\bsnm{{Lotti}}, \binits{S.}},
\oauthor{\bsnm{{de Plaa}}, \binits{J.}},
\oauthor{\bsnm{{Pointecouteau}}, \binits{E.}},
\oauthor{\bsnm{{Schmid}}, \binits{C.}},
\oauthor{\bsnm{{Akamatsu}}, \binits{H.}},
\oauthor{\bsnm{{Bagliani}}, \binits{D.}},
\oauthor{\bsnm{{Bandler}}, \binits{S.}},
\oauthor{\bsnm{{Barbera}}, \binits{M.}},
\oauthor{\bsnm{{Bastia}}, \binits{P.}},
\oauthor{\bsnm{{Biasotti}}, \binits{M.}},
\oauthor{\bsnm{{Branco}}, \binits{M.}},
\oauthor{\bsnm{{Camon}}, \binits{A.}},
\oauthor{\bsnm{{Cara}}, \binits{C.}},
\oauthor{\bsnm{{Cobo}}, \binits{B.}},
\oauthor{\bsnm{{Colasanti}}, \binits{L.}},
\oauthor{\bsnm{{Costa-Kramer}}, \binits{J.L.}},
\oauthor{\bsnm{{Corcione}}, \binits{L.}},
\oauthor{\bsnm{{Doriese}}, \binits{W.}},
\oauthor{\bsnm{{Duval}}, \binits{J.M.}},
\oauthor{\bsnm{{Fabrega}}, \binits{L.}},
\oauthor{\bsnm{{Gatti}}, \binits{F.}},
\oauthor{\bsnm{{de Gerone}}, \binits{M.}},
\oauthor{\bsnm{{Guttridge}}, \binits{P.}},
\oauthor{\bsnm{{Kelley}}, \binits{R.}},
\oauthor{\bsnm{{Kilbourne}}, \binits{C.}},
\oauthor{\bsnm{{van der Kuur}}, \binits{J.}},
\oauthor{\bsnm{{Mineo}}, \binits{T.}},
\oauthor{\bsnm{{Mitsuda}}, \binits{K.}},
\oauthor{\bsnm{{Natalucci}}, \binits{L.}},
\oauthor{\bsnm{{Ohashi}}, \binits{T.}},
\oauthor{\bsnm{{Peille}}, \binits{P.}},
\oauthor{\bsnm{{Perinati}}, \binits{E.}},
\oauthor{\bsnm{{Pigot}}, \binits{C.}},
\oauthor{\bsnm{{Pizzigoni}}, \binits{G.}},
\oauthor{\bsnm{{Pobes}}, \binits{C.}},
\oauthor{\bsnm{{Porter}}, \binits{F.}},
\oauthor{\bsnm{{Renotte}}, \binits{E.}},
\oauthor{\bsnm{{Sauvageot}}, \binits{J.L.}},
\oauthor{\bsnm{{Sciortino}}, \binits{S.}},
\oauthor{\bsnm{{Torrioli}}, \binits{G.}},
\oauthor{\bsnm{{Valenziano}}, \binits{L.}},
\oauthor{\bsnm{{Willingale}}, \binits{D.}},
\oauthor{\bsnm{{de Vries}}, \binits{C.}},
\oauthor{\bsnm{{van Weers}}, \binits{H.}}:
{The Hot and Energetic Universe: The X-ray Integral Field Unit (X-IFU) for Athena+}.
arXiv e-prints,
1308--6784
(2013)
{\href{https://arxiv.org/abs/1308.6784}{{arXiv:1308.6784}}}
{[astro-ph.IM]}
\end{botherref}
\endbibitem

\bibitem[\protect\citeauthoryear{{Ravera} et~al.}{2014}]{Ravera_2014SPIE.9144E..2LR}
\begin{bchapter}
\bauthor{\bsnm{{Ravera}}, \binits{L.}},
\bauthor{\bsnm{{Barret}}, \binits{D.}},
\bauthor{\bsnm{{den Herder}}, \binits{J.W.}},
\bauthor{\bsnm{{Piro}}, \binits{L.}},
\bauthor{\bsnm{{Cl{\'e}dassou}}, \binits{R.}},
\bauthor{\bsnm{{Pointecouteau}}, \binits{E.}},
\bauthor{\bsnm{{Peille}}, \binits{P.}},
\bauthor{\bsnm{{Pajot}}, \binits{F.}},
\bauthor{\bsnm{{Arnaud}}, \binits{M.}},
\bauthor{\bsnm{{Pigot}}, \binits{C.}},
\bauthor{\bsnm{{Duband}}, \binits{L.}},
\bauthor{\bsnm{{Cara}}, \binits{C.}},
\bauthor{\bsnm{{den Hartog}}, \binits{R.H.}},
\bauthor{\bsnm{{Gottardi}}, \binits{L.}},
\bauthor{\bsnm{{Akamatsu}}, \binits{H.}},
\bauthor{\bsnm{{van der Kuur}}, \binits{J.}},
\bauthor{\bsnm{{van Weers}}, \binits{H.J.}},
\bauthor{\bsnm{{de Plaa}}, \binits{J.}},
\bauthor{\bsnm{{Macculi}}, \binits{C.}},
\bauthor{\bsnm{{Lotti}}, \binits{S.}},
\bauthor{\bsnm{{Torrioli}}, \binits{G.}},
\bauthor{\bsnm{{Gatti}}, \binits{F.}},
\bauthor{\bsnm{{Valenziano}}, \binits{L.}},
\bauthor{\bsnm{{Barbera}}, \binits{M.}},
\bauthor{\bsnm{{Barcons}}, \binits{X.}},
\bauthor{\bsnm{{Ceballos}}, \binits{M.T.}},
\bauthor{\bsnm{{F{\`a}brega}}, \binits{L.}},
\bauthor{\bsnm{{Mas-Hesse}}, \binits{J.M.}},
\bauthor{\bsnm{{Page}}, \binits{M.J.}},
\bauthor{\bsnm{{Guttridge}}, \binits{P.R.}},
\bauthor{\bsnm{{Willingale}}, \binits{R.}},
\bauthor{\bsnm{{Paltani}}, \binits{S.}},
\bauthor{\bsnm{{Genolet}}, \binits{L.}},
\bauthor{\bsnm{{Bozzo}}, \binits{E.}},
\bauthor{\bsnm{{Rauw}}, \binits{G.}},
\bauthor{\bsnm{{Renotte}}, \binits{E.}},
\bauthor{\bsnm{{Wilms}}, \binits{J.}},
\bauthor{\bsnm{{Schmid}}, \binits{C.}}:
\bctitle{{The X-ray Integral Field Unit (X-IFU) for Athena}}.
In: \beditor{\bsnm{{Takahashi}}, \binits{T.}},
\beditor{\bsnm{{den Herder}}, \binits{J.-W.A.}},
\beditor{\bsnm{{Bautz}}, \binits{M.}} (eds.)
\bbtitle{Space Telescopes and Instrumentation 2014: Ultraviolet to Gamma Ray}.
\bsertitle{Society of Photo-Optical Instrumentation Engineers (SPIE) Conference Series},
vol. \bseriesno{9144},
p. \bfpage{91442}
(\byear{2014}).
\doiurl{10.1117/12.2055884}
\end{bchapter}
\endbibitem

\bibitem[\protect\citeauthoryear{{Barret} et~al.}{2016}]{Barret_2016SPIE.9905E..2FB}
\begin{bchapter}
\bauthor{\bsnm{{Barret}}, \binits{D.}},
\bauthor{\bsnm{{Lam Trong}}, \binits{T.}},
\bauthor{\bsnm{{den Herder}}, \binits{J.-W.}},
\bauthor{\bsnm{{Piro}}, \binits{L.}},
\bauthor{\bsnm{{Barcons}}, \binits{X.}},
\bauthor{\bsnm{{Huovelin}}, \binits{J.}},
\bauthor{\bsnm{{Kelley}}, \binits{R.}},
\bauthor{\bsnm{{Mas-Hesse}}, \binits{J.M.}},
\bauthor{\bsnm{{Mitsuda}}, \binits{K.}},
\bauthor{\bsnm{{Paltani}}, \binits{S.}},
\bauthor{\bsnm{{Rauw}}, \binits{G.}},
\bauthor{\bsnm{{Ro{\.Z}anska}}, \binits{A.}},
\bauthor{\bsnm{{Wilms}}, \binits{J.}},
\bauthor{\bsnm{{Barbera}}, \binits{M.}},
\bauthor{\bsnm{{Bozzo}}, \binits{E.}},
\bauthor{\bsnm{{Ceballos}}, \binits{M.T.}},
\bauthor{\bsnm{{Charles}}, \binits{I.}},
\bauthor{\bsnm{{Decourchelle}}, \binits{A.}},
\bauthor{\bsnm{{den Hartog}}, \binits{R.}},
\bauthor{\bsnm{{Duval}}, \binits{J.-M.}},
\bauthor{\bsnm{{Fiore}}, \binits{F.}},
\bauthor{\bsnm{{Gatti}}, \binits{F.}},
\bauthor{\bsnm{{Goldwurm}}, \binits{A.}},
\bauthor{\bsnm{{Jackson}}, \binits{B.}},
\bauthor{\bsnm{{Jonker}}, \binits{P.}},
\bauthor{\bsnm{{Kilbourne}}, \binits{C.}},
\bauthor{\bsnm{{Macculi}}, \binits{C.}},
\bauthor{\bsnm{{Mendez}}, \binits{M.}},
\bauthor{\bsnm{{Molendi}}, \binits{S.}},
\bauthor{\bsnm{{Orleanski}}, \binits{P.}},
\bauthor{\bsnm{{Pajot}}, \binits{F.}},
\bauthor{\bsnm{{Pointecouteau}}, \binits{E.}},
\bauthor{\bsnm{{Porter}}, \binits{F.}},
\bauthor{\bsnm{{Pratt}}, \binits{G.W.}},
\bauthor{\bsnm{{Pr{\^e}le}}, \binits{D.}},
\bauthor{\bsnm{{Ravera}}, \binits{L.}},
\bauthor{\bsnm{{Renotte}}, \binits{E.}},
\bauthor{\bsnm{{Schaye}}, \binits{J.}},
\bauthor{\bsnm{{Shinozaki}}, \binits{K.}},
\bauthor{\bsnm{{Valenziano}}, \binits{L.}},
\bauthor{\bsnm{{Vink}}, \binits{J.}},
\bauthor{\bsnm{{Webb}}, \binits{N.}},
\bauthor{\bsnm{{Yamasaki}}, \binits{N.}},
\bauthor{\bsnm{{Delcelier-Douchin}}, \binits{F.}},
\bauthor{\bsnm{{Le Du}}, \binits{M.}},
\bauthor{\bsnm{{Mesnager}}, \binits{J.-M.}},
\bauthor{\bsnm{{Pradines}}, \binits{A.}},
\bauthor{\bsnm{{Branduardi-Raymont}}, \binits{G.}},
\bauthor{\bsnm{{Dadina}}, \binits{M.}},
\bauthor{\bsnm{{Finoguenov}}, \binits{A.}},
\bauthor{\bsnm{{Fukazawa}}, \binits{Y.}},
\bauthor{\bsnm{{Janiuk}}, \binits{A.}},
\bauthor{\bsnm{{Miller}}, \binits{J.}},
\bauthor{\bsnm{{Naz{\'e}}}, \binits{Y.}},
\bauthor{\bsnm{{Nicastro}}, \binits{F.}},
\bauthor{\bsnm{{Sciortino}}, \binits{S.}},
\bauthor{\bsnm{{Torrejon}}, \binits{J.M.}},
\bauthor{\bsnm{{Geoffray}}, \binits{H.}},
\bauthor{\bsnm{{Hernandez}}, \binits{I.}},
\bauthor{\bsnm{{Luno}}, \binits{L.}},
\bauthor{\bsnm{{Peille}}, \binits{P.}},
\bauthor{\bsnm{{Andr{\'e}}}, \binits{J.}},
\bauthor{\bsnm{{Daniel}}, \binits{C.}},
\bauthor{\bsnm{{Etcheverry}}, \binits{C.}},
\bauthor{\bsnm{{Gloaguen}}, \binits{E.}},
\bauthor{\bsnm{{Hassin}}, \binits{J.}},
\bauthor{\bsnm{{Hervet}}, \binits{G.}},
\bauthor{\bsnm{{Maussang}}, \binits{I.}},
\bauthor{\bsnm{{Moueza}}, \binits{J.}},
\bauthor{\bsnm{{Paillet}}, \binits{A.}},
\bauthor{\bsnm{{Vella}}, \binits{B.}},
\bauthor{\bsnm{{Campos Garrido}}, \binits{G.}},
\bauthor{\bsnm{{Damery}}, \binits{J.-C.}},
\bauthor{\bsnm{{Panem}}, \binits{C.}},
\bauthor{\bsnm{{Panh}}, \binits{J.}},
\bauthor{\bsnm{{Bandler}}, \binits{S.}},
\bauthor{\bsnm{{Biffi}}, \binits{J.-M.}},
\bauthor{\bsnm{{Boyce}}, \binits{K.}},
\bauthor{\bsnm{{Cl{\'e}net}}, \binits{A.}},
\bauthor{\bsnm{{DiPirro}}, \binits{M.}},
\bauthor{\bsnm{{Jamotton}}, \binits{P.}},
\bauthor{\bsnm{{Lotti}}, \binits{S.}},
\bauthor{\bsnm{{Schwander}}, \binits{D.}},
\bauthor{\bsnm{{Smith}}, \binits{S.}},
\bauthor{\bsnm{{van Leeuwen}}, \binits{B.-J.}},
\bauthor{\bsnm{{van Weers}}, \binits{H.}},
\bauthor{\bsnm{{Brand}}, \binits{T.}},
\bauthor{\bsnm{{Cobo}}, \binits{B.}},
\bauthor{\bsnm{{Dauser}}, \binits{T.}},
\bauthor{\bsnm{{de Plaa}}, \binits{J.}},
\bauthor{\bsnm{{Cucchetti}}, \binits{E.}}:
\bctitle{{The Athena X-ray Integral Field Unit (X-IFU)}}.
In: \beditor{\bsnm{{den Herder}}, \binits{J.-W.A.}},
\beditor{\bsnm{{Takahashi}}, \binits{T.}},
\beditor{\bsnm{{Bautz}}, \binits{M.}} (eds.)
\bbtitle{Space Telescopes and Instrumentation 2016: Ultraviolet to Gamma Ray}.
\bsertitle{Society of Photo-Optical Instrumentation Engineers (SPIE) Conference Series},
vol. \bseriesno{9905},
p. \bfpage{99052}
(\byear{2016}).
\doiurl{10.1117/12.2232432}
\end{bchapter}
\endbibitem

\bibitem[\protect\citeauthoryear{{Barret} et~al.}{2018}]{Barret_2018SPIE10699E..1GB}
\begin{bchapter}
\bauthor{\bsnm{{Barret}}, \binits{D.}},
\bauthor{\bsnm{{Lam Trong}}, \binits{T.}},
\bauthor{\bsnm{{den Herder}}, \binits{J.-W.}},
\bauthor{\bsnm{{Piro}}, \binits{L.}},
\bauthor{\bsnm{{Cappi}}, \binits{M.}},
\bauthor{\bsnm{{Houvelin}}, \binits{J.}},
\bauthor{\bsnm{{Kelley}}, \binits{R.}},
\bauthor{\bsnm{{Mas-Hesse}}, \binits{J.M.}},
\bauthor{\bsnm{{Mitsuda}}, \binits{K.}},
\bauthor{\bsnm{{Paltani}}, \binits{S.}},
\bauthor{\bsnm{{Rauw}}, \binits{G.}},
\bauthor{\bsnm{{Rozanska}}, \binits{A.}},
\bauthor{\bsnm{{Wilms}}, \binits{J.}},
\bauthor{\bsnm{{Bandler}}, \binits{S.}},
\bauthor{\bsnm{{Barbera}}, \binits{M.}},
\bauthor{\bsnm{{Barcons}}, \binits{X.}},
\bauthor{\bsnm{{Bozzo}}, \binits{E.}},
\bauthor{\bsnm{{Ceballos}}, \binits{M.T.}},
\bauthor{\bsnm{{Charles}}, \binits{I.}},
\bauthor{\bsnm{{Costantini}}, \binits{E.}},
\bauthor{\bsnm{{Decourchelle}}, \binits{A.}},
\bauthor{\bsnm{{den Hartog}}, \binits{R.}},
\bauthor{\bsnm{{Duband}}, \binits{L.}},
\bauthor{\bsnm{{Duval}}, \binits{J.-M.}},
\bauthor{\bsnm{{Fiore}}, \binits{F.}},
\bauthor{\bsnm{{Gatti}}, \binits{F.}},
\bauthor{\bsnm{{Goldwurm}}, \binits{A.}},
\bauthor{\bsnm{{Jackson}}, \binits{B.}},
\bauthor{\bsnm{{Jonker}}, \binits{P.}},
\bauthor{\bsnm{{Kilbourne}}, \binits{C.}},
\bauthor{\bsnm{{Macculi}}, \binits{C.}},
\bauthor{\bsnm{{Mendez}}, \binits{M.}},
\bauthor{\bsnm{{Molendi}}, \binits{S.}},
\bauthor{\bsnm{{Orleanski}}, \binits{P.}},
\bauthor{\bsnm{{Pajot}}, \binits{F.}},
\bauthor{\bsnm{{Pointecouteau}}, \binits{E.}},
\bauthor{\bsnm{{Porter}}, \binits{F.}},
\bauthor{\bsnm{{Pratt}}, \binits{G.W.}},
\bauthor{\bsnm{{Pr{\^e}le}}, \binits{D.}},
\bauthor{\bsnm{{Ravera}}, \binits{L.}},
\bauthor{\bsnm{{Sato}}, \binits{K.}},
\bauthor{\bsnm{{Schaye}}, \binits{J.}},
\bauthor{\bsnm{{Shinozaki}}, \binits{K.}},
\bauthor{\bsnm{{Thibert}}, \binits{T.}},
\bauthor{\bsnm{{Valenziano}}, \binits{L.}},
\bauthor{\bsnm{{Valette}}, \binits{V.}},
\bauthor{\bsnm{{Vink}}, \binits{J.}},
\bauthor{\bsnm{{Webb}}, \binits{N.}},
\bauthor{\bsnm{{Wise}}, \binits{M.}},
\bauthor{\bsnm{{Yamasaki}}, \binits{N.}},
\bauthor{\bsnm{{Douchin}}, \binits{F.}},
\bauthor{\bsnm{{Mesnager}}, \binits{J.-M.}},
\bauthor{\bsnm{{Pontet}}, \binits{B.}},
\bauthor{\bsnm{{Pradines}}, \binits{A.}},
\bauthor{\bsnm{{Branduardi-Raymont}}, \binits{G.}},
\bauthor{\bsnm{{Bulbul}}, \binits{E.}},
\bauthor{\bsnm{{Dadina}}, \binits{M.}},
\bauthor{\bsnm{{Ettori}}, \binits{S.}},
\bauthor{\bsnm{{Finoguenov}}, \binits{A.}},
\bauthor{\bsnm{{Fukazawa}}, \binits{Y.}},
\bauthor{\bsnm{{Janiuk}}, \binits{A.}},
\bauthor{\bsnm{{Kaastra}}, \binits{J.}},
\bauthor{\bsnm{{Mazzotta}}, \binits{P.}},
\bauthor{\bsnm{{Miller}}, \binits{J.}},
\bauthor{\bsnm{{Miniutti}}, \binits{G.}},
\bauthor{\bsnm{{Naze}}, \binits{Y.}},
\bauthor{\bsnm{{Nicastro}}, \binits{F.}},
\bauthor{\bsnm{{Scioritino}}, \binits{S.}},
\bauthor{\bsnm{{Simonescu}}, \binits{A.}},
\bauthor{\bsnm{{Torrejon}}, \binits{J.M.}},
\bauthor{\bsnm{{Frezouls}}, \binits{B.}},
\bauthor{\bsnm{{Geoffray}}, \binits{H.}},
\bauthor{\bsnm{{Peille}}, \binits{P.}},
\bauthor{\bsnm{{Aicardi}}, \binits{C.}},
\bauthor{\bsnm{{Andr{\'e}}}, \binits{J.}},
\bauthor{\bsnm{{Daniel}}, \binits{C.}},
\bauthor{\bsnm{{Cl{\'e}net}}, \binits{A.}},
\bauthor{\bsnm{{Etcheverry}}, \binits{C.}},
\bauthor{\bsnm{{Gloaguen}}, \binits{E.}},
\bauthor{\bsnm{{Hervet}}, \binits{G.}},
\bauthor{\bsnm{{Jolly}}, \binits{A.}},
\bauthor{\bsnm{{Ledot}}, \binits{A.}},
\bauthor{\bsnm{{Paillet}}, \binits{I.}},
\bauthor{\bsnm{{Schmisser}}, \binits{R.}},
\bauthor{\bsnm{{Vella}}, \binits{B.}},
\bauthor{\bsnm{{Damery}}, \binits{J.-C.}},
\bauthor{\bsnm{{Boyce}}, \binits{K.}},
\bauthor{\bsnm{{Dipirro}}, \binits{M.}},
\bauthor{\bsnm{{Lotti}}, \binits{S.}},
\bauthor{\bsnm{{Schwander}}, \binits{D.}},
\bauthor{\bsnm{{Smith}}, \binits{S.}},
\bauthor{\bsnm{{Van Leeuwen}}, \binits{B.-J.}},
\bauthor{\bsnm{{van Weers}}, \binits{H.}},
\bauthor{\bsnm{{Clerc}}, \binits{N.}},
\bauthor{\bsnm{{Cobo}}, \binits{B.}},
\bauthor{\bsnm{{Dauser}}, \binits{T.}},
\bauthor{\bsnm{{Kirsch}}, \binits{C.}},
\bauthor{\bsnm{{Cucchetti}}, \binits{E.}},
\bauthor{\bsnm{{Eckart}}, \binits{M.}},
\bauthor{\bsnm{{Ferrando}}, \binits{P.}},
\bauthor{\bsnm{{Natalucci}}, \binits{L.}}:
\bctitle{{The ATHENA X-ray Integral Field Unit (X-IFU)}}.
In: \beditor{\bsnm{{den Herder}}, \binits{J.-W.A.}},
\beditor{\bsnm{{Nikzad}}, \binits{S.}},
\beditor{\bsnm{{Nakazawa}}, \binits{K.}} (eds.)
\bbtitle{Space Telescopes and Instrumentation 2018: Ultraviolet to Gamma Ray}.
\bsertitle{Society of Photo-Optical Instrumentation Engineers (SPIE) Conference Series},
vol. \bseriesno{10699},
p. \bfpage{106991}
(\byear{2018}).
\doiurl{10.1117/12.2312409}
\end{bchapter}
\endbibitem

\bibitem[\protect\citeauthoryear{{Pajot} et~al.}{2018}]{Pajot_2018JLTP..193..901P}
\begin{barticle}
\bauthor{\bsnm{{Pajot}}, \binits{F.}},
\bauthor{\bsnm{{Barret}}, \binits{D.}},
\bauthor{\bsnm{{Lam-Trong}}, \binits{T.}},
\bauthor{\bsnm{{den Herder}}, \binits{J.-W.}},
\bauthor{\bsnm{{Piro}}, \binits{L.}},
\bauthor{\bsnm{{Cappi}}, \binits{M.}},
\bauthor{\bsnm{{Huovelin}}, \binits{J.}},
\bauthor{\bsnm{{Kelley}}, \binits{R.}},
\bauthor{\bsnm{{Mas-Hesse}}, \binits{J.M.}},
\bauthor{\bsnm{{Mitsuda}}, \binits{K.}},
\bauthor{\bsnm{{Paltani}}, \binits{S.}},
\bauthor{\bsnm{{Rauw}}, \binits{G.}},
\bauthor{\bsnm{{Rozanska}}, \binits{A.}},
\bauthor{\bsnm{{Wilms}}, \binits{J.}},
\bauthor{\bsnm{{Barbera}}, \binits{M.}},
\bauthor{\bsnm{{Douchin}}, \binits{F.}},
\bauthor{\bsnm{{Geoffray}}, \binits{H.}},
\bauthor{\bsnm{{den Hartog}}, \binits{R.}},
\bauthor{\bsnm{{Kilbourne}}, \binits{C.}},
\bauthor{\bsnm{{Le Du}}, \binits{M.}},
\bauthor{\bsnm{{Macculi}}, \binits{C.}},
\bauthor{\bsnm{{Mesnager}}, \binits{J.-M.}},
\bauthor{\bsnm{{Peille}}, \binits{P.}}:
\batitle{{The Athena X-ray Integral Field Unit (X-IFU)}}.
\bjtitle{Journal of Low Temperature Physics}
\bvolume{193}(\bissue{5-6}),
\bfpage{901}--\blpage{907}
(\byear{2018})
\doiurl{10.1007/s10909-018-1904-5}
\end{barticle}
\endbibitem

\bibitem[\protect\citeauthoryear{Meidinger et~al.}{2020}]{Meidinger2020WFI}
\begin{bchapter}
\bauthor{\bsnm{Meidinger}, \binits{N.}},
\bauthor{\bsnm{Albrecht}, \binits{S.}},
\bauthor{\bsnm{Beitler}, \binits{C.}},
\bauthor{\bsnm{Bonholzer}, \binits{M.}},
\bauthor{\bsnm{Emberger}, \binits{V.}},
\bauthor{\bsnm{Frank}, \binits{J.}},
\bauthor{\bsnm{Lederhuber}, \binits{A.}},
\bauthor{\bsnm{M{\"u}ller-Seidlitz}, \binits{J.}},
\bauthor{\bsnm{Nandra}, \binits{K.}},
\bauthor{\bsnm{Oser}, \binits{J.}},
\bauthor{\bsnm{Ott}, \binits{S.}},
\bauthor{\bsnm{Plattner}, \binits{M.}},
\bauthor{\bsnm{Strecker}, \binits{R.}}:
\bctitle{{Development status of the wide field imager instrument for Athena}}.
In: \beditor{\bsnm{Herder}, \binits{J.-W.A.}},
\beditor{\bsnm{Nikzad}, \binits{S.}},
\beditor{\bsnm{Nakazawa}, \binits{K.}} (eds.)
\bbtitle{Space Telescopes and Instrumentation 2020: Ultraviolet to Gamma Ray},
vol. \bseriesno{11444},
p. \bfpage{114440}.
\bpublisher{SPIE}, \blocation{???}
(\byear{2020}).
\doiurl{10.1117/12.2560507} .
\bcomment{International Society for Optics and Photonics}
\end{bchapter}
\endbibitem

\bibitem[\protect\citeauthoryear{Macculi et~al.}{2023}]{Macculi2023}
\begin{botherref}
\oauthor{\bsnm{Macculi}, \binits{C.}},
\oauthor{\bsnm{Argan}, \binits{A.}},
\oauthor{\bsnm{D’Andrea}, \binits{M.}},
\oauthor{\bsnm{Lotti}, \binits{S.}},
\oauthor{\bsnm{Minervini}, \binits{G.}},
\oauthor{\bsnm{Piro}, \binits{L.}},
\oauthor{\bsnm{Ferrari~Barusso}, \binits{L.}},
\oauthor{\bsnm{Boragno}, \binits{C.}},
\oauthor{\bsnm{Celasco}, \binits{E.}},
\oauthor{\bsnm{Gallucci}, \binits{G.}},
\oauthor{\bsnm{Gatti}, \binits{F.}},
\oauthor{\bsnm{Grosso}, \binits{D.}},
\oauthor{\bsnm{Rigano}, \binits{M.}},
\oauthor{\bsnm{Chiarello}, \binits{F.}},
\oauthor{\bsnm{Torrioli}, \binits{G.}},
\oauthor{\bsnm{Fiorini}, \binits{M.}},
\oauthor{\bsnm{Uslenghi}, \binits{M.}},
\oauthor{\bsnm{Brienza}, \binits{D.}},
\oauthor{\bsnm{Cavazzuti}, \binits{E.}},
\oauthor{\bsnm{Puccetti}, \binits{S.}},
\oauthor{\bsnm{Volpe}, \binits{A.}},
\oauthor{\bsnm{Bastia}, \binits{P.}}:
The cryogenic anticoincidence detector for the newathena x-ifu instrument: A program overview.
Condensed Matter
\textbf{8}(4)
(2023)
\doiurl{10.3390/condmat8040108}
\end{botherref}
\endbibitem

\bibitem[\protect\citeauthoryear{{D'Andrea} et~al.}{2024}]{Andrea2024}
\begin{barticle}
\bauthor{\bsnm{{D'Andrea}}, \binits{M.}},
\bauthor{\bsnm{{Macculi}}, \binits{C.}},
\bauthor{\bsnm{{Lotti}}, \binits{S.}},
\bauthor{\bsnm{{Piro}}, \binits{L.}},
\bauthor{\bsnm{{Argan}}, \binits{A.}},
\bauthor{\bsnm{{Minervini}}, \binits{G.}},
\bauthor{\bsnm{{Torrioli}}, \binits{G.}},
\bauthor{\bsnm{{Chiarello}}, \binits{F.}},
\bauthor{\bsnm{{Ferrari Barusso}}, \binits{L.}},
\bauthor{\bsnm{{Celasco}}, \binits{E.}},
\bauthor{\bsnm{{Gallucci}}, \binits{G.}},
\bauthor{\bsnm{{Gatti}}, \binits{F.}},
\bauthor{\bsnm{{Grosso}}, \binits{D.}},
\bauthor{\bsnm{{Rigano}}, \binits{M.}},
\bauthor{\bsnm{{Brienza}}, \binits{D.}},
\bauthor{\bsnm{{Cavazzuti}}, \binits{E.}},
\bauthor{\bsnm{{Volpe}}, \binits{A.}}:
\batitle{{The TES-based Cryogenic AntiCoincidence Detector (CryoAC) of ATHENA X-IFU: A Large Area Silicon Microcalorimeter for Background Particles Detection}}.
\bjtitle{Journal of Low Temperature Physics}
\bvolume{214}(\bissue{3-4}),
\bfpage{164}--\blpage{172}
(\byear{2024})
\doiurl{10.1007/s10909-023-03034-5}
{\href{https://arxiv.org/abs/2401.10827}{{arXiv:2401.10827}}}
{[astro-ph.IM]}
\end{barticle}
\endbibitem

\bibitem[\protect\citeauthoryear{{Tauber} et~al.}{2010}]{Planck}
\begin{barticle}
\bauthor{\bsnm{{Tauber}}, \binits{J.A.}},
\bauthor{\bsnm{{Mandolesi}}, \binits{N.}},
\bauthor{\bsnm{{Puget}}, \binits{J.-L.}},
\bauthor{\bsnm{{Banos}}, \binits{T.}},
\bauthor{\bsnm{{Bersanelli}}, \binits{M.}},
\bauthor{\bsnm{{Bouchet}}, \binits{F.R.}},
\bauthor{\bsnm{{Butler}}, \binits{R.C.}},
\bauthor{\bsnm{{Charra}}, \binits{J.}},
\bauthor{\bsnm{{Crone}}, \binits{G.}},
\bauthor{\bsnm{{Dodsworth}}, \binits{J.}},
\bauthor{\bsnm{{Efstathiou}}, \binits{G.}},
\bauthor{\bsnm{{Gispert}}, \binits{R.}},
\bauthor{\bsnm{{Guyot}}, \binits{G.}},
\bauthor{\bsnm{{Gregorio}}, \binits{A.}},
\bauthor{\bsnm{{Juillet}}, \binits{J.J.}},
\bauthor{\bsnm{{Lamarre}}, \binits{J.-M.}},
\bauthor{\bsnm{{Laureijs}}, \binits{R.J.}},
\bauthor{\bsnm{{Lawrence}}, \binits{C.R.}},
\bauthor{\bsnm{{N{\o}rgaard-Nielsen}}, \binits{H.U.}},
\bauthor{\bsnm{{Passvogel}}, \binits{T.}},
\bauthor{\bsnm{{Reix}}, \binits{J.M.}},
\bauthor{\bsnm{{Texier}}, \binits{D.}},
\bauthor{\bsnm{{Vibert}}, \binits{L.}},
\bauthor{\bsnm{{Zacchei}}, \binits{A.}},
\bauthor{\bsnm{{Ade}}, \binits{P.A.R.}},
\bauthor{\bsnm{{Aghanim}}, \binits{N.}},
\bauthor{\bsnm{{Aja}}, \binits{B.}},
\bauthor{\bsnm{{Alippi}}, \binits{E.}},
\bauthor{\bsnm{{Aloy}}, \binits{L.}},
\bauthor{\bsnm{{Armand}}, \binits{P.}},
\bauthor{\bsnm{{Arnaud}}, \binits{M.}},
\bauthor{\bsnm{{Arondel}}, \binits{A.}},
\bauthor{\bsnm{{Arreola-Villanueva}}, \binits{A.}},
\bauthor{\bsnm{{Artal}}, \binits{E.}},
\bauthor{\bsnm{{Artina}}, \binits{E.}},
\bauthor{\bsnm{{Arts}}, \binits{A.}},
\bauthor{\bsnm{{Ashdown}}, \binits{M.}},
\bauthor{\bsnm{{Aumont}}, \binits{J.}},
\bauthor{\bsnm{{Azzaro}}, \binits{M.}},
\bauthor{\bsnm{{Bacchetta}}, \binits{A.}},
\bauthor{\bsnm{{Baccigalupi}}, \binits{C.}},
\bauthor{\bsnm{{Baker}}, \binits{M.}},
\bauthor{\bsnm{{Balasini}}, \binits{M.}},
\bauthor{\bsnm{{Balbi}}, \binits{A.}},
\bauthor{\bsnm{{Banday}}, \binits{A.J.}},
\bauthor{\bsnm{{Barbier}}, \binits{G.}},
\bauthor{\bsnm{{Barreiro}}, \binits{R.B.}},
\bauthor{\bsnm{{Bartelmann}}, \binits{M.}},
\bauthor{\bsnm{{Battaglia}}, \binits{P.}},
\bauthor{\bsnm{{Battaner}}, \binits{E.}},
\bauthor{\bsnm{{Benabed}}, \binits{K.}},
\bauthor{\bsnm{{Beney}}, \binits{J.-L.}},
\bauthor{\bsnm{{Beneyton}}, \binits{R.}},
\bauthor{\bsnm{{Bennett}}, \binits{K.}},
\bauthor{\bsnm{{Benoit}}, \binits{A.}},
\bauthor{\bsnm{{Bernard}}, \binits{J.-P.}},
\bauthor{\bsnm{{Bhandari}}, \binits{P.}},
\bauthor{\bsnm{{Bhatia}}, \binits{R.}},
\bauthor{\bsnm{{Biggi}}, \binits{M.}},
\bauthor{\bsnm{{Biggins}}, \binits{R.}},
\bauthor{\bsnm{{Billig}}, \binits{G.}},
\bauthor{\bsnm{{Blanc}}, \binits{Y.}},
\bauthor{\bsnm{{Blavot}}, \binits{H.}},
\bauthor{\bsnm{{Bock}}, \binits{J.J.}},
\bauthor{\bsnm{{Bonaldi}}, \binits{A.}},
\bauthor{\bsnm{{Bond}}, \binits{R.}},
\bauthor{\bsnm{{Bonis}}, \binits{J.}},
\bauthor{\bsnm{{Borders}}, \binits{J.}},
\bauthor{\bsnm{{Borrill}}, \binits{J.}},
\bauthor{\bsnm{{Boschini}}, \binits{L.}},
\bauthor{\bsnm{{Boulanger}}, \binits{F.}},
\bauthor{\bsnm{{Bouvier}}, \binits{J.}},
\bauthor{\bsnm{{Bouzit}}, \binits{M.}},
\bauthor{\bsnm{{Bowman}}, \binits{R.}},
\bauthor{\bsnm{{Br{\'e}elle}}, \binits{E.}},
\bauthor{\bsnm{{Bradshaw}}, \binits{T.}},
\bauthor{\bsnm{{Braghin}}, \binits{M.}},
\bauthor{\bsnm{{Bremer}}, \binits{M.}},
\bauthor{\bsnm{{Brienza}}, \binits{D.}},
\bauthor{\bsnm{{Broszkiewicz}}, \binits{D.}},
\bauthor{\bsnm{{Burigana}}, \binits{C.}},
\bauthor{\bsnm{{Burkhalter}}, \binits{M.}},
\bauthor{\bsnm{{Cabella}}, \binits{P.}},
\bauthor{\bsnm{{Cafferty}}, \binits{T.}},
\bauthor{\bsnm{{Cairola}}, \binits{M.}},
\bauthor{\bsnm{{Caminade}}, \binits{S.}},
\bauthor{\bsnm{{Camus}}, \binits{P.}},
\bauthor{\bsnm{{Cantalupo}}, \binits{C.M.}},
\bauthor{\bsnm{{Cappellini}}, \binits{B.}},
\bauthor{\bsnm{{Cardoso}}, \binits{J.-F.}},
\bauthor{\bsnm{{Carr}}, \binits{R.}},
\bauthor{\bsnm{{Catalano}}, \binits{A.}},
\bauthor{\bsnm{{Cay{\'o}n}}, \binits{L.}},
\bauthor{\bsnm{{Cesa}}, \binits{M.}},
\bauthor{\bsnm{{Chaigneau}}, \binits{M.}},
\bauthor{\bsnm{{Challinor}}, \binits{A.}},
\bauthor{\bsnm{{Chamballu}}, \binits{A.}},
\bauthor{\bsnm{{Chambelland}}, \binits{J.P.}},
\bauthor{\bsnm{{Charra}}, \binits{M.}},
\bauthor{\bsnm{{Chiang}}, \binits{L.-Y.}},
\bauthor{\bsnm{{Chlewicki}}, \binits{G.}},
\bauthor{\bsnm{{Christensen}}, \binits{P.R.}},
\bauthor{\bsnm{{Church}}, \binits{S.}},
\bauthor{\bsnm{{Ciancietta}}, \binits{E.}},
\bauthor{\bsnm{{Cibrario}}, \binits{M.}},
\bauthor{\bsnm{{Cizeron}}, \binits{R.}},
\bauthor{\bsnm{{Clements}}, \binits{D.}},
\bauthor{\bsnm{{Collaudin}}, \binits{B.}},
\bauthor{\bsnm{{Colley}}, \binits{J.-M.}},
\bauthor{\bsnm{{Colombi}}, \binits{S.}},
\bauthor{\bsnm{{Colombo}}, \binits{A.}},
\bauthor{\bsnm{{Colombo}}, \binits{F.}},
\bauthor{\bsnm{{Corre}}, \binits{O.}},
\bauthor{\bsnm{{Couchot}}, \binits{F.}},
\bauthor{\bsnm{{Cougrand}}, \binits{B.}},
\bauthor{\bsnm{{Coulais}}, \binits{A.}},
\bauthor{\bsnm{{Couzin}}, \binits{P.}},
\bauthor{\bsnm{{Crane}}, \binits{B.}},
\bauthor{\bsnm{{Crill}}, \binits{B.}},
\bauthor{\bsnm{{Crook}}, \binits{M.}},
\bauthor{\bsnm{{Crumb}}, \binits{D.}},
\bauthor{\bsnm{{Cuttaia}}, \binits{F.}},
\bauthor{\bsnm{{D{\"o}rl}}, \binits{U.}},
\bauthor{\bsnm{{da Silva}}, \binits{P.}},
\bauthor{\bsnm{{Daddato}}, \binits{R.}},
\bauthor{\bsnm{{Damasio}}, \binits{C.}},
\bauthor{\bsnm{{Danese}}, \binits{L.}},
\bauthor{\bsnm{{D'Aquino}}, \binits{G.}},
\bauthor{\bsnm{{D'Arcangelo}}, \binits{O.}},
\bauthor{\bsnm{{Dassas}}, \binits{K.}},
\bauthor{\bsnm{{Davies}}, \binits{R.D.}},
\bauthor{\bsnm{{Davies}}, \binits{W.}},
\bauthor{\bsnm{{Davis}}, \binits{R.J.}},
\bauthor{\bsnm{{de Bernardis}}, \binits{P.}},
\bauthor{\bsnm{{de Chambure}}, \binits{D.}},
\bauthor{\bsnm{{de Gasperis}}, \binits{G.}},
\bauthor{\bsnm{{de La Fuente}}, \binits{M.L.}},
\bauthor{\bsnm{{de Paco}}, \binits{P.}},
\bauthor{\bsnm{{de Rosa}}, \binits{A.}},
\bauthor{\bsnm{{de Troia}}, \binits{G.}},
\bauthor{\bsnm{{de Zotti}}, \binits{G.}},
\bauthor{\bsnm{{Dehamme}}, \binits{M.}},
\bauthor{\bsnm{{Delabrouille}}, \binits{J.}},
\bauthor{\bsnm{{Delouis}}, \binits{J.-M.}},
\bauthor{\bsnm{{D{\'e}sert}}, \binits{F.-X.}},
\bauthor{\bsnm{{di Girolamo}}, \binits{G.}},
\bauthor{\bsnm{{Dickinson}}, \binits{C.}},
\bauthor{\bsnm{{Doelling}}, \binits{E.}},
\bauthor{\bsnm{{Dolag}}, \binits{K.}},
\bauthor{\bsnm{{Domken}}, \binits{I.}},
\bauthor{\bsnm{{Douspis}}, \binits{M.}},
\bauthor{\bsnm{{Doyle}}, \binits{D.}},
\bauthor{\bsnm{{Du}}, \binits{S.}},
\bauthor{\bsnm{{Dubruel}}, \binits{D.}},
\bauthor{\bsnm{{Dufour}}, \binits{C.}},
\bauthor{\bsnm{{Dumesnil}}, \binits{C.}},
\bauthor{\bsnm{{Dupac}}, \binits{X.}},
\bauthor{\bsnm{{Duret}}, \binits{P.}},
\bauthor{\bsnm{{Eder}}, \binits{C.}},
\bauthor{\bsnm{{Elfving}}, \binits{A.}},
\bauthor{\bsnm{{En{\ss}lin}}, \binits{T.A.}},
\bauthor{\bsnm{{Eng}}, \binits{P.}},
\bauthor{\bsnm{{English}}, \binits{K.}},
\bauthor{\bsnm{{Eriksen}}, \binits{H.K.}},
\bauthor{\bsnm{{Estaria}}, \binits{P.}},
\bauthor{\bsnm{{Falvella}}, \binits{M.C.}},
\bauthor{\bsnm{{Ferrari}}, \binits{F.}},
\bauthor{\bsnm{{Finelli}}, \binits{F.}},
\bauthor{\bsnm{{Fishman}}, \binits{A.}},
\bauthor{\bsnm{{Fogliani}}, \binits{S.}},
\bauthor{\bsnm{{Foley}}, \binits{S.}},
\bauthor{\bsnm{{Fonseca}}, \binits{A.}},
\bauthor{\bsnm{{Forma}}, \binits{G.}},
\bauthor{\bsnm{{Forni}}, \binits{O.}},
\bauthor{\bsnm{{Fosalba}}, \binits{P.}},
\bauthor{\bsnm{{Fourmond}}, \binits{J.-J.}},
\bauthor{\bsnm{{Frailis}}, \binits{M.}},
\bauthor{\bsnm{{Franceschet}}, \binits{C.}},
\bauthor{\bsnm{{Franceschi}}, \binits{E.}},
\bauthor{\bsnm{{Fran{\c{c}}ois}}, \binits{S.}},
\bauthor{\bsnm{{Frerking}}, \binits{M.}},
\bauthor{\bsnm{{G{\'o}mez-Re{\~n}asco}}, \binits{M.F.}},
\bauthor{\bsnm{{G{\'o}rski}}, \binits{K.M.}},
\bauthor{\bsnm{{Gaier}}, \binits{T.C.}},
\bauthor{\bsnm{{Galeotta}}, \binits{S.}},
\bauthor{\bsnm{{Ganga}}, \binits{K.}},
\bauthor{\bsnm{{Garc{\'\i}a L{\'a}zaro}}, \binits{J.}},
\bauthor{\bsnm{{Garnica}}, \binits{A.}},
\bauthor{\bsnm{{Gaspard}}, \binits{M.}},
\bauthor{\bsnm{{Gavila}}, \binits{E.}},
\bauthor{\bsnm{{Giard}}, \binits{M.}},
\bauthor{\bsnm{{Giardino}}, \binits{G.}},
\bauthor{\bsnm{{Gienger}}, \binits{G.}},
\bauthor{\bsnm{{Giraud-Heraud}}, \binits{Y.}},
\bauthor{\bsnm{{Glorian}}, \binits{J.-M.}},
\bauthor{\bsnm{{Griffin}}, \binits{M.}},
\bauthor{\bsnm{{Gruppuso}}, \binits{A.}},
\bauthor{\bsnm{{Guglielmi}}, \binits{L.}},
\bauthor{\bsnm{{Guichon}}, \binits{D.}},
\bauthor{\bsnm{{Guillaume}}, \binits{B.}},
\bauthor{\bsnm{{Guillouet}}, \binits{P.}},
\bauthor{\bsnm{{Haissinski}}, \binits{J.}},
\bauthor{\bsnm{{Hansen}}, \binits{F.K.}},
\bauthor{\bsnm{{Hardy}}, \binits{J.}},
\bauthor{\bsnm{{Harrison}}, \binits{D.}},
\bauthor{\bsnm{{Hazell}}, \binits{A.}},
\bauthor{\bsnm{{Hechler}}, \binits{M.}},
\bauthor{\bsnm{{Heckenauer}}, \binits{V.}},
\bauthor{\bsnm{{Heinzer}}, \binits{D.}},
\bauthor{\bsnm{{Hell}}, \binits{R.}},
\bauthor{\bsnm{{Henrot-Versill{\'e}}}, \binits{S.}},
\bauthor{\bsnm{{Hern{\'a}ndez-Monteagudo}}, \binits{C.}},
\bauthor{\bsnm{{Herranz}}, \binits{D.}},
\bauthor{\bsnm{{Herreros}}, \binits{J.M.}},
\bauthor{\bsnm{{Hervier}}, \binits{V.}},
\bauthor{\bsnm{{Heske}}, \binits{A.}},
\bauthor{\bsnm{{Heurtel}}, \binits{A.}},
\bauthor{\bsnm{{Hildebrandt}}, \binits{S.R.}},
\bauthor{\bsnm{{Hills}}, \binits{R.}},
\bauthor{\bsnm{{Hivon}}, \binits{E.}},
\bauthor{\bsnm{{Hobson}}, \binits{M.}},
\bauthor{\bsnm{{Hollert}}, \binits{D.}},
\bauthor{\bsnm{{Holmes}}, \binits{W.}},
\bauthor{\bsnm{{Hornstrup}}, \binits{A.}},
\bauthor{\bsnm{{Hovest}}, \binits{W.}},
\bauthor{\bsnm{{Hoyland}}, \binits{R.J.}},
\bauthor{\bsnm{{Huey}}, \binits{G.}},
\bauthor{\bsnm{{Huffenberger}}, \binits{K.M.}},
\bauthor{\bsnm{{Hughes}}, \binits{N.}},
\bauthor{\bsnm{{Israelsson}}, \binits{U.}},
\bauthor{\bsnm{{Jackson}}, \binits{B.}},
\bauthor{\bsnm{{Jaffe}}, \binits{A.}},
\bauthor{\bsnm{{Jaffe}}, \binits{T.R.}},
\bauthor{\bsnm{{Jagemann}}, \binits{T.}},
\bauthor{\bsnm{{Jessen}}, \binits{N.C.}},
\bauthor{\bsnm{{Jewell}}, \binits{J.}},
\bauthor{\bsnm{{Jones}}, \binits{W.}},
\bauthor{\bsnm{{Juvela}}, \binits{M.}},
\bauthor{\bsnm{{Kaplan}}, \binits{J.}},
\bauthor{\bsnm{{Karlman}}, \binits{P.}},
\bauthor{\bsnm{{Keck}}, \binits{F.}},
\bauthor{\bsnm{{Keih{\"a}nen}}, \binits{E.}},
\bauthor{\bsnm{{King}}, \binits{M.}},
\bauthor{\bsnm{{Kisner}}, \binits{T.S.}},
\bauthor{\bsnm{{Kletzkine}}, \binits{P.}},
\bauthor{\bsnm{{Kneissl}}, \binits{R.}},
\bauthor{\bsnm{{Knoche}}, \binits{J.}},
\bauthor{\bsnm{{Knox}}, \binits{L.}},
\bauthor{\bsnm{{Koch}}, \binits{T.}},
\bauthor{\bsnm{{Krassenburg}}, \binits{M.}},
\bauthor{\bsnm{{Kurki-Suonio}}, \binits{H.}},
\bauthor{\bsnm{{L{\"a}hteenm{\"a}ki}}, \binits{A.}},
\bauthor{\bsnm{{Lagache}}, \binits{G.}},
\bauthor{\bsnm{{Lagorio}}, \binits{E.}},
\bauthor{\bsnm{{Lami}}, \binits{P.}},
\bauthor{\bsnm{{Lande}}, \binits{J.}},
\bauthor{\bsnm{{Lange}}, \binits{A.}},
\bauthor{\bsnm{{Langlet}}, \binits{F.}},
\bauthor{\bsnm{{Lapini}}, \binits{R.}},
\bauthor{\bsnm{{Lapolla}}, \binits{M.}},
\bauthor{\bsnm{{Lasenby}}, \binits{A.}},
\bauthor{\bsnm{{Le Jeune}}, \binits{M.}},
\bauthor{\bsnm{{Leahy}}, \binits{J.P.}},
\bauthor{\bsnm{{Lefebvre}}, \binits{M.}},
\bauthor{\bsnm{{Legrand}}, \binits{F.}},
\bauthor{\bsnm{{Le Meur}}, \binits{G.}},
\bauthor{\bsnm{{Leonardi}}, \binits{R.}},
\bauthor{\bsnm{{Leriche}}, \binits{B.}},
\bauthor{\bsnm{{Leroy}}, \binits{C.}},
\bauthor{\bsnm{{Leutenegger}}, \binits{P.}},
\bauthor{\bsnm{{Levin}}, \binits{S.M.}},
\bauthor{\bsnm{{Lilje}}, \binits{P.B.}},
\bauthor{\bsnm{{Lindensmith}}, \binits{C.}},
\bauthor{\bsnm{{Linden-V{\o}rnle}}, \binits{M.}},
\bauthor{\bsnm{{Loc}}, \binits{A.}},
\bauthor{\bsnm{{Longval}}, \binits{Y.}},
\bauthor{\bsnm{{Lubin}}, \binits{P.M.}},
\bauthor{\bsnm{{Luchik}}, \binits{T.}},
\bauthor{\bsnm{{Luthold}}, \binits{I.}},
\bauthor{\bsnm{{Macias-Perez}}, \binits{J.F.}},
\bauthor{\bsnm{{Maciaszek}}, \binits{T.}},
\bauthor{\bsnm{{MacTavish}}, \binits{C.}},
\bauthor{\bsnm{{Madden}}, \binits{S.}},
\bauthor{\bsnm{{Maffei}}, \binits{B.}},
\bauthor{\bsnm{{Magneville}}, \binits{C.}},
\bauthor{\bsnm{{Maino}}, \binits{D.}},
\bauthor{\bsnm{{Mambretti}}, \binits{A.}},
\bauthor{\bsnm{{Mansoux}}, \binits{B.}},
\bauthor{\bsnm{{Marchioro}}, \binits{D.}},
\bauthor{\bsnm{{Maris}}, \binits{M.}},
\bauthor{\bsnm{{Marliani}}, \binits{F.}},
\bauthor{\bsnm{{Marrucho}}, \binits{J.-C.}},
\bauthor{\bsnm{{Mart{\'\i}-Canales}}, \binits{J.}},
\bauthor{\bsnm{{Mart{\'\i}nez-Gonz{\'a}lez}}, \binits{E.}},
\bauthor{\bsnm{{Mart{\'\i}n-Polegre}}, \binits{A.}},
\bauthor{\bsnm{{Martin}}, \binits{P.}},
\bauthor{\bsnm{{Marty}}, \binits{C.}},
\bauthor{\bsnm{{Marty}}, \binits{W.}},
\bauthor{\bsnm{{Masi}}, \binits{S.}},
\bauthor{\bsnm{{Massardi}}, \binits{M.}},
\bauthor{\bsnm{{Matarrese}}, \binits{S.}},
\bauthor{\bsnm{{Matthai}}, \binits{F.}},
\bauthor{\bsnm{{Mazzotta}}, \binits{P.}},
\bauthor{\bsnm{{McDonald}}, \binits{A.}},
\bauthor{\bsnm{{McGrath}}, \binits{P.}},
\bauthor{\bsnm{{Mediavilla}}, \binits{A.}},
\bauthor{\bsnm{{Meinhold}}, \binits{P.R.}},
\bauthor{\bsnm{{M{\'e}lin}}, \binits{J.-B.}},
\bauthor{\bsnm{{Melot}}, \binits{F.}},
\bauthor{\bsnm{{Mendes}}, \binits{L.}},
\bauthor{\bsnm{{Mennella}}, \binits{A.}},
\bauthor{\bsnm{{Mervier}}, \binits{C.}},
\bauthor{\bsnm{{Meslier}}, \binits{L.}},
\bauthor{\bsnm{{Miccolis}}, \binits{M.}},
\bauthor{\bsnm{{Miville-Deschenes}}, \binits{M.-A.}},
\bauthor{\bsnm{{Moneti}}, \binits{A.}},
\bauthor{\bsnm{{Montet}}, \binits{D.}},
\bauthor{\bsnm{{Montier}}, \binits{L.}},
\bauthor{\bsnm{{Mora}}, \binits{J.}},
\bauthor{\bsnm{{Morgante}}, \binits{G.}},
\bauthor{\bsnm{{Morigi}}, \binits{G.}},
\bauthor{\bsnm{{Morinaud}}, \binits{G.}},
\bauthor{\bsnm{{Morisset}}, \binits{N.}},
\bauthor{\bsnm{{Mortlock}}, \binits{D.}},
\bauthor{\bsnm{{Mottet}}, \binits{S.}},
\bauthor{\bsnm{{Mulder}}, \binits{J.}},
\bauthor{\bsnm{{Munshi}}, \binits{D.}},
\bauthor{\bsnm{{Murphy}}, \binits{A.}},
\bauthor{\bsnm{{Murphy}}, \binits{P.}},
\bauthor{\bsnm{{Musi}}, \binits{P.}},
\bauthor{\bsnm{{Narbonne}}, \binits{J.}},
\bauthor{\bsnm{{Naselsky}}, \binits{P.}},
\bauthor{\bsnm{{Nash}}, \binits{A.}},
\bauthor{\bsnm{{Nati}}, \binits{F.}},
\bauthor{\bsnm{{Natoli}}, \binits{P.}},
\bauthor{\bsnm{{Netterfield}}, \binits{B.}},
\bauthor{\bsnm{{Newell}}, \binits{J.}},
\bauthor{\bsnm{{Nexon}}, \binits{M.}},
\bauthor{\bsnm{{Nicolas}}, \binits{C.}},
\bauthor{\bsnm{{Nielsen}}, \binits{P.H.}},
\bauthor{\bsnm{{Ninane}}, \binits{N.}},
\bauthor{\bsnm{{Noviello}}, \binits{F.}},
\bauthor{\bsnm{{Novikov}}, \binits{D.}},
\bauthor{\bsnm{{Novikov}}, \binits{I.}},
\bauthor{\bsnm{{O'Dwyer}}, \binits{I.J.}},
\bauthor{\bsnm{{Oldeman}}, \binits{P.}},
\bauthor{\bsnm{{Olivier}}, \binits{P.}},
\bauthor{\bsnm{{Ouchet}}, \binits{L.}},
\bauthor{\bsnm{{Oxborrow}}, \binits{C.A.}},
\bauthor{\bsnm{{P{\'e}rez-Cuevas}}, \binits{L.}},
\bauthor{\bsnm{{Pagan}}, \binits{L.}},
\bauthor{\bsnm{{Paine}}, \binits{C.}},
\bauthor{\bsnm{{Pajot}}, \binits{F.}},
\bauthor{\bsnm{{Paladini}}, \binits{R.}},
\bauthor{\bsnm{{Pancher}}, \binits{F.}},
\bauthor{\bsnm{{Panh}}, \binits{J.}},
\bauthor{\bsnm{{Parks}}, \binits{G.}},
\bauthor{\bsnm{{Parnaudeau}}, \binits{P.}},
\bauthor{\bsnm{{Partridge}}, \binits{B.}},
\bauthor{\bsnm{{Parvin}}, \binits{B.}},
\bauthor{\bsnm{{Pascual}}, \binits{J.P.}},
\bauthor{\bsnm{{Pasian}}, \binits{F.}},
\bauthor{\bsnm{{Pearson}}, \binits{D.P.}},
\bauthor{\bsnm{{Pearson}}, \binits{T.}},
\bauthor{\bsnm{{Pecora}}, \binits{M.}},
\bauthor{\bsnm{{Perdereau}}, \binits{O.}},
\bauthor{\bsnm{{Perotto}}, \binits{L.}},
\bauthor{\bsnm{{Perrotta}}, \binits{F.}},
\bauthor{\bsnm{{Piacentini}}, \binits{F.}},
\bauthor{\bsnm{{Piat}}, \binits{M.}},
\bauthor{\bsnm{{Pierpaoli}}, \binits{E.}},
\bauthor{\bsnm{{Piersanti}}, \binits{O.}},
\bauthor{\bsnm{{Plaige}}, \binits{E.}},
\bauthor{\bsnm{{Plaszczynski}}, \binits{S.}},
\bauthor{\bsnm{{Platania}}, \binits{P.}},
\bauthor{\bsnm{{Pointecouteau}}, \binits{E.}},
\bauthor{\bsnm{{Polenta}}, \binits{G.}},
\bauthor{\bsnm{{Ponthieu}}, \binits{N.}},
\bauthor{\bsnm{{Popa}}, \binits{L.}},
\bauthor{\bsnm{{Poulleau}}, \binits{G.}},
\bauthor{\bsnm{{Poutanen}}, \binits{T.}},
\bauthor{\bsnm{{Pr{\'e}zeau}}, \binits{G.}},
\bauthor{\bsnm{{Pradell}}, \binits{L.}},
\bauthor{\bsnm{{Prina}}, \binits{M.}},
\bauthor{\bsnm{{Prunet}}, \binits{S.}},
\bauthor{\bsnm{{Rachen}}, \binits{J.P.}},
\bauthor{\bsnm{{Rambaud}}, \binits{D.}},
\bauthor{\bsnm{{Rame}}, \binits{F.}},
\bauthor{\bsnm{{Rasmussen}}, \binits{I.}},
\bauthor{\bsnm{{Rautakoski}}, \binits{J.}},
\bauthor{\bsnm{{Reach}}, \binits{W.T.}},
\bauthor{\bsnm{{Rebolo}}, \binits{R.}},
\bauthor{\bsnm{{Reinecke}}, \binits{M.}},
\bauthor{\bsnm{{Reiter}}, \binits{J.}},
\bauthor{\bsnm{{Renault}}, \binits{C.}},
\bauthor{\bsnm{{Ricciardi}}, \binits{S.}},
\bauthor{\bsnm{{Rideau}}, \binits{P.}},
\bauthor{\bsnm{{Riller}}, \binits{T.}},
\bauthor{\bsnm{{Ristorcelli}}, \binits{I.}},
\bauthor{\bsnm{{Riti}}, \binits{J.B.}},
\bauthor{\bsnm{{Rocha}}, \binits{G.}},
\bauthor{\bsnm{{Roche}}, \binits{Y.}},
\bauthor{\bsnm{{Pons}}, \binits{R.}},
\bauthor{\bsnm{{Rohlfs}}, \binits{R.}},
\bauthor{\bsnm{{Romero}}, \binits{D.}},
\bauthor{\bsnm{{Roose}}, \binits{S.}},
\bauthor{\bsnm{{Rosset}}, \binits{C.}},
\bauthor{\bsnm{{Rouberol}}, \binits{S.}},
\bauthor{\bsnm{{Rowan-Robinson}}, \binits{M.}},
\bauthor{\bsnm{{Rubi{\~n}o-Mart{\'\i}n}}, \binits{J.A.}},
\bauthor{\bsnm{{Rusconi}}, \binits{P.}},
\bauthor{\bsnm{{Rusholme}}, \binits{B.}},
\bauthor{\bsnm{{Salama}}, \binits{M.}},
\bauthor{\bsnm{{Salerno}}, \binits{E.}},
\bauthor{\bsnm{{Sandri}}, \binits{M.}},
\bauthor{\bsnm{{Santos}}, \binits{D.}},
\bauthor{\bsnm{{Sanz}}, \binits{J.L.}},
\bauthor{\bsnm{{Sauter}}, \binits{L.}},
\bauthor{\bsnm{{Sauvage}}, \binits{F.}},
\bauthor{\bsnm{{Savini}}, \binits{G.}},
\bauthor{\bsnm{{Schmelzel}}, \binits{M.}},
\bauthor{\bsnm{{Schnorhk}}, \binits{A.}},
\bauthor{\bsnm{{Schwarz}}, \binits{W.}},
\bauthor{\bsnm{{Scott}}, \binits{D.}},
\bauthor{\bsnm{{Seiffert}}, \binits{M.D.}},
\bauthor{\bsnm{{Shellard}}, \binits{P.}},
\bauthor{\bsnm{{Shih}}, \binits{C.}},
\bauthor{\bsnm{{Sias}}, \binits{M.}},
\bauthor{\bsnm{{Silk}}, \binits{J.I.}},
\bauthor{\bsnm{{Silvestri}}, \binits{R.}},
\bauthor{\bsnm{{Sippel}}, \binits{R.}},
\bauthor{\bsnm{{Smoot}}, \binits{G.F.}},
\bauthor{\bsnm{{Starck}}, \binits{J.-L.}},
\bauthor{\bsnm{{Stassi}}, \binits{P.}},
\bauthor{\bsnm{{Sternberg}}, \binits{J.}},
\bauthor{\bsnm{{Stivoli}}, \binits{F.}},
\bauthor{\bsnm{{Stolyarov}}, \binits{V.}},
\bauthor{\bsnm{{Stompor}}, \binits{R.}},
\bauthor{\bsnm{{Stringhetti}}, \binits{L.}},
\bauthor{\bsnm{{Strommen}}, \binits{D.}},
\bauthor{\bsnm{{Stute}}, \binits{T.}},
\bauthor{\bsnm{{Sudiwala}}, \binits{R.}},
\bauthor{\bsnm{{Sugimura}}, \binits{R.}},
\bauthor{\bsnm{{Sunyaev}}, \binits{R.}},
\bauthor{\bsnm{{Sygnet}}, \binits{J.-F.}},
\bauthor{\bsnm{{T{\"u}rler}}, \binits{M.}},
\bauthor{\bsnm{{Taddei}}, \binits{E.}},
\bauthor{\bsnm{{Tallon}}, \binits{J.}},
\bauthor{\bsnm{{Tamiatto}}, \binits{C.}},
\bauthor{\bsnm{{Taurigna}}, \binits{M.}},
\bauthor{\bsnm{{Taylor}}, \binits{D.}},
\bauthor{\bsnm{{Terenzi}}, \binits{L.}},
\bauthor{\bsnm{{Thuerey}}, \binits{S.}},
\bauthor{\bsnm{{Tillis}}, \binits{J.}},
\bauthor{\bsnm{{Tofani}}, \binits{G.}},
\bauthor{\bsnm{{Toffolatti}}, \binits{L.}},
\bauthor{\bsnm{{Tommasi}}, \binits{E.}},
\bauthor{\bsnm{{Tomasi}}, \binits{M.}},
\bauthor{\bsnm{{Tonazzini}}, \binits{E.}},
\bauthor{\bsnm{{Torre}}, \binits{J.-P.}},
\bauthor{\bsnm{{Tosti}}, \binits{S.}},
\bauthor{\bsnm{{Touze}}, \binits{F.}},
\bauthor{\bsnm{{Tristram}}, \binits{M.}},
\bauthor{\bsnm{{Tuovinen}}, \binits{J.}},
\bauthor{\bsnm{{Tuttlebee}}, \binits{M.}},
\bauthor{\bsnm{{Umana}}, \binits{G.}},
\bauthor{\bsnm{{Valenziano}}, \binits{L.}},
\bauthor{\bsnm{{Vall{\'e}e}}, \binits{D.}},
\bauthor{\bsnm{{van der Vlis}}, \binits{M.}},
\bauthor{\bsnm{{van Leeuwen}}, \binits{F.}},
\bauthor{\bsnm{{Vanel}}, \binits{J.-C.}},
\bauthor{\bsnm{{van-Tent}}, \binits{B.}},
\bauthor{\bsnm{{Varis}}, \binits{J.}},
\bauthor{\bsnm{{Vassallo}}, \binits{E.}},
\bauthor{\bsnm{{Vescovi}}, \binits{C.}},
\bauthor{\bsnm{{Vezzu}}, \binits{F.}},
\bauthor{\bsnm{{Vibert}}, \binits{D.}},
\bauthor{\bsnm{{Vielva}}, \binits{P.}},
\bauthor{\bsnm{{Vierra}}, \binits{J.}},
\bauthor{\bsnm{{Villa}}, \binits{F.}},
\bauthor{\bsnm{{Vittorio}}, \binits{N.}},
\bauthor{\bsnm{{Vuerli}}, \binits{C.}},
\bauthor{\bsnm{{Wade}}, \binits{L.A.}},
\bauthor{\bsnm{{Walker}}, \binits{A.R.}},
\bauthor{\bsnm{{Wandelt}}, \binits{B.D.}},
\bauthor{\bsnm{{Watson}}, \binits{C.}},
\bauthor{\bsnm{{Werner}}, \binits{D.}},
\bauthor{\bsnm{{White}}, \binits{M.}},
\bauthor{\bsnm{{White}}, \binits{S.D.M.}},
\bauthor{\bsnm{{Wilkinson}}, \binits{A.}},
\bauthor{\bsnm{{Wilson}}, \binits{P.}},
\bauthor{\bsnm{{Woodcraft}}, \binits{A.}},
\bauthor{\bsnm{{Yoffo}}, \binits{B.}},
\bauthor{\bsnm{{Yun}}, \binits{M.}},
\bauthor{\bsnm{{Yurchenko}}, \binits{V.}},
\bauthor{\bsnm{{Yvon}}, \binits{D.}},
\bauthor{\bsnm{{Zhang}}, \binits{B.}},
\bauthor{\bsnm{{Zimmermann}}, \binits{O.}},
\bauthor{\bsnm{{Zonca}}, \binits{A.}},
\bauthor{\bsnm{{Zorita}}, \binits{D.}}:
\batitle{{Planck pre-launch status: The Planck mission}}.
\bjtitle{Astron. Astrophys.}
\bvolume{520},
\bfpage{1}
(\byear{2010})
\doiurl{10.1051/0004-6361/200912983}
\end{barticle}
\endbibitem

\bibitem[\protect\citeauthoryear{{Morgante} et~al.}{2022}]{Ariel}
\begin{barticle}
\bauthor{\bsnm{{Morgante}}, \binits{G.}},
\bauthor{\bsnm{{Terenzi}}, \binits{L.}},
\bauthor{\bsnm{{Desjonqueres}}, \binits{L.}},
\bauthor{\bsnm{{Eccleston}}, \binits{P.}},
\bauthor{\bsnm{{Bishop}}, \binits{G.}},
\bauthor{\bsnm{{Caldwell}}, \binits{A.}},
\bauthor{\bsnm{{Crook}}, \binits{M.}},
\bauthor{\bsnm{{Drummond}}, \binits{R.}},
\bauthor{\bsnm{{Hills}}, \binits{M.}},
\bauthor{\bsnm{{Hunt}}, \binits{T.}},
\bauthor{\bsnm{{Rust}}, \binits{D.}},
\bauthor{\bsnm{{Puig}}, \binits{L.}},
\bauthor{\bsnm{{Tirolien}}, \binits{T.}},
\bauthor{\bsnm{{Focardi}}, \binits{M.}},
\bauthor{\bsnm{{Zuppella}}, \binits{P.}},
\bauthor{\bsnm{{Holmes}}, \binits{W.}},
\bauthor{\bsnm{{Amiaux}}, \binits{J.}},
\bauthor{\bsnm{{Czupalla}}, \binits{M.}},
\bauthor{\bsnm{{Rataj}}, \binits{M.}},
\bauthor{\bsnm{{Jessen}}, \binits{N.C.}},
\bauthor{\bsnm{{Pedersen}}, \binits{S.M.}},
\bauthor{\bsnm{{Pascale}}, \binits{E.}},
\bauthor{\bsnm{{Pace}}, \binits{E.}},
\bauthor{\bsnm{{Malaguti}}, \binits{G.}},
\bauthor{\bsnm{{Micela}}, \binits{G.}}:
\batitle{{The thermal architecture of the ESA ARIEL payload at the end of phase B1}}.
\bjtitle{Experimental Astronomy}
\bvolume{53}(\bissue{2}),
\bfpage{905}--\blpage{944}
(\byear{2022})
\doiurl{10.1007/s10686-022-09851-y}
\end{barticle}
\endbibitem

\bibitem[\protect\citeauthoryear{{Jackson} et~al.}{2016}]{Jackson_2016SPIE.9905E..2IJ}
\begin{bchapter}
\bauthor{\bsnm{{Jackson}}, \binits{B.D.}},
\bauthor{\bsnm{{van Weers}}, \binits{H.}},
\bauthor{\bsnm{{van der Kuur}}, \binits{J.}},
\bauthor{\bsnm{{den Hartog}}, \binits{R.}},
\bauthor{\bsnm{{Akamatsu}}, \binits{H.}},
\bauthor{\bsnm{{Argan}}, \binits{A.}},
\bauthor{\bsnm{{Bandler}}, \binits{S.R.}},
\bauthor{\bsnm{{Barbera}}, \binits{M.}},
\bauthor{\bsnm{{Barret}}, \binits{D.}},
\bauthor{\bsnm{{Bruijn}}, \binits{M.P.}},
\bauthor{\bsnm{{Chervenak}}, \binits{J.A.}},
\bauthor{\bsnm{{Dercksen}}, \binits{J.}},
\bauthor{\bsnm{{Gatti}}, \binits{F.}},
\bauthor{\bsnm{{Gottardi}}, \binits{L.}},
\bauthor{\bsnm{{Haas}}, \binits{D.}},
\bauthor{\bsnm{{den Herder}}, \binits{J.-W.}},
\bauthor{\bsnm{{Kilbourne}}, \binits{C.A.}},
\bauthor{\bsnm{{Kiviranta}}, \binits{M.}},
\bauthor{\bsnm{{Lam-Trong}}, \binits{T.}},
\bauthor{\bsnm{{van Leeuwen}}, \binits{B.-J.}},
\bauthor{\bsnm{{Macculi}}, \binits{C.}},
\bauthor{\bsnm{{Piro}}, \binits{L.}},
\bauthor{\bsnm{{Smith}}, \binits{S.J.}}:
\bctitle{{The focal plane assembly for the Athena X-ray Integral Field Unit instrument}}.
In: \beditor{\bsnm{{den Herder}}, \binits{J.-W.A.}},
\beditor{\bsnm{{Takahashi}}, \binits{T.}},
\beditor{\bsnm{{Bautz}}, \binits{M.}} (eds.)
\bbtitle{Space Telescopes and Instrumentation 2016: Ultraviolet to Gamma Ray}.
\bsertitle{Society of Photo-Optical Instrumentation Engineers (SPIE) Conference Series},
vol. \bseriesno{9905},
p. \bfpage{99052}
(\byear{2016}).
\doiurl{10.1117/12.2232544}
\end{bchapter}
\endbibitem

\bibitem[\protect\citeauthoryear{{van Weers} et~al.}{2024}]{Weers2024}
\begin{bchapter}
\bauthor{\bsnm{{van Weers}}, \binits{H.J.}},
\bauthor{\bsnm{Dercksen}, \binits{J.P.C.}},
\bauthor{\bsnm{Leeman}, \binits{M.}},
\bauthor{\bsnm{{van der Hulst}}, \binits{P.}},
\bauthor{\bsnm{{van der Linden}}, \binits{A.J.}},
\bauthor{\bsnm{Bennebroek}, \binits{A.Q.}},
\bauthor{\bsnm{{den Hartog}}, \binits{R.J.}},
\bauthor{\bsnm{Khosropanah}, \binits{P.}},
\bauthor{\bsnm{Jackson}, \binits{B.D.}},
\bauthor{\bsnm{Roelfsema}, \binits{P.}}:
\bctitle{{X-IFU focal plane assembly development model design upgrade and critical technology developments}}.
In: \beditor{\bsnm{Herder}, \binits{J.-W.A.}},
\beditor{\bsnm{Nikzad}, \binits{S.}},
\beditor{\bsnm{Nakazawa}, \binits{K.}} (eds.)
\bbtitle{Space Telescopes and Instrumentation 2024: Ultraviolet to Gamma Ray},
vol. \bseriesno{13093},
p. \bfpage{130930}.
\bpublisher{SPIE}, \blocation{???}
(\byear{2024}).
\doiurl{10.1117/12.3020114} .
\bcomment{International Society for Optics and Photonics}
\end{bchapter}
\endbibitem

\bibitem[\protect\citeauthoryear{{Duval} et~al.}{2024}]{Duval2024}
\begin{botherref}
\oauthor{\bsnm{{Duval}}, \binits{J.-M.}},
\oauthor{\bsnm{{Bancel}}, \binits{F.}},
\oauthor{\bsnm{{Charles}}, \binits{I.}},
\oauthor{\bsnm{{Durand}}, \binits{J.-L.}},
\oauthor{\bsnm{{Martin}}, \binits{S.}},
\oauthor{\bsnm{{Prouvé}}, \binits{T.}},
\oauthor{\bsnm{{Marin}}, \binits{C.}}:
{5-stage ADR cooler for the Athena space mission: design and preliminary characterization}.
To be published in Cryocooler 23
(2024)
\end{botherref}
\endbibitem

\bibitem[\protect\citeauthoryear{{Bozzo} et~al.}{2016}]{Bozzo_2016arXiv160903776B}
\begin{botherref}
\oauthor{\bsnm{{Bozzo}}, \binits{E.}},
\oauthor{\bsnm{{Barbera}}, \binits{M.}},
\oauthor{\bsnm{{Genolet}}, \binits{L.}},
\oauthor{\bsnm{{Paltani}}, \binits{S.}},
\oauthor{\bsnm{{Sordet}}, \binits{M.}},
\oauthor{\bsnm{{Branduardi-Raymont}}, \binits{G.}},
\oauthor{\bsnm{{Rauw}}, \binits{G.}},
\oauthor{\bsnm{{Sciortino}}, \binits{S.}},
\oauthor{\bsnm{{Barret}}, \binits{D.}},
\oauthor{\bsnm{{Den Herder}}, \binits{J.W.}}:
{The Filter Wheel and Filters development for the X-IFU instrument on-board Athena}.
arXiv e-prints,
1609--03776
(2016)
{\href{https://arxiv.org/abs/1609.03776}{{arXiv:1609.03776}}}
{[astro-ph.IM]}
\end{botherref}
\endbibitem

\bibitem[\protect\citeauthoryear{{Durkin} et~al.}{2021}]{Durkin_2021ITAS...3165279D}
\begin{barticle}
\bauthor{\bsnm{{Durkin}}, \binits{M.}},
\bauthor{\bsnm{{Adams}}, \binits{J.S.}},
\bauthor{\bsnm{{Bandler}}, \binits{S.R.}},
\bauthor{\bsnm{{Chervenak}}, \binits{J.A.}},
\bauthor{\bsnm{{Denison}}, \binits{E.V.}},
\bauthor{\bsnm{{Doriese}}, \binits{W.B.}},
\bauthor{\bsnm{{Duff}}, \binits{S.M.}},
\bauthor{\bsnm{{Finkbeiner}}, \binits{F.M.}},
\bauthor{\bsnm{{Fowler}}, \binits{J.W.}},
\bauthor{\bsnm{{Gard}}, \binits{J.D.}},
\bauthor{\bsnm{{Hilton}}, \binits{G.C.}},
\bauthor{\bsnm{{Hummatov}}, \binits{R.}},
\bauthor{\bsnm{{Irwin}}, \binits{K.D.}},
\bauthor{\bsnm{{Joe}}, \binits{Y.I.}},
\bauthor{\bsnm{{Kelley}}, \binits{R.L.}},
\bauthor{\bsnm{{Kilbourne}}, \binits{C.A.}},
\bauthor{\bsnm{{Miniussi}}, \binits{A.R.}},
\bauthor{\bsnm{{Morgan}}, \binits{K.M.}},
\bauthor{\bsnm{{O'Neil}}, \binits{G.C.}},
\bauthor{\bsnm{{Pappas}}, \binits{C.G.}},
\bauthor{\bsnm{{Porter}}, \binits{F.S.}},
\bauthor{\bsnm{{Reintsema}}, \binits{C.D.}},
\bauthor{\bsnm{{Rudman}}, \binits{D.A.}},
\bauthor{\bsnm{{Sakai}}, \binits{K.}},
\bauthor{\bsnm{{Smith}}, \binits{S.J.}},
\bauthor{\bsnm{{Stevens}}, \binits{R.W.}},
\bauthor{\bsnm{{Swetz}}, \binits{D.S.}},
\bauthor{\bsnm{{Szypryt}}, \binits{P.}},
\bauthor{\bsnm{{Ullom}}, \binits{J.N.}},
\bauthor{\bsnm{{Vale}}, \binits{L.R.}},
\bauthor{\bsnm{{Wakeham}}, \binits{N.}}:
\batitle{{Mitigation of Finite Bandwidth Effects in Time-Division-Multiplexed SQUID Readout of TES Arrays}}.
\bjtitle{IEEE Transactions on Applied Superconductivity}
\bvolume{31}(\bissue{5}),
\bfpage{3065279}
(\byear{2021})
\doiurl{10.1109/TASC.2021.3065279}
\end{barticle}
\endbibitem

\bibitem[\protect\citeauthoryear{Durkin et~al.}{2023}]{Durkin2023}
\begin{barticle}
\bauthor{\bsnm{Durkin}, \binits{M.}},
\bauthor{\bsnm{Backhaus}, \binits{S.}},
\bauthor{\bsnm{Bandler}, \binits{S.R.}},
\bauthor{\bsnm{Chervenak}, \binits{J.A.}},
\bauthor{\bsnm{Denison}, \binits{E.V.}},
\bauthor{\bsnm{Doriese}, \binits{W.B.}},
\bauthor{\bsnm{Gard}, \binits{J.D.}},
\bauthor{\bsnm{Hilton}, \binits{G.C.}},
\bauthor{\bsnm{Lew}, \binits{R.A.}},
\bauthor{\bsnm{Lucas}, \binits{T.J.}},
\bauthor{\bsnm{Reintsema}, \binits{C.D.}},
\bauthor{\bsnm{Schmidt}, \binits{D.R.}},
\bauthor{\bsnm{Smith}, \binits{S.J.}},
\bauthor{\bsnm{Ullom}, \binits{J.N.}},
\bauthor{\bsnm{Vale}, \binits{L.R.}},
\bauthor{\bsnm{Vissers}, \binits{M.R.}},
\bauthor{\bsnm{Wakeham}, \binits{N.A.}}:
\batitle{Symmetric time-division-multiplexed squid readout with two-layer switches for future tes observatories}.
\bjtitle{IEEE Transactions on Applied Superconductivity}
\bvolume{33}(\bissue{5}),
\bfpage{1}--\blpage{5}
(\byear{2023})
\doiurl{10.1109/TASC.2023.3264175}
\end{barticle}
\endbibitem

\bibitem[\protect\citeauthoryear{{Kiviranta} et~al.}{2021}]{Kiviranta_2021ITAS...3160356K}
\begin{barticle}
\bauthor{\bsnm{{Kiviranta}}, \binits{M.}},
\bauthor{\bsnm{{Gronberg}}, \binits{L.}},
\bauthor{\bsnm{{Puranen}}, \binits{T.}},
\bauthor{\bsnm{{van der Kuur}}, \binits{J.}},
\bauthor{\bsnm{{Beev}}, \binits{N.}},
\bauthor{\bsnm{{Salonen}}, \binits{J.}},
\bauthor{\bsnm{{Hazra}}, \binits{D.}},
\bauthor{\bsnm{{Korpela}}, \binits{S.}}:
\batitle{{Two-Stage SQUID Amplifier for the Frequency Multiplexed Readout of the X-IFU X-Ray Camera}}.
\bjtitle{IEEE Transactions on Applied Superconductivity}
\bvolume{31}(\bissue{5}),
\bfpage{3060356}
(\byear{2021})
\doiurl{10.1109/TASC.2021.3060356}
\end{barticle}
\endbibitem

\bibitem[\protect\citeauthoryear{{Smith} et~al.}{2021}]{Smith_2021ITAS...3161918S}
\begin{barticle}
\bauthor{\bsnm{{Smith}}, \binits{S.J.}},
\bauthor{\bsnm{{Adams}}, \binits{J.S.}},
\bauthor{\bsnm{{Bandler}}, \binits{S.R.}},
\bauthor{\bsnm{{Beaumont}}, \binits{S.}},
\bauthor{\bsnm{{Chervenak}}, \binits{J.A.}},
\bauthor{\bsnm{{Denison}}, \binits{E.V.}},
\bauthor{\bsnm{{Doriese}}, \binits{W.B.}},
\bauthor{\bsnm{{Durkin}}, \binits{M.}},
\bauthor{\bsnm{{Finkbeiner}}, \binits{F.M.}},
\bauthor{\bsnm{{Fowler}}, \binits{J.W.}},
\bauthor{\bsnm{{Hilton}}, \binits{G.C.}},
\bauthor{\bsnm{{Hummatov}}, \binits{R.}},
\bauthor{\bsnm{{Irwin}}, \binits{K.D.}},
\bauthor{\bsnm{{Kelley}}, \binits{R.L.}},
\bauthor{\bsnm{{Kilbourne}}, \binits{C.A.}},
\bauthor{\bsnm{{Leutenegger}}, \binits{M.A.}},
\bauthor{\bsnm{{Miniussi}}, \binits{A.R.}},
\bauthor{\bsnm{{Porter}}, \binits{F.S.}},
\bauthor{\bsnm{{Reintsema}}, \binits{C.D.}},
\bauthor{\bsnm{{Sadleir}}, \binits{J.E.}},
\bauthor{\bsnm{{Sakai}}, \binits{K.}},
\bauthor{\bsnm{{Swetz}}, \binits{D.S.}},
\bauthor{\bsnm{{Ullom}}, \binits{J.N.}},
\bauthor{\bsnm{{Vale}}, \binits{L.R.}},
\bauthor{\bsnm{{Wakeham}}, \binits{N.A.}},
\bauthor{\bsnm{{Wassell}}, \binits{E.J.}},
\bauthor{\bsnm{{Witthoeft}}, \binits{M.C.}}:
\batitle{{Performance of a Broad-Band, High-Resolution, Transition-Edge Sensor Spectrometer for X-ray Astrophysics}}.
\bjtitle{IEEE Transactions on Applied Superconductivity}
\bvolume{31}(\bissue{5}),
\bfpage{3061918}
(\byear{2021})
\doiurl{10.1109/TASC.2021.3061918}
\end{barticle}
\endbibitem

\bibitem[\protect\citeauthoryear{{Wakeham} et~al.}{2023}]{Wakeham2023}
\begin{barticle}
\bauthor{\bsnm{{Wakeham}}, \binits{N.A.}},
\bauthor{\bsnm{{Adams}}, \binits{J.S.}},
\bauthor{\bsnm{{Bandler}}, \binits{S.R.}},
\bauthor{\bsnm{{Beaumont}}, \binits{S.}},
\bauthor{\bsnm{{Chervenak}}, \binits{J.A.}},
\bauthor{\bsnm{{Cumbee}}, \binits{R.S.}},
\bauthor{\bsnm{{Finkbeiner}}, \binits{F.M.}},
\bauthor{\bsnm{{Ha}}, \binits{J.Y.}},
\bauthor{\bsnm{{Hull}}, \binits{S.}},
\bauthor{\bsnm{{Kelley}}, \binits{R.L.}},
\bauthor{\bsnm{{Kilbourne}}, \binits{C.A.}},
\bauthor{\bsnm{{Porter}}, \binits{F.S.}},
\bauthor{\bsnm{{Sakai}}, \binits{K.}},
\bauthor{\bsnm{{Smith}}, \binits{S.J.}},
\bauthor{\bsnm{{Wassell}}, \binits{E.J.}},
\bauthor{\bsnm{{Yoon}}, \binits{S.}}:
\batitle{{Refinement of Transition-Edge Sensor Dimensions for the X-Ray Integral Field Unit on ATHENA}}.
\bjtitle{IEEE Transactions on Applied Superconductivity}
\bvolume{33}(\bissue{5}),
\bfpage{3253067}
(\byear{2023})
\doiurl{10.1109/TASC.2023.3253067}
\end{barticle}
\endbibitem

\bibitem[\protect\citeauthoryear{Geoffray et~al.}{2024}]{Geoffray2024}
\begin{bchapter}
\bauthor{\bsnm{Geoffray}, \binits{H.}},
\bauthor{\bsnm{Jackson}, \binits{B.D.}},
\bauthor{\bsnm{Bandler}, \binits{S.R.}},
\bauthor{\bsnm{Smith}, \binits{S.J.}},
\bauthor{\bsnm{Doriese}, \binits{W.B.}},
\bauthor{\bsnm{Durkin}, \binits{M.}},
\bauthor{\bsnm{{van der Kuur}}, \binits{J.}},
\bauthor{\bsnm{{van der Hulst}}, \binits{P.}},
\bauthor{\bsnm{Kiviranta}, \binits{M.}},
\bauthor{\bsnm{Pr{\^e}le}, \binits{D.}},
\bauthor{\bsnm{Gonzalez}, \binits{M.}},
\bauthor{\bsnm{Ravera}, \binits{L.}},
\bauthor{\bsnm{Parot}, \binits{Y.}},
\bauthor{\bsnm{{den Hartog}}, \binits{R.}},
\bauthor{\bsnm{Taralli}, \binits{E.}},
\bauthor{\bsnm{Vaccaro}, \binits{D.}},
\bauthor{\bsnm{{van Weers}}, \binits{H.}},
\bauthor{\bsnm{Adams}, \binits{J.}},
\bauthor{\bsnm{Chervenak}, \binits{J.A.}},
\bauthor{\bsnm{Hull}, \binits{S.}},
\bauthor{\bsnm{Sakai}, \binits{K.}},
\bauthor{\bsnm{Wakeham}, \binits{N.}},
\bauthor{\bsnm{Reintsema}, \binits{C.D.}},
\bauthor{\bsnm{Ullom}, \binits{J.N.}},
\bauthor{\bsnm{Beaumont}, \binits{S.}},
\bauthor{\bsnm{Brachet}, \binits{F.}},
\bauthor{\bsnm{Cenac-Morthe}, \binits{C.}},
\bauthor{\bsnm{Cucchetti}, \binits{E.}},
\bauthor{\bsnm{Daniel}, \binits{C.}},
\bauthor{\bsnm{Peille}, \binits{P.}},
\bauthor{\bsnm{Soucek}, \binits{J.}},
\bauthor{\bsnm{Argan}, \binits{A.}},
\bauthor{\bsnm{Macculi}, \binits{C.}}:
\bctitle{{The detection chain for Athena X-IFU: a status on the design and demonstrations}}.
In: \beditor{\bsnm{Herder}, \binits{J.-W.A.}},
\beditor{\bsnm{Nikzad}, \binits{S.}},
\beditor{\bsnm{Nakazawa}, \binits{K.}} (eds.)
\bbtitle{Space Telescopes and Instrumentation 2024: Ultraviolet to Gamma Ray},
vol. \bseriesno{13093},
p. \bfpage{130930}.
\bpublisher{SPIE}, \blocation{???}
(\byear{2024}).
\doiurl{10.1117/12.3020011} .
\bcomment{International Society for Optics and Photonics}
\end{bchapter}
\endbibitem

\bibitem[\protect\citeauthoryear{Pr\^ele et~al.}{2024}]{Prele:2024msq}
\begin{barticle}
\bauthor{\bsnm{Pr\^ele}, \binits{D.}},
\bauthor{\bsnm{Gonzalez}, \binits{M.}},
\bauthor{\bsnm{Arnaldi}, \binits{L.H.}},
\bauthor{\bsnm{Chen}, \binits{S.}},
\bauthor{\bsnm{Courty}, \binits{B.}},
\bauthor{\bsnm{Dambrauskas}, \binits{E.}},
\bauthor{\bsnm{Givaudan}, \binits{A.}},
\bauthor{\bsnm{Le~Cam}, \binits{M.}},
\bauthor{\bsnm{Lesrel}, \binits{J.}},
\bauthor{\bsnm{Punch}, \binits{M.}}:
\batitle{{X-IFU warm front-end electronics demonstrator model}}.
\bjtitle{Proc. SPIE Int. Soc. Opt. Eng.}
\bvolume{13093},
\bfpage{130934}
(\byear{2024})
\doiurl{10.1117/12.3020767}
\end{barticle}
\endbibitem

\bibitem[\protect\citeauthoryear{{Murat} et~al.}{2024}]{Murat2024}
\begin{bchapter}
\bauthor{\bsnm{{Murat}}, \binits{D.}},
\bauthor{\bsnm{{Ravera}}, \binits{L.}},
\bauthor{\bsnm{{Parot}}, \binits{Y.}},
\bauthor{\bsnm{{Camus}}, \binits{T.}},
\bauthor{\bsnm{{Coeur-Joly}}, \binits{O.}},
\bauthor{\bsnm{{Guerin}}, \binits{C.}}:
\bctitle{{The DEMUX module of the ATHENA/X-IFU digital readout electronics demonstration model}}.
In: \beditor{\bsnm{{den Herder}}, \binits{J.-W.A.}},
\beditor{\bsnm{{Nikzad}}, \binits{S.}},
\beditor{\bsnm{{Nakazawa}}, \binits{K.}} (eds.)
\bbtitle{Space Telescopes and Instrumentation 2024: Ultraviolet to Gamma Ray}.
\bsertitle{Society of Photo-Optical Instrumentation Engineers (SPIE) Conference Series},
vol. \bseriesno{13093},
p. \bfpage{130934}
(\byear{2024}).
\doiurl{10.1117/12.3020045}
\end{bchapter}
\endbibitem

\bibitem[\protect\citeauthoryear{Smith et~al.}{2024}]{Smith2024}
\begin{bchapter}
\bauthor{\bsnm{Smith}, \binits{S.J.}},
\bauthor{\bsnm{Adams}, \binits{J.S.}},
\bauthor{\bsnm{Bandler}, \binits{S.R.}},
\bauthor{\bsnm{Borrelli}, \binits{R.B.}},
\bauthor{\bsnm{Chervenak}, \binits{J.A.}},
\bauthor{\bsnm{Colazo-Petit}, \binits{F.A.}},
\bauthor{\bsnm{Cumbee}, \binits{R.S.}},
\bauthor{\bsnm{Finkbeiner}, \binits{F.M.}},
\bauthor{\bsnm{Fuhrman}, \binits{J.D.}},
\bauthor{\bsnm{Hull}, \binits{S.V.}},
\bauthor{\bsnm{Kelley}, \binits{R.L.}},
\bauthor{\bsnm{Kilbourne}, \binits{C.A.}},
\bauthor{\bsnm{Porter}, \binits{F.S.}},
\bauthor{\bsnm{Rani}, \binits{A.}},
\bauthor{\bsnm{Sakai}, \binits{K.}},
\bauthor{\bsnm{Wakeham}, \binits{N.A.}},
\bauthor{\bsnm{Wassell}, \binits{E.J.}},
\bauthor{\bsnm{Witthoeft}, \binits{M.C.}},
\bauthor{\bsnm{Yoon}, \binits{S.H.}}:
\bctitle{{Development of the microcalorimeter detector for the Athena/X-ray Integral Field Unit}}.
In: \beditor{\bsnm{Herder}, \binits{J.-W.A.}},
\beditor{\bsnm{Nikzad}, \binits{S.}},
\beditor{\bsnm{Nakazawa}, \binits{K.}} (eds.)
\bbtitle{Space Telescopes and Instrumentation 2024: Ultraviolet to Gamma Ray},
vol. \bseriesno{13093},
p. \bfpage{130930}.
\bpublisher{SPIE}, \blocation{???}
(\byear{2024}).
\doiurl{10.1117/12.3019198} .
\bcomment{International Society for Optics and Photonics}
\end{bchapter}
\endbibitem

\bibitem[\protect\citeauthoryear{{Peille} et~al.}{2018a}]{Peille_2018JLTP..193..940P}
\begin{barticle}
\bauthor{\bsnm{{Peille}}, \binits{P.}},
\bauthor{\bsnm{{Dauser}}, \binits{T.}},
\bauthor{\bsnm{{Kirsch}}, \binits{C.}},
\bauthor{\bsnm{{den Hartog}}, \binits{R.}},
\bauthor{\bsnm{{Cucchetti}}, \binits{E.}},
\bauthor{\bsnm{{Wilms}}, \binits{J.}},
\bauthor{\bsnm{{Barret}}, \binits{D.}},
\bauthor{\bsnm{{den Herder}}, \binits{J.-W.}},
\bauthor{\bsnm{{Piro}}, \binits{L.}}:
\batitle{{The Performance of the Athena X-ray Integral Field Unit at Very High Count Rates}}.
\bjtitle{Journal of Low Temperature Physics}
\bvolume{193}(\bissue{5-6}),
\bfpage{940}--\blpage{948}
(\byear{2018})
\doiurl{10.1007/s10909-018-1964-6}
\end{barticle}
\endbibitem

\bibitem[\protect\citeauthoryear{{Peille} et~al.}{2018b}]{Peille_2018SPIE10699E..4KP}
\begin{bchapter}
\bauthor{\bsnm{{Peille}}, \binits{P.}},
\bauthor{\bsnm{{den Hartog}}, \binits{R.}},
\bauthor{\bsnm{{Macculi}}, \binits{C.}},
\bauthor{\bsnm{{Barbera}}, \binits{M.}},
\bauthor{\bsnm{{Lotti}}, \binits{S.}},
\bauthor{\bsnm{{Cucchetti}}, \binits{E.}},
\bauthor{\bsnm{{Kirsch}}, \binits{C.}},
\bauthor{\bsnm{{Dauser}}, \binits{T.}},
\bauthor{\bsnm{{Wilms}}, \binits{J.}},
\bauthor{\bsnm{{Bandler}}, \binits{S.R.}},
\bauthor{\bsnm{{Smith}}, \binits{S.J.}},
\bauthor{\bsnm{{Jackson}}, \binits{B.}},
\bauthor{\bsnm{{Geoffray}}, \binits{H.}},
\bauthor{\bsnm{{Mesnager}}, \binits{J.-M.}},
\bauthor{\bsnm{{Pajot}}, \binits{F.}},
\bauthor{\bsnm{{Barret}}, \binits{D.}},
\bauthor{\bsnm{{Lam-Trong}}, \binits{T.}},
\bauthor{\bsnm{{den Herder}}, \binits{J.-W.}},
\bauthor{\bsnm{{Piro}}, \binits{L.}}:
\bctitle{{The performance of the ATHENA X-ray Integral Field Unit}}.
In: \beditor{\bsnm{{den Herder}}, \binits{J.-W.A.}},
\beditor{\bsnm{{Nikzad}}, \binits{S.}},
\beditor{\bsnm{{Nakazawa}}, \binits{K.}} (eds.)
\bbtitle{Space Telescopes and Instrumentation 2018: Ultraviolet to Gamma Ray}.
\bsertitle{Society of Photo-Optical Instrumentation Engineers (SPIE) Conference Series},
vol. \bseriesno{10699},
p. \bfpage{106994}
(\byear{2018}).
\doiurl{10.1117/12.2313720}
\end{bchapter}
\endbibitem

\bibitem[\protect\citeauthoryear{{Tashiro} et~al.}{2020}]{Tashiro2020}
\begin{bchapter}
\bauthor{\bsnm{{Tashiro}}, \binits{M.}},
\bauthor{\bsnm{{Maejima}}, \binits{H.}},
\bauthor{\bsnm{{Toda}}, \binits{K.}},
\bauthor{\bsnm{{Kelley}}, \binits{R.}},
\bauthor{\bsnm{{Reichenthal}}, \binits{L.}},
\bauthor{\bsnm{{Hartz}}, \binits{L.}},
\bauthor{\bsnm{{Petre}}, \binits{R.}},
\bauthor{\bsnm{{Williams}}, \binits{B.}},
\bauthor{\bsnm{{Guainazzi}}, \binits{M.}},
\bauthor{\bsnm{{Costantini}}, \binits{E.}},
\bauthor{\bsnm{{Fujimoto}}, \binits{R.}},
\bauthor{\bsnm{{Hayashida}}, \binits{K.}},
\bauthor{\bsnm{{Henegar-Leon}}, \binits{J.}},
\bauthor{\bsnm{{Holland}}, \binits{M.}},
\bauthor{\bsnm{{Ishisaki}}, \binits{Y.}},
\bauthor{\bsnm{{Kilbourne}}, \binits{C.}},
\bauthor{\bsnm{{Loewenstein}}, \binits{M.}},
\bauthor{\bsnm{{Matsushita}}, \binits{K.}},
\bauthor{\bsnm{{Mori}}, \binits{K.}},
\bauthor{\bsnm{{Okajima}}, \binits{T.}},
\bauthor{\bsnm{{Porter}}, \binits{F.S.}},
\bauthor{\bsnm{{Sneiderman}}, \binits{G.}},
\bauthor{\bsnm{{Takei}}, \binits{Y.}},
\bauthor{\bsnm{{Terada}}, \binits{Y.}},
\bauthor{\bsnm{{Tomida}}, \binits{H.}},
\bauthor{\bsnm{{Yamaguchi}}, \binits{H.}},
\bauthor{\bsnm{{Watanabe}}, \binits{S.}},
\bauthor{\bsnm{{Akamatsu}}, \binits{H.}},
\bauthor{\bsnm{{Arai}}, \binits{Y.}},
\bauthor{\bsnm{{Audard}}, \binits{M.}},
\bauthor{\bsnm{{Awaki}}, \binits{H.}},
\bauthor{\bsnm{{Babyk}}, \binits{I.}},
\bauthor{\bsnm{{Bamba}}, \binits{A.}},
\bauthor{\bsnm{{Bando}}, \binits{N.}},
\bauthor{\bsnm{{Behar}}, \binits{E.}},
\bauthor{\bsnm{{Bialas}}, \binits{T.}},
\bauthor{\bsnm{{Boissay-Malaquin}}, \binits{R.}},
\bauthor{\bsnm{{Brenneman}}, \binits{L.}},
\bauthor{\bsnm{{Brown}}, \binits{G.}},
\bauthor{\bsnm{{Canavan}}, \binits{E.}},
\bauthor{\bsnm{{Chiao}}, \binits{M.}},
\bauthor{\bsnm{{Comber}}, \binits{B.}},
\bauthor{\bsnm{{Corrales}}, \binits{L.}},
\bauthor{\bsnm{{Cumbee}}, \binits{R.}},
\bauthor{\bsnm{{de Vries}}, \binits{C.}},
\bauthor{\bsnm{{den Herder}}, \binits{J.-W.}},
\bauthor{\bsnm{{Dercksen}}, \binits{J.}},
\bauthor{\bsnm{{Diaz-Trigo}}, \binits{M.}},
\bauthor{\bsnm{{DiPirro}}, \binits{M.}},
\bauthor{\bsnm{{Done}}, \binits{C.}},
\bauthor{\bsnm{{Dotani}}, \binits{T.}},
\bauthor{\bsnm{{Ebisawa}}, \binits{K.}},
\bauthor{\bsnm{{Eckart}}, \binits{M.}},
\bauthor{\bsnm{{Eckert}}, \binits{D.}},
\bauthor{\bsnm{{Eguchi}}, \binits{S.}},
\bauthor{\bsnm{{Enoto}}, \binits{T.}},
\bauthor{\bsnm{{Ezoe}}, \binits{Y.}},
\bauthor{\bsnm{{Ferrigno}}, \binits{C.}},
\bauthor{\bsnm{{Fujita}}, \binits{Y.}},
\bauthor{\bsnm{{Fukazawa}}, \binits{Y.}},
\bauthor{\bsnm{{Furuzawa}}, \binits{A.}},
\bauthor{\bsnm{{Gallo}}, \binits{L.}},
\bauthor{\bsnm{{Gorter}}, \binits{N.}},
\bauthor{\bsnm{{Grim}}, \binits{M.}},
\bauthor{\bsnm{{Gu}}, \binits{L.}},
\bauthor{\bsnm{{Hagino}}, \binits{K.}},
\bauthor{\bsnm{{Hamaguchi}}, \binits{K.}},
\bauthor{\bsnm{{Hatsukade}}, \binits{I.}},
\bauthor{\bsnm{{Hawthorn}}, \binits{D.}},
\bauthor{\bsnm{{Hayashi}}, \binits{K.}},
\bauthor{\bsnm{{Hell}}, \binits{N.}},
\bauthor{\bsnm{{Hiraga}}, \binits{J.}},
\bauthor{\bsnm{{Hodges-Kluck}}, \binits{E.}},
\bauthor{\bsnm{{Horiuchi}}, \binits{T.}},
\bauthor{\bsnm{{Hornschemeier}}, \binits{A.}},
\bauthor{\bsnm{{Hoshino}}, \binits{A.}},
\bauthor{\bsnm{{Ichinohe}}, \binits{Y.}},
\bauthor{\bsnm{{Iga}}, \binits{S.}},
\bauthor{\bsnm{{Iizuka}}, \binits{R.}},
\bauthor{\bsnm{{Ishida}}, \binits{M.}},
\bauthor{\bsnm{{Ishihama}}, \binits{N.}},
\bauthor{\bsnm{{Ishikawa}}, \binits{K.}},
\bauthor{\bsnm{{Ishimura}}, \binits{K.}},
\bauthor{\bsnm{{Jaffe}}, \binits{T.}},
\bauthor{\bsnm{{Kaastra}}, \binits{J.}},
\bauthor{\bsnm{{Kallman}}, \binits{T.}},
\bauthor{\bsnm{{Kara}}, \binits{E.}},
\bauthor{\bsnm{{Katsuda}}, \binits{S.}},
\bauthor{\bsnm{{Kenyon}}, \binits{S.}},
\bauthor{\bsnm{{Kimball}}, \binits{M.}},
\bauthor{\bsnm{{Kitaguchi}}, \binits{T.}},
\bauthor{\bsnm{{Kitamoto}}, \binits{S.}},
\bauthor{\bsnm{{Kobayashi}}, \binits{S.}},
\bauthor{\bsnm{{Kobayashi}}, \binits{A.}},
\bauthor{\bsnm{{Kohmura}}, \binits{T.}},
\bauthor{\bsnm{{Kubota}}, \binits{A.}},
\bauthor{\bsnm{{Leutenegger}}, \binits{M.}},
\bauthor{\bsnm{{Li}}, \binits{M.}},
\bauthor{\bsnm{{Lockard}}, \binits{T.}},
\bauthor{\bsnm{{Maeda}}, \binits{Y.}},
\bauthor{\bsnm{{Markevitch}}, \binits{M.}},
\bauthor{\bsnm{{Martz}}, \binits{C.}},
\bauthor{\bsnm{{Matsumoto}}, \binits{H.}},
\bauthor{\bsnm{{Matsuzaki}}, \binits{K.}},
\bauthor{\bsnm{{McCammon}}, \binits{D.}},
\bauthor{\bsnm{{McLaughlin}}, \binits{B.}},
\bauthor{\bsnm{{McNamara}}, \binits{B.}},
\bauthor{\bsnm{{Miko}}, \binits{J.}},
\bauthor{\bsnm{{Miller}}, \binits{E.}},
\bauthor{\bsnm{{Miller}}, \binits{J.}},
\bauthor{\bsnm{{Minesugi}}, \binits{K.}},
\bauthor{\bsnm{{Mitani}}, \binits{S.}},
\bauthor{\bsnm{{Mitsuishi}}, \binits{I.}},
\bauthor{\bsnm{{Mizumoto}}, \binits{M.}},
\bauthor{\bsnm{{Mizuno}}, \binits{T.}},
\bauthor{\bsnm{{Mukai}}, \binits{K.}},
\bauthor{\bsnm{{Murakami}}, \binits{H.}},
\bauthor{\bsnm{{Mushotzky}}, \binits{R.}},
\bauthor{\bsnm{{Nakajima}}, \binits{H.}},
\bauthor{\bsnm{{Nakamura}}, \binits{H.}},
\bauthor{\bsnm{{Nakazawa}}, \binits{K.}},
\bauthor{\bsnm{{Natsukari}}, \binits{C.}},
\bauthor{\bsnm{{Nigo}}, \binits{K.}},
\bauthor{\bsnm{{Nishioka}}, \binits{Y.}},
\bauthor{\bsnm{{Nobukawa}}, \binits{K.}},
\bauthor{\bsnm{{Nobukawa}}, \binits{M.}},
\bauthor{\bsnm{{Noda}}, \binits{H.}},
\bauthor{\bsnm{{Odaka}}, \binits{H.}},
\bauthor{\bsnm{{Ogawa}}, \binits{M.}},
\bauthor{\bsnm{{Ohashi}}, \binits{T.}},
\bauthor{\bsnm{{Ohno}}, \binits{M.}},
\bauthor{\bsnm{{Ohta}}, \binits{M.}},
\bauthor{\bsnm{{Okamoto}}, \binits{A.}},
\bauthor{\bsnm{{Ota}}, \binits{N.}},
\bauthor{\bsnm{{Ozaki}}, \binits{M.}},
\bauthor{\bsnm{{Paltani}}, \binits{S.}},
\bauthor{\bsnm{{Plucinsky}}, \binits{P.}},
\bauthor{\bsnm{{Pottschmidt}}, \binits{K.}},
\bauthor{\bsnm{{Sampson}}, \binits{M.}},
\bauthor{\bsnm{{Sasaki}}, \binits{T.}},
\bauthor{\bsnm{{Sato}}, \binits{K.}},
\bauthor{\bsnm{{Sato}}, \binits{R.}},
\bauthor{\bsnm{{Sato}}, \binits{T.}},
\bauthor{\bsnm{{Sawada}}, \binits{M.}},
\bauthor{\bsnm{{Seta}}, \binits{H.}},
\bauthor{\bsnm{{Shibano}}, \binits{Y.}},
\bauthor{\bsnm{{Shida}}, \binits{M.}},
\bauthor{\bsnm{{Shidatsu}}, \binits{M.}},
\bauthor{\bsnm{{Shigeto}}, \binits{S.}},
\bauthor{\bsnm{{Shinozaki}}, \binits{K.}},
\bauthor{\bsnm{{Shirron}}, \binits{P.}},
\bauthor{\bsnm{{Simionescu}}, \binits{A.}},
\bauthor{\bsnm{{Smith}}, \binits{R.}},
\bauthor{\bsnm{{Someya}}, \binits{K.}},
\bauthor{\bsnm{{Soong}}, \binits{Y.}},
\bauthor{\bsnm{{Sugawara}}, \binits{K.}},
\bauthor{\bsnm{{Sugawara}}, \binits{Y.}},
\bauthor{\bsnm{{Szymkowiak}}, \binits{A.}},
\bauthor{\bsnm{{Takahashi}}, \binits{H.}},
\bauthor{\bsnm{{Takeshima}}, \binits{T.}},
\bauthor{\bsnm{{Tamagawa}}, \binits{T.}},
\bauthor{\bsnm{{Tamura}}, \binits{K.}},
\bauthor{\bsnm{{Tanaka}}, \binits{T.}},
\bauthor{\bsnm{{Tanimoto}}, \binits{A.}},
\bauthor{\bsnm{{Terashima}}, \binits{Y.}},
\bauthor{\bsnm{{Tsuboi}}, \binits{Y.}},
\bauthor{\bsnm{{Tsujimoto}}, \binits{M.}},
\bauthor{\bsnm{{Tsunemi}}, \binits{H.}},
\bauthor{\bsnm{{Tsuru}}, \binits{T.}},
\bauthor{\bsnm{{Uchida}}, \binits{H.}},
\bauthor{\bsnm{{Uchida}}, \binits{Y.}},
\bauthor{\bsnm{{Uchiyama}}, \binits{H.}},
\bauthor{\bsnm{{Ueda}}, \binits{Y.}},
\bauthor{\bsnm{{Uno}}, \binits{S.}},
\bauthor{\bsnm{{Vink}}, \binits{J.}},
\bauthor{\bsnm{{Watanabe}}, \binits{T.}},
\bauthor{\bsnm{{Witthoeft}}, \binits{M.}},
\bauthor{\bsnm{{Wolfs}}, \binits{R.}},
\bauthor{\bsnm{{Yamada}}, \binits{S.}},
\bauthor{\bsnm{{Yamaoka}}, \binits{K.}},
\bauthor{\bsnm{{Yamasaki}}, \binits{N.}},
\bauthor{\bsnm{{Yamauchi}}, \binits{M.}},
\bauthor{\bsnm{{Yamauchi}}, \binits{S.}},
\bauthor{\bsnm{{Yanagase}}, \binits{K.}},
\bauthor{\bsnm{{Yaqoob}}, \binits{T.}},
\bauthor{\bsnm{{Yasuda}}, \binits{S.}},
\bauthor{\bsnm{{Yoshida}}, \binits{T.}},
\bauthor{\bsnm{{Yoshioka}}, \binits{N.}},
\bauthor{\bsnm{{Zhuravleva}}, \binits{I.}}:
\bctitle{{Status of x-ray imaging and spectroscopy mission (XRISM)}}.
In: \beditor{\bsnm{{den Herder}}, \binits{J.-W.A.}},
\beditor{\bsnm{{Nikzad}}, \binits{S.}},
\beditor{\bsnm{{Nakazawa}}, \binits{K.}} (eds.)
\bbtitle{Space Telescopes and Instrumentation 2020: Ultraviolet to Gamma Ray}.
\bsertitle{Society of Photo-Optical Instrumentation Engineers (SPIE) Conference Series},
vol. \bseriesno{11444},
p. \bfpage{1144422}
(\byear{2020}).
\doiurl{10.1117/12.2565812}
\end{bchapter}
\endbibitem

\bibitem[\protect\citeauthoryear{{Ishisaki} et~al.}{2022}]{Ishisaki2022}
\begin{bchapter}
\bauthor{\bsnm{{Ishisaki}}, \binits{Y.}},
\bauthor{\bsnm{{Kelley}}, \binits{R.L.}},
\bauthor{\bsnm{{Awaki}}, \binits{H.}},
\bauthor{\bsnm{{Balleza}}, \binits{J.C.}},
\bauthor{\bsnm{{Barnstable}}, \binits{K.R.}},
\bauthor{\bsnm{{Bialas}}, \binits{T.G.}},
\bauthor{\bsnm{{Boissay-Malaquin}}, \binits{R.}},
\bauthor{\bsnm{{Brown}}, \binits{G.V.}},
\bauthor{\bsnm{{Canavan}}, \binits{E.R.}},
\bauthor{\bsnm{{Cumbee}}, \binits{R.S.}},
\bauthor{\bsnm{{Carnahan}}, \binits{T.M.}},
\bauthor{\bsnm{{Chiao}}, \binits{M.P.}},
\bauthor{\bsnm{{Comber}}, \binits{B.J.}},
\bauthor{\bsnm{{Costantini}}, \binits{E.}},
\bauthor{\bsnm{{den Herder}}, \binits{J.-W.}},
\bauthor{\bsnm{{Dercksen}}, \binits{J.}},
\bauthor{\bsnm{{de Vries}}, \binits{C.P.}},
\bauthor{\bsnm{{DiPirro}}, \binits{M.J.}},
\bauthor{\bsnm{{Eckart}}, \binits{M.E.}},
\bauthor{\bsnm{{Ezoe}}, \binits{Y.}},
\bauthor{\bsnm{{Ferrigno}}, \binits{C.}},
\bauthor{\bsnm{{Fujimoto}}, \binits{R.}},
\bauthor{\bsnm{{Gorter}}, \binits{N.}},
\bauthor{\bsnm{{Graham}}, \binits{S.M.}},
\bauthor{\bsnm{{Grim}}, \binits{M.}},
\bauthor{\bsnm{{Hartz}}, \binits{L.S.}},
\bauthor{\bsnm{{Hayakawa}}, \binits{R.}},
\bauthor{\bsnm{{Hayashi}}, \binits{T.}},
\bauthor{\bsnm{{Hell}}, \binits{N.}},
\bauthor{\bsnm{{Hoshino}}, \binits{A.}},
\bauthor{\bsnm{{Ichinohe}}, \binits{Y.}},
\bauthor{\bsnm{{Ishida}}, \binits{M.}},
\bauthor{\bsnm{{Ishikawa}}, \binits{K.}},
\bauthor{\bsnm{{James}}, \binits{B.L.}},
\bauthor{\bsnm{{Kenyon}}, \binits{S.J.}},
\bauthor{\bsnm{{Kilbourne}}, \binits{C.A.}},
\bauthor{\bsnm{{Kimball}}, \binits{M.O.}},
\bauthor{\bsnm{{Kitamoto}}, \binits{S.}},
\bauthor{\bsnm{{Leutenegger}}, \binits{M.A.}},
\bauthor{\bsnm{{Maeda}}, \binits{Y.}},
\bauthor{\bsnm{{McCammon}}, \binits{D.}},
\bauthor{\bsnm{{Miko}}, \binits{J.J.}},
\bauthor{\bsnm{{Mizumoto}}, \binits{M.}},
\bauthor{\bsnm{{Okajima}}, \binits{T.}},
\bauthor{\bsnm{{Okamoto}}, \binits{A.}},
\bauthor{\bsnm{{Paltani}}, \binits{S.}},
\bauthor{\bsnm{{Porter}}, \binits{F.S.}},
\bauthor{\bsnm{{Sato}}, \binits{K.}},
\bauthor{\bsnm{{Sato}}, \binits{T.}},
\bauthor{\bsnm{{Sawada}}, \binits{M.}},
\bauthor{\bsnm{{Shinozaki}}, \binits{K.}},
\bauthor{\bsnm{{Shipman}}, \binits{R.}},
\bauthor{\bsnm{{Shirron}}, \binits{P.J.}},
\bauthor{\bsnm{{Sneiderman}}, \binits{G.A.}},
\bauthor{\bsnm{{Soong}}, \binits{Y.}},
\bauthor{\bsnm{{Szymkiewicz}}, \binits{R.}},
\bauthor{\bsnm{{Szymkowiak}}, \binits{A.E.}},
\bauthor{\bsnm{{Takei}}, \binits{Y.}},
\bauthor{\bsnm{{Tamura}}, \binits{K.}},
\bauthor{\bsnm{{Tsujimoto}}, \binits{M.}},
\bauthor{\bsnm{{Uchida}}, \binits{Y.}},
\bauthor{\bsnm{{Wasserzug}}, \binits{S.}},
\bauthor{\bsnm{{Witthoeft}}, \binits{M.C.}},
\bauthor{\bsnm{{Wolfs}}, \binits{R.}},
\bauthor{\bsnm{{Yamada}}, \binits{S.}},
\bauthor{\bsnm{{Yasuda}}, \binits{S.}}:
\bctitle{{Status of resolve instrument onboard X-Ray Imaging and Spectroscopy Mission (XRISM)}}.
In: \beditor{\bsnm{{den Herder}}, \binits{J.-W.A.}},
\beditor{\bsnm{{Nikzad}}, \binits{S.}},
\beditor{\bsnm{{Nakazawa}}, \binits{K.}} (eds.)
\bbtitle{Space Telescopes and Instrumentation 2022: Ultraviolet to Gamma Ray}.
\bsertitle{Society of Photo-Optical Instrumentation Engineers (SPIE) Conference Series},
vol. \bseriesno{12181},
p. \bfpage{121811}
(\byear{2022}).
\doiurl{10.1117/12.2630654}
\end{bchapter}
\endbibitem

\bibitem[\protect\citeauthoryear{{Kelley} et~al.}{2016}]{Kelley2016}
\begin{bchapter}
\bauthor{\bsnm{{Kelley}}, \binits{R.L.}},
\bauthor{\bsnm{{Akamatsu}}, \binits{H.}},
\bauthor{\bsnm{{Azzarello}}, \binits{P.}},
\bauthor{\bsnm{{Bialas}}, \binits{T.}},
\bauthor{\bsnm{{Boyce}}, \binits{K.R.}},
\bauthor{\bsnm{{Brown}}, \binits{G.V.}},
\bauthor{\bsnm{{Canavan}}, \binits{E.}},
\bauthor{\bsnm{{Chiao}}, \binits{M.P.}},
\bauthor{\bsnm{{Costantini}}, \binits{E.}},
\bauthor{\bsnm{{DiPirro}}, \binits{M.J.}},
\bauthor{\bsnm{{Eckart}}, \binits{M.E.}},
\bauthor{\bsnm{{Ezoe}}, \binits{Y.}},
\bauthor{\bsnm{{Fujimoto}}, \binits{R.}},
\bauthor{\bsnm{{Haas}}, \binits{D.}},
\bauthor{\bsnm{{den Herder}}, \binits{J.-W.}},
\bauthor{\bsnm{{Hoshino}}, \binits{A.}},
\bauthor{\bsnm{{Ishikawa}}, \binits{K.}},
\bauthor{\bsnm{{Ishisaki}}, \binits{Y.}},
\bauthor{\bsnm{{Iyomoto}}, \binits{N.}},
\bauthor{\bsnm{{Kilbourne}}, \binits{C.A.}},
\bauthor{\bsnm{{Kimball}}, \binits{M.O.}},
\bauthor{\bsnm{{Kitamoto}}, \binits{S.}},
\bauthor{\bsnm{{Konami}}, \binits{S.}},
\bauthor{\bsnm{{Koyama}}, \binits{S.}},
\bauthor{\bsnm{{Leutenegger}}, \binits{M.A.}},
\bauthor{\bsnm{{McCammon}}, \binits{D.}},
\bauthor{\bsnm{{Mitsuda}}, \binits{K.}},
\bauthor{\bsnm{{Mitsuishi}}, \binits{I.}},
\bauthor{\bsnm{{Moseley}}, \binits{H.}},
\bauthor{\bsnm{{Murakami}}, \binits{H.}},
\bauthor{\bsnm{{Murakami}}, \binits{M.}},
\bauthor{\bsnm{{Noda}}, \binits{H.}},
\bauthor{\bsnm{{Ogawa}}, \binits{M.}},
\bauthor{\bsnm{{Ohashi}}, \binits{T.}},
\bauthor{\bsnm{{Okamoto}}, \binits{A.}},
\bauthor{\bsnm{{Ota}}, \binits{N.}},
\bauthor{\bsnm{{Paltani}}, \binits{S.}},
\bauthor{\bsnm{{Porter}}, \binits{F.S.}},
\bauthor{\bsnm{{Sakai}}, \binits{K.}},
\bauthor{\bsnm{{Sato}}, \binits{K.}},
\bauthor{\bsnm{{Sato}}, \binits{Y.}},
\bauthor{\bsnm{{Sawada}}, \binits{M.}},
\bauthor{\bsnm{{Seta}}, \binits{H.}},
\bauthor{\bsnm{{Shinozaki}}, \binits{K.}},
\bauthor{\bsnm{{Shirron}}, \binits{P.J.}},
\bauthor{\bsnm{{Sneiderman}}, \binits{G.A.}},
\bauthor{\bsnm{{Sugita}}, \binits{H.}},
\bauthor{\bsnm{{Szymkowiak}}, \binits{A.E.}},
\bauthor{\bsnm{{Takei}}, \binits{Y.}},
\bauthor{\bsnm{{Tamagawa}}, \binits{T.}},
\bauthor{\bsnm{{Tashiro}}, \binits{M.}},
\bauthor{\bsnm{{Terada}}, \binits{Y.}},
\bauthor{\bsnm{{Tsujimoto}}, \binits{M.}},
\bauthor{\bsnm{{de Vries}}, \binits{C.P.}},
\bauthor{\bsnm{{Yamada}}, \binits{S.}},
\bauthor{\bsnm{{Yamasaki}}, \binits{N.Y.}},
\bauthor{\bsnm{{Yatsu}}, \binits{Y.}}:
\bctitle{{The Astro-H high resolution soft x-ray spectrometer}}.
In: \beditor{\bsnm{{den Herder}}, \binits{J.-W.A.}},
\beditor{\bsnm{{Takahashi}}, \binits{T.}},
\beditor{\bsnm{{Bautz}}, \binits{M.}} (eds.)
\bbtitle{Space Telescopes and Instrumentation 2016: Ultraviolet to Gamma Ray}.
\bsertitle{Society of Photo-Optical Instrumentation Engineers (SPIE) Conference Series},
vol. \bseriesno{9905},
p. \bfpage{99050}
(\byear{2016}).
\doiurl{10.1117/12.2232509}
\end{bchapter}
\endbibitem

\bibitem[\protect\citeauthoryear{{Arnaud}}{1996}]{Arnaud1996}
\begin{bchapter}
\bauthor{\bsnm{{Arnaud}}, \binits{K.A.}}:
\bctitle{{XSPEC: The First Ten Years}}.
In: \beditor{\bsnm{{Jacoby}}, \binits{G.H.}},
\beditor{\bsnm{{Barnes}}, \binits{J.}} (eds.)
\bbtitle{Astronomical Data Analysis Software and Systems V}.
\bsertitle{Astronomical Society of the Pacific Conference Series},
vol. \bseriesno{101},
p. \bfpage{17}
(\byear{1996})
\end{bchapter}
\endbibitem

\bibitem[\protect\citeauthoryear{{Wilms} et~al.}{2023}]{2023WilmsSIXTE}
\begin{bchapter}
\bauthor{\bsnm{{Wilms}}, \binits{J.}},
\bauthor{\bsnm{{Dauser}}, \binits{T.}},
\bauthor{\bsnm{{Dauner}}, \binits{L.}},
\bauthor{\bsnm{{Kirsch}}, \binits{C.}},
\bauthor{\bsnm{{Koenig}}, \binits{O.}},
\bauthor{\bsnm{{Lorenz}}, \binits{M.}},
\bauthor{\bsnm{{Peille}}, \binits{P.}},
\bauthor{\bsnm{{Cucchetti}}, \binits{E.}},
\bauthor{\bsnm{{Cobo}}, \binits{B.}},
\bauthor{\bsnm{{Ceballos}}, \binits{M.}},
\bauthor{\bsnm{{Rau}}, \binits{A.}},
\bauthor{\bsnm{{Ptak}}, \binits{A.}},
\bauthor{\bsnm{{Pottschmidt}}, \binits{K.}},
\bauthor{\bsnm{{Tzanavaris}}, \binits{P.}}:
\bctitle{{SIXTE - a generic X-ray mission simulator}}.
In: \bbtitle{AAS/High Energy Astrophysics Division}.
\bsertitle{AAS/High Energy Astrophysics Division},
vol. \bseriesno{20},
pp. \bfpage{103}--\blpage{71}
(\byear{2023})
\end{bchapter}
\endbibitem

\bibitem[\protect\citeauthoryear{Peille et~al.}{2016}]{Peille2016}
\begin{bchapter}
\bauthor{\bsnm{Peille}, \binits{P.}},
\bauthor{\bsnm{Ceballos}, \binits{M.T.}},
\bauthor{\bsnm{Cobo}, \binits{B.}},
\bauthor{\bsnm{Wilms}, \binits{J.}},
\bauthor{\bsnm{Bandler}, \binits{S.}},
\bauthor{\bsnm{Smith}, \binits{S.J.}},
\bauthor{\bsnm{Dauser}, \binits{T.}},
\bauthor{\bsnm{Brand}, \binits{T.}},
\bauthor{\bsnm{Hartog}, \binits{R.}},
\bauthor{\bsnm{Plaa}, \binits{J.}},
\bauthor{\bsnm{Barret}, \binits{D.}},
\bauthor{\bsnm{Herder}, \binits{J.-W.}},
\bauthor{\bsnm{Piro}, \binits{L.}},
\bauthor{\bsnm{Barcons}, \binits{X.}},
\bauthor{\bsnm{Pointecouteau}, \binits{E.}}:
\bctitle{{Performance assessment of different pulse reconstruction algorithms for the ATHENA X-ray Integral Field Unit}}.
In: \beditor{\bsnm{Herder}, \binits{J.-W.A.}},
\beditor{\bsnm{Takahashi}, \binits{T.}},
\beditor{\bsnm{Bautz}, \binits{M.}} (eds.)
\bbtitle{Space Telescopes and Instrumentation 2016: Ultraviolet to Gamma Ray},
vol. \bseriesno{9905},
p. \bfpage{99055}.
\bpublisher{SPIE}, \blocation{???}
(\byear{2016}).
\doiurl{10.1117/12.2232011} .
\bcomment{International Society for Optics and Photonics}
\end{bchapter}
\endbibitem

\bibitem[\protect\citeauthoryear{Kirsch et~al.}{2022}]{Kirsch2022}
\begin{barticle}
\bauthor{\bsnm{Kirsch}, \binits{C.}},
\bauthor{\bsnm{Lorenz}, \binits{M.}},
\bauthor{\bsnm{Peille}, \binits{P.}},
\bauthor{\bsnm{Dauser}, \binits{T.}},
\bauthor{\bsnm{Ceballos}, \binits{M.T.}},
\bauthor{\bsnm{Cobo}, \binits{B.}},
\bauthor{\bsnm{Merino-Alonso}, \binits{P.E.}},
\bauthor{\bsnm{Cucchetti}, \binits{E.}},
\bauthor{\bsnm{Smith}, \binits{S.J.}},
\bauthor{\bsnm{Gottardi}, \binits{L.}},
\bauthor{\bsnm{Hartog}, \binits{R.H.d.}},
\bauthor{\bsnm{Miniussi}, \binits{A.}},
\bauthor{\bsnm{Durkin}, \binits{M.}},
\bauthor{\bsnm{Pr{\^e}le}, \binits{D.}},
\bauthor{\bsnm{Wilms}, \binits{J.}}:
\batitle{The athena x-ifu instrument simulator xifusim}.
\bjtitle{Journal of Low Temperature Physics}
\bvolume{209}(\bissue{5}),
\bfpage{988}--\blpage{997}
(\byear{2022})
\doiurl{10.1007/s10909-022-02700-4}
\end{barticle}
\endbibitem

\bibitem[\protect\citeauthoryear{{Castellani} et~al.}{2022}]{Castellani2022}
\begin{bchapter}
\bauthor{\bsnm{{Castellani}}, \binits{F.}},
\bauthor{\bsnm{{Beaumont}}, \binits{S.}},
\bauthor{\bsnm{{Pajot}}, \binits{F.}},
\bauthor{\bsnm{{Roudil}}, \binits{G.}},
\bauthor{\bsnm{{Adams}}, \binits{J.}},
\bauthor{\bsnm{{Bandler}}, \binits{S.}},
\bauthor{\bsnm{{Chervenak}}, \binits{J.}},
\bauthor{\bsnm{{Daniel}}, \binits{C.}},
\bauthor{\bsnm{{Denison}}, \binits{E.V.}},
\bauthor{\bsnm{{Doriese}}, \binits{W.B.}},
\bauthor{\bsnm{{Dupieux}}, \binits{M.}},
\bauthor{\bsnm{{Durkin}}, \binits{M.}},
\bauthor{\bsnm{{Geoffray}}, \binits{H.}},
\bauthor{\bsnm{{Hilton}}, \binits{G.C.}},
\bauthor{\bsnm{{Murat}}, \binits{D.}},
\bauthor{\bsnm{{Parot}}, \binits{Y.}},
\bauthor{\bsnm{{Peille}}, \binits{P.}},
\bauthor{\bsnm{{Pr{\^e}le}}, \binits{D.}},
\bauthor{\bsnm{{Ravera}}, \binits{L.}},
\bauthor{\bsnm{{Reintsema}}, \binits{C.D.}},
\bauthor{\bsnm{{Sakai}}, \binits{K.}},
\bauthor{\bsnm{{Stevens}}, \binits{R.W.}},
\bauthor{\bsnm{{Ullom}}, \binits{J.N.}},
\bauthor{\bsnm{{Vale}}, \binits{L.R.}},
\bauthor{\bsnm{{Wakeham}}, \binits{N.}}:
\bctitle{{A 50 mK test bench for demonstration of the readout chain of Athena/X-IFU}}.
In: \beditor{\bsnm{{den Herder}}, \binits{J.-W.A.}},
\beditor{\bsnm{{Nikzad}}, \binits{S.}},
\beditor{\bsnm{{Nakazawa}}, \binits{K.}} (eds.)
\bbtitle{Space Telescopes and Instrumentation 2022: Ultraviolet to Gamma Ray}.
\bsertitle{Society of Photo-Optical Instrumentation Engineers (SPIE) Conference Series},
vol. \bseriesno{12181},
p. \bfpage{1218144}
(\byear{2022}).
\doiurl{10.1117/12.2630323}
\end{bchapter}
\endbibitem

\bibitem[\protect\citeauthoryear{{Beaumont} et~al.}{2022}]{Beaumont2022}
\begin{barticle}
\bauthor{\bsnm{{Beaumont}}, \binits{S.}},
\bauthor{\bsnm{{Pajot}}, \binits{F.}},
\bauthor{\bsnm{{Roudil}}, \binits{G.}},
\bauthor{\bsnm{{Adams}}, \binits{J.S.}},
\bauthor{\bsnm{{Bandler}}, \binits{S.R.}},
\bauthor{\bsnm{{Bertrand}}, \binits{B.}},
\bauthor{\bsnm{{Betancourt-Martinez}}, \binits{G.}},
\bauthor{\bsnm{{Castellani}}, \binits{F.}},
\bauthor{\bsnm{{Chervenak}}, \binits{J.A.}},
\bauthor{\bsnm{{Daniel}}, \binits{C.}},
\bauthor{\bsnm{{Denison}}, \binits{E.V.}},
\bauthor{\bsnm{{Doriese}}, \binits{W.B.}},
\bauthor{\bsnm{{Dupieux}}, \binits{M.}},
\bauthor{\bsnm{{Durkin}}, \binits{M.}},
\bauthor{\bsnm{{Geoffray}}, \binits{H.}},
\bauthor{\bsnm{{Hilton}}, \binits{G.C.}},
\bauthor{\bsnm{{Parot}}, \binits{Y.}},
\bauthor{\bsnm{{Peille}}, \binits{P.}},
\bauthor{\bsnm{{Pr{\^e}le}}, \binits{D.}},
\bauthor{\bsnm{{Ravera}}, \binits{L.}},
\bauthor{\bsnm{{Reintsema}}, \binits{C.D.}},
\bauthor{\bsnm{{Sakai}}, \binits{K.}},
\bauthor{\bsnm{{Smith}}, \binits{S.J.}},
\bauthor{\bsnm{{Stevens}}, \binits{R.W.}},
\bauthor{\bsnm{{Ullom}}, \binits{J.N.}},
\bauthor{\bsnm{{Vale}}, \binits{L.R.}},
\bauthor{\bsnm{{Wakeham}}, \binits{N.A.}}:
\batitle{{Development of an End-to-End Demonstration Readout Chain for Athena/X-IFU}}.
\bjtitle{Journal of Low Temperature Physics}
\bvolume{209}(\bissue{3-4}),
\bfpage{718}--\blpage{725}
(\byear{2022})
\doiurl{10.1007/s10909-022-02779-9}
\end{barticle}
\endbibitem

\bibitem[\protect\citeauthoryear{Molin et~al.}{2024}]{Molin2024}
\begin{bchapter}
\bauthor{\bsnm{Molin}, \binits{A.}},
\bauthor{\bsnm{Pajot}, \binits{F.}},
\bauthor{\bsnm{Audard}, \binits{M.}},
\bauthor{\bsnm{Barbera}, \binits{M.}},
\bauthor{\bsnm{Beaumont}, \binits{S.}},
\bauthor{\bsnm{Cuchetti}, \binits{E.}},
\bauthor{\bsnm{D'Andrea}, \binits{M.}},
\bauthor{\bsnm{Daniel}, \binits{C.}},
\bauthor{\bsnm{Hartog}, \binits{R.}},
\bauthor{\bsnm{Eckart}, \binits{M.}},
\bauthor{\bsnm{Ferrando}, \binits{P.}},
\bauthor{\bsnm{Kammoun}, \binits{E.}},
\bauthor{\bsnm{Leutenegger}, \binits{M.}},
\bauthor{\bsnm{Lotti}, \binits{S.}},
\bauthor{\bsnm{Mesnager}, \binits{J.-M.}},
\bauthor{\bsnm{Natalucci}, \binits{L.}},
\bauthor{\bsnm{Peille}, \binits{P.}},
\bauthor{\bsnm{Plaa}, \binits{J.}},
\bauthor{\bsnm{Pointecouteau}, \binits{E.}},
\bauthor{\bsnm{Porter}, \binits{F.}},
\bauthor{\bsnm{Sato}, \binits{K.}},
\bauthor{\bsnm{Wilms}, \binits{J.}},
\bauthor{\bsnm{Gottardi}, \binits{L.}},
\bauthor{\bsnm{Albouys}, \binits{V.}},
\bauthor{\bsnm{Barret}, \binits{D.}},
\bauthor{\bsnm{Cappi}, \binits{M.}},
\bauthor{\bsnm{Herder}, \binits{J.-W.}},
\bauthor{\bsnm{Piro}, \binits{L.}}:
\bctitle{{Ground calibration plan for the Athena/X-IFU microcalorimeter spectrometer}}.
In: \beditor{\bsnm{Herder}, \binits{J.-W.A.}},
\beditor{\bsnm{Nikzad}, \binits{S.}},
\beditor{\bsnm{Nakazawa}, \binits{K.}} (eds.)
\bbtitle{Space Telescopes and Instrumentation 2024: Ultraviolet to Gamma Ray},
vol. \bseriesno{13093},
p. \bfpage{1309310}.
\bpublisher{SPIE}, \blocation{???}
(\byear{2024}).
\doiurl{10.1117/12.3020082} .
\bcomment{International Society for Optics and Photonics}
\end{bchapter}
\endbibitem

\bibitem[\protect\citeauthoryear{{Barret} et~al.}{2024}]{Barret2024ExA....57...19B}
\begin{barticle}
\bauthor{\bsnm{{Barret}}, \binits{D.}},
\bauthor{\bsnm{{Albouys}}, \binits{V.}},
\bauthor{\bsnm{{Kn{\"o}dlseder}}, \binits{J.}},
\bauthor{\bsnm{{Loizillon}}, \binits{X.}},
\bauthor{\bsnm{{D'Andrea}}, \binits{M.}},
\bauthor{\bsnm{{Ardellier}}, \binits{F.}},
\bauthor{\bsnm{{Bandler}}, \binits{S.}},
\bauthor{\bsnm{{Dieleman}}, \binits{P.}},
\bauthor{\bsnm{{Duband}}, \binits{L.}},
\bauthor{\bsnm{{Dubbeldam}}, \binits{L.}},
\bauthor{\bsnm{{Macculi}}, \binits{C.}},
\bauthor{\bsnm{{Medinaceli}}, \binits{E.}},
\bauthor{\bsnm{{Pajot}}, \binits{F.}},
\bauthor{\bsnm{{Pr{\^e}le}}, \binits{D.}},
\bauthor{\bsnm{{Ravera}}, \binits{L.}},
\bauthor{\bsnm{{Thibert}}, \binits{T.}},
\bauthor{\bsnm{{Trallero}}, \binits{I.V.}},
\bauthor{\bsnm{{Webb}}, \binits{N.}}:
\batitle{{Life cycle assessment of the Athena X-ray integral field unit}}.
\bjtitle{Experimental Astronomy}
\bvolume{57}(\bissue{3}),
\bfpage{19}
(\byear{2024})
\doiurl{10.1007/s10686-024-09939-7}
{\href{https://arxiv.org/abs/2404.15122}{{arXiv:2404.15122}}}
{[astro-ph.IM]}
\end{barticle}
\endbibitem

\end{thebibliography}

\end{document}